\documentclass[authoryear,preprint,review,12pt]{elsarticle}

\usepackage{amssymb,amsmath}
\usepackage{mathabx}
\usepackage{morefloats}
\usepackage{rotating}
\journal{Icarus}

\begin{document}

\begin{frontmatter}

  \title{Star catalog position and proper motion corrections in asteroid astrometry}

  \author[jpl]{D. Farnocchia}
  \ead{Davide.Farnocchia@jpl.nasa.gov}
  \author[jpl]{S.~R. Chesley}
  \author[jpl]{A.~B. Chamberlin}
   \author[ifa]{D.~J. Tholen}

  \address[jpl]{Jet Propulsion Laboratory, California Institute of
    Technology, Pasadena, CA 91109, USA} 
  \address[ifa]{Institute for Astronomy, University of Hawaii,
    Honolulu, HI 96822, USA}

\begin{abstract}
  We provide a scheme to correct asteroid astrometric observations for
  star catalog systematic errors due to inaccurate star positions and
  proper motions. As reference we select the most accurate stars in
  the PPMXL catalog, i.e., those based on 2MASS astrometry.  We
  compute position and proper motion corrections for 19 of the most
  used star catalogs. The use of these corrections provides better
  ephemeris predictions and improves the error statistics of
  astrometric observations, e.g., by removing most of the regional
  systematic errors previously seen in Pan-STARRS PS1 asteroid
  astrometry. The correction table is publicly available at
  ftp://ssd.jpl.nasa.gov/pub/ssd/debias/debias\_2014.tgz and can be
  freely used in orbit determination algorithms to obtain more
  reliable asteroid trajectories. 
\end{abstract}


\end{frontmatter}

\section{Introduction}
Whenever we compute an asteroid's orbit, it comes with an uncertainty
region due to the limited accuracy of the available observations.  In
other words, orbits are only known in a statistical sense and the
accuracy of the related probabilistic interpretation relies heavily on
the observation accuracy and error modeling. Therefore, it is
important to apply an appropriate statistical treatment to the
observations used to compute the orbit.

The vast majority of asteroid astrometry is given by optical
observations, i.e., each observation provides two angular
measurements, typically right ascension (RA) and declination (DEC) in
the equatorial reference frame J2000, describing the position of an
asteroid on the celestial sphere at a specified time. Such
measurements are obtained with respect to nearby reference stars,
whose positions are provided by a reference star catalog.  In general,
the more accurate the star catalog, the more accurate the observation.

Despite the common assumption that observation errors have zero mean,
\citet{carpino03} show that asteroid astrometry is significantly
biased and suggest the reason is
the presence of systematic errors in the star catalogs used to reduce
the astrometry.

\citet{cbm10} computed star catalog systematic errors for USNO-A1.0
\citep{usnoa1}, USNO-A2.0 \citep{usnoa2}, USNO-B1.0 \citep{usnob1},
UCAC2 \citep{ucac2}, and Tycho-2 \citep{tycho2} by comparing each of
these catalogs to 2MASS \citep{2mass}. Despite the lack of proper
motions, 2MASS was chosen as the reference catalog because of its very
accurate star positions at epoch J2000.0 and high spatial
density. \citet{cbm10} showed that correcting asteroid astrometry
using their computed biases leads to significantly lower systematic
errors and statistically better ephemeris predictions.

Pan-STARRS PS1 \citep{ps1} is one of the most accurate asteroid
surveys with an astrometric quality of the order of 0.1''. Although this
survey uses 2MASS as reference catalog for the astrometric reduction,
\citet{milani12} found that Pan-STARRS PS1 data have surprisingly
high biases on the order of 0.05--0.1'' with a strong regional
dependence. \citet{tholen_2mass} show that the lack of proper motion
in 2MASS is likely to be the cause of the Pan-STARRS PS1 astrometry
systematic errors and signatures. Moreover, they suggest that PPMXL
\citep{ppmxl} be used as reference catalog because of its spatial
density, accuracy comparable to that of 2MASS, and availability of
proper motion information.

Since the lack of proper motion can be significant for high quality
observations, in this paper we describe how to correct asteroid
observations for both position and proper motion errors.  Moreover, we
perform this analysis for a more comprehensive list of star catalogs
than that considered in \citet{cbm10}.

\section{Asteroid astrometry}
As of January 2014 more than 600,000 asteroids have been designated,
$\sim$60\% of which are numbered. The number of asteroid optical
observations is already larger than 100,000,000 and increases every
day. Observers submit their observations to the Minor Planet Center
(MPC)\footnote{http://www.minorplanetcenter.net/} and usually provide
information on the catalog used to perform the astrometric
reduction. The MPC in turn makes the catalog information publicly
available by using an alphabetical
flag\footnote{http://www.minorplanetcenter.net/iau/info/CatalogueCodes.html}.

Table~\ref{t:catalogs} shows the MPC flag, the number of stars, and
the number of asteroid observations different catalogs. We only
consider the catalogs for which the number of asteroid observations
reported to the MPC with the corresponding catalog flag was larger
than 40,000 as of January 2014.  We also included the GSC-1.2
\citep{gsc1.2}
catalog to complete the GSC-1 catalog series. The most used catalog is
USNO-A2.0, with more than 40,000,000 asteroid observations. 2MASS,
which was used as the reference catalog by \citet{cbm10}, is the
fourth most used catalog and the related astrometry is dominated by
Pan-STARRS PS1 observations (more than 75\% of the
sample). Observations reported with code `z' were reduced with one of
the GSC catalogs, but we do not know which one.

\begin{table}\small
\begin{center}
\begin{tabular}{lccccc}
\hline
Catalog & MPC & Number & \multicolumn{2}{c}{Asteroid observations} & Reference\\
             & flag  & of stars & Count &  \%   & \\
\hline
USNO-A2.0 & c & 526,280,881 & 40,408,360 & 38.47 & \citet{usnoa2}\\
UCAC-2 & r & 48,330,571 & 29,793,925 & 28.37 & \citet{ucac2}\\
USNO-B1.0 & o & 1,045,175,762 & 12,834,999 & 12.22 & \citet{usnob1}\\
2MASS & L & 470,992,970 & 8,136,250 & 7.75 & \citet{2mass}\\
UCAC-4 & q & 113,780,093 & 2,629,456 & 2.50 &\citet{ucac4}\\
UCAC-3 & u & 100,766,420 & 2,228,325 & 2.12 &\citet{ucac3}\\
USNO-A1.0 & a & 488,006,860 & 2,193,938 & 2.08 &\citet{usnoa1}\\
USNO-SA2.0 & d & 55,368,239 & 1,698,129 & 1.62 &\citet{usnoa2}\\
GSC-1.1 & i & 18,836,912 & 614,617 & 0.59 &\citet{gsc1.1}\\
UCAC-1 & e & 27,425,433 & 501,774 & 0.48 &\citet{ucac1}\\
SDSS-DR7 & N & 357,175,411 & 479,914 & 0.46 &\citet{sloan7}\\
GSC-ACT & m & 18,836,912 & 404,473 & 0.39 &\citet{gsc_act}\\
CMC-14 & w & 95,858,475 & 361,928 & 0.34 &\citet{cmc14}\\
Tycho-2 & g & 2,430,468 & 355,813 & 0.34 &\citet{tycho2}\\
USNO-SA1.0 & b & 54,787,624 & 337,561 & 0.32 &\citet{usnoa1}\\
GSC (unspecified) & z & N/A & 288,156 & 0.27 &N/A\\
ACT & l & 988,758 & 117,638 & 0.11 &\citet{act}\\
PPMXL & t & 910,468,688 & 88,328 & 0.08 &\citet{ppmxl}\\
NOMAD & v & 1,117,612,732 & 58,266 & 0.06 &\citet{nomad}\\
PPM & p & 378,910& 41,468 & 0.04 &\citet{ppm}\\
GSC-1.2 & j & 18,841,548 & 16,975 & 0.02 & \citet{gsc1.2}\\
\hline
\end{tabular}
\end{center}
\caption{Star catalogs and MPC flags. The number
  of asteroid observations for each catalog account for all the astrometry
  available up to January 7, 2014.}
\label{t:catalogs}
\end{table}

\section{Star catalog position and proper motion corrections}
To correct asteroid optical astrometry for star catalog systematic
errors, we need to select a reference for comparison with the other
catalogs. Such a selection is far from easy. Hipparcos
\citep{hipparcos} and Tycho2 are space-based, so they are not subject
to differential refraction corrections as ground-based observations
are, possibly making them the best available catalogs. However, a
reference catalog should be both dense and accurate and neither
Tycho-2 nor Hipparcos are dense enough. As shown in
Table~\ref{t:catalogs}, the catalogs with the largest number of stars
include USNO-A1.0 \citep{usnoa1}, USNO-A2.0 \citep{usnoa2}, USNO-B1.0
\citep{usnob1}, 2MASS \citep{2mass}, PPMXL \citep{ppmxl}, and NOMAD
\citep{nomad}.  \citet{cbm10} proved that the USNO catalogs are
affected by systematic errors in position as large as 1--2''. NOMAD is
a simple merge of the Hipparcos, Tycho-2, UCAC2, and USNO-B1.0 and is
therefore still affected by the biases present in
USNO-B1.0. \citet{tholen_2mass} show that 2MASS is not the appropriate
choice because of the lack of proper motion. PPMXL \citep{ppmxl} is
also a merge of 2MASS and USNO-B1.0, but it includes proper motions
and a critical reprocessing of star positions from 2MASS and
USNO-B1.0. Therefore, PPXML seems a sensible choice for a reference
catalog. However, tests similar to one presented in
Sec.~\ref{s:pred_test} were not satisfactory as we found that
correcting with respect to PPMXL rather than 2MASS \citep[as
in][]{cbm10} can provide less accurate predictions. As described by
\citet{ppmxl}, more than 50\% of PPMXL stars are based on USNO-B1.0
and are not accurate enough for our purposes. To fix this problem, we
selected as a reference catalog the subset of PPMXL corresponding to
over 400 millions stars derived from 2MASS. This reference benefits
from the accuracy of 2MASS star positions and yet accounts for proper
motions.

As in \citet{cbm10}, to compare the different star catalogs to our
reference catalog we divided the celestial sphere into 49,152
equal-area tiles ($\sim$0.8 deg$^2$) using the JPL HEALPix package
\citep{healpix}.  For all the catalogs analyzed, we took star
positions at epoch J2000.0. To identify stars in common within a given
tile we used a spatial correlation of 2''. Whenever more than one
identification with the same star is possible, we need to be careful
and avoid spurious identifications. If $d_i, i=1,N$ are the distances
between the considered star and the matches in the reference catalog, as
a safety measure we selected the identification $j$ only if $d_j < 0.2
d_i$ for $i=1, N$ and $i \ne j$. If none of the identifications met this
condition we rejected all the identification to avoid including
spurious matches in our analysis.

We also made sure that stars in the reference catalog were not paired
to more than one star. For each tile we computed the average
correction in position and proper motion for both right ascension and
declination. Because of the present biases, for some catalogs the 2''
spatial correlation may not be enough to find matching
stars. Therefore, we applied the procedure iteratively, i.e., we
corrected the stars in the catalog to be debiased by subtracting the
systematic error for the corresponding tile found at the previous
iteration.

At the end of the process, for each given tile and catalog we have a
correction in RA and DEC at epoch J2000.0,
$(\Delta \text{RA}_{2000}, \Delta \text{DEC}_{2000})$, and proper motion
corrections $(\Delta \mu_{\text{RA}}, \Delta \mu_{\text{DEC}})$. These numbers
can be used to correct asteroid astrometric observations by subtracting the
following quantities:
\begin{eqnarray*}
\Delta\text{RA} & = & \Delta \text{RA}_{2000} + \Delta \mu_{\text{RA}} (t - 2000.0)\\
\Delta\text{DEC} & = & \Delta \text{DEC}_{2000} + \Delta \mu_{\text{DEC}} (t - 2000.0)\\
\end{eqnarray*}
where $t$ is the observation epoch, and $\Delta \text{RA}_{2000}$,
$\Delta \text{DEC}_{2000}$, $\Delta \mu_{\text{RA}}$, and $\Delta
\mu_{\text{RA}}$ are the position and proper motion corrections for
the tile containing the astrometric observation. Note that
$\Delta{\text{RA}}$, $\Delta{\text{RA}}_{2000}$, and $\Delta
\mu_{\text{RA}}$ account for the spherical metric factor
$\cos\text{DEC}$. Of course, the successful application of these
corrections relies on the accuracy of the catalog information provided
in the MPC observation database. The star position and
proper motion correction table is publicly available at
ftp://ssd.jpl.nasa.gov/pub/ssd/debias/debias\_2014.tgz. The main
differences with respect to the \citet{cbm10} debiasing scheme are:
\begin{itemize}
\item our reference catalog is not 2MASS but the subset of PPMXL
  based on 2MASS astrometry;
\item the present debiasing scheme accounts for both position and
  proper motion errors, while \citet{cbm10} only considered position
  errors;
\item we compute corrections for a more comprehensive list of star
  catalogs.
\end{itemize}

For each analyzed catalog, Table~\ref{t:cat_rms} reports the average
size of the corrections in terms of RMS, e.g.
\begin{equation}
\overline{\Delta {\text{RA}}}_{2000} = \sqrt{\frac{1}{n_{tiles}}\sum_{i=1}^{n_{tiles}}
 \left(\Delta {\text{RA}}_{2000}\right)_i^2}.
\end{equation}

\begin{table}
\begin{center}
\begin{tabular}{l|cc|ccc|c}
  \hline
Catalog & $\overline{\Delta{\text{RA}}}_{2000}$ &
$\overline{\Delta{\text{DEC}}}_{2000}$ & PM & $\overline{\Delta\mu}_{\text{RA}}$ &
$\overline{\Delta\mu}_{\text{DEC}}$ & Sky\\
& [arcsec] & [arcsec] & inc. & [mas/yr] & [mas/yr] & coverage\\
%
\hline
Tycho-2 & 0.02 & 0.02 & Yes & 0.7 & 0.7 & 100\%\\
ACT & 0.02 & 0.02 & Yes & 1.5 & 1.4 & 100\%\\
2MASS & 0.03 & 0.02 & No & 5.8 & 6.4 & 100\%\\
USNO-A1.0 & 0.45 & 0.37 & No & 5.1 & 5.7 & 100\%\\
USNO-SA1.0 & 0.45 & 0.37 & No & 5.0 & 5.5 & 100\%\\
USNO-A2.0 & 0.21 & 0.24 & No & 5.1 & 5.7 & 100\%\\
USNO-SA2.0 & 0.21 & 0.24 & No & 5.0 & 5.6 & 100\%\\
USNO-B1.0 & 0.12 & 0.17 & Yes & 4.4 & 4.9 & 100\%\\
UCAC-1 & 0.03 & 0.03 & Yes & 5.8 & 7.4 & 39\%\\ 
UCAC-2 & 0.01 & 0.01 & Yes & 2.5 & 2.2 & 88\%\\ 
UCAC-3 & 0.02 & 0.02 & Yes & 5.0 & 4.6 & 100\%\\ 
UCAC-4 & 0.02 & 0.02 & Yes & 2.2 & 2.5 & 100\%\\ 
GSC-1.1 & 0.47 & 0.38 & No & 6.8 & 6.6 & 100\%\\
GSC-1.2 & 0.20 & 0.18 & No & 6.7 & 6.6 & 100\%\\
GSC-ACT & 0.15 & 0.13 & No & 6.7 & 6.6 & 100\%\\
NOMAD & 0.10 & 0.15 & Yes & 3.7 & 4.3 & 100\%\\
PPM & 0.23 & 0.24 & Yes & 4.1 & 4.2 & 100\%\\
CMC-14 & 0.03 & 0.04 & No & 6.3 & 7.0 & 62\%\\
SDSS-DR7 & 0.05 & 0.07 & Yes & 2.6 & 3.2 & 31\%\\
\hline
\end{tabular}
\end{center}
\caption{For each analyzed catalog columns are: average corrections in
  position (right ascension and declination), information on whether
  or not the catalog includes proper motions, average corrections in
  proper motion (right ascension and declination), and the fraction of
  the sky covered by the catalog.}
\label{t:cat_rms}
\end{table}

Figures~\ref{f:2mass}--\ref{f:sdss7} depict sky maps of the position
and proper motion corrections for the analyzed catalogs, which we
discuss in more detail in the following subsections. Note that the
color scale is not the same for all catalogs to reveal the regional
structures of the position and proper motion corrections.

As shown in Table~\ref{t:counts}, the right ascension and declination
corrections are not available for 1.65\% of the reported astrometry.
In most of these cases ($>$ 1,000,000 observations) we cannot apply
corrections because there is no catalog information. Moreover, almost
290,000 observations were reported as reduced using a GSC catalog,
without specifying which specific GSC catalog was used. Finally, about
100,000 observations were reduced with catalogs not included in our
analysis.

\begin{sidewaystable}\small
\begin{center}
\begin{tabular}{lc|c|cc|c|c}
\hline
Type & MPC flag & Unknown catalog & \multicolumn{2}{c|}{Known Catalog}
& Total & Dates\\
 & & & Corrected & Not corrected & & \\
\hline
CCD & C & 530,374 & 99,938,430 & 279,375 & 100,748,179 & 1986--2014\\
Corrected CCD & c & 440 & 1,214,291 & 1,778 & 1,216,509 & 1991--2013\\
Former B1950 & A & 554,220 & 10,036 & 82,519 & 646,775 & 1802--1999\\
Photographic & ' ' & 207,812 & 118,743 & 27,258 & 353,813 & 1898--2012\\
Meridian or transit circle & T & 25,357 & 1,611 & 0 & 26,968 & 1984--2005\\
Micrometer & M & 0 & 13,008 & 0 & 13,008 & 1845--1954\\
Suppressed & X & 6,538 & 4,289 & 2,183 & 13,010 & 1891--2010\\
Suppressed & x & 5 & 3,763 & 0 & 3,768 & 1996--2010 \\
Hipparcos & H & 5,494 & 0 & 0 & 5,494 & 1989--1993\\
Occultation & E & 1,005 & 871 & 68 & 1,944 & 1961--2013\\ 
Satellite & S & 4,331 & 1,996,679 & 1,335 & 2,002,345 & 1994--2013\\
Roving observer & V & 102 & 92 & 0 & 194 & 2000--2012\\
Normal place & N & 0 & 37 & 0 & 37 & 1906--1923\\
Mini-normal place & n & 0 & 273 & 0 & 273 & 2009--2013\\
Encoder & e & 1 & 14 & 1 & 16 & 1993--1995\\
\hline
Total & & 1,335,679 & 103,302,137 & 394,517 & 105,032,333\\
& & 1.27\% & 98.35\% & 0.38\% & \\
\hline
\end{tabular}
\end{center}
\caption{For each optical observation type, number of observations
  with unknown catalog, with known catalog and computed correction
  tables, and with known catalog and no corrections available. As
  suggested by the MPC, X and x-type observations should not be used
  in the orbital fits. Observations are as of Jan 7, 2014.}
\label{t:counts}
\end{sidewaystable}

\subsection{PPMXL and 2MASS}
Since our reference catalog is a subset of PPMXL, the comparison
yields no corrections for PPMXL. Still, it is worth pointing out that
astrometry reduced with PPMXL can suffer from the lower accuracy of
USNO-B1.0 based stars.

\begin{figure}
\centerline{\includegraphics[width=0.6\textwidth]{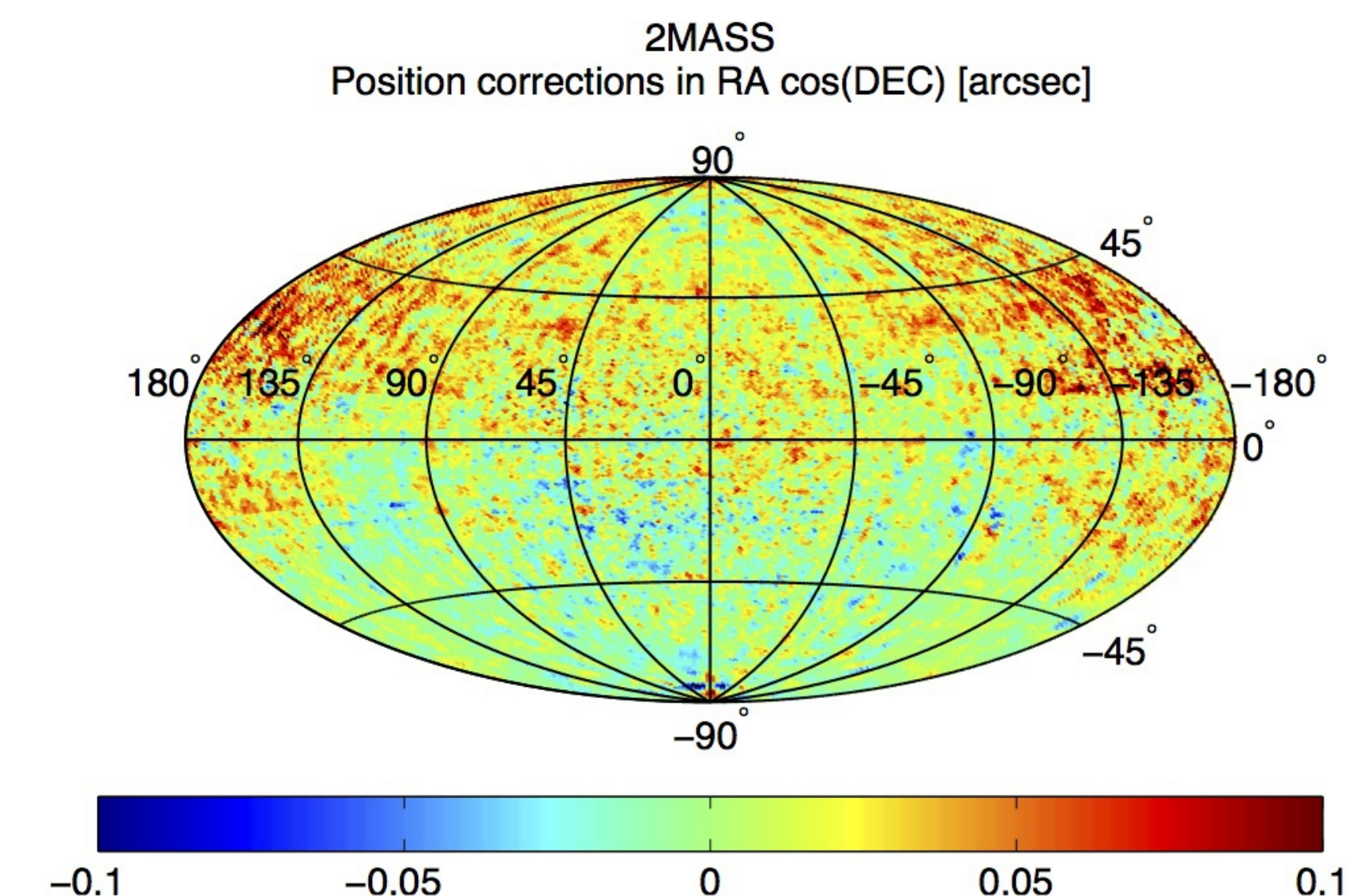}
\includegraphics[width=0.6\textwidth]{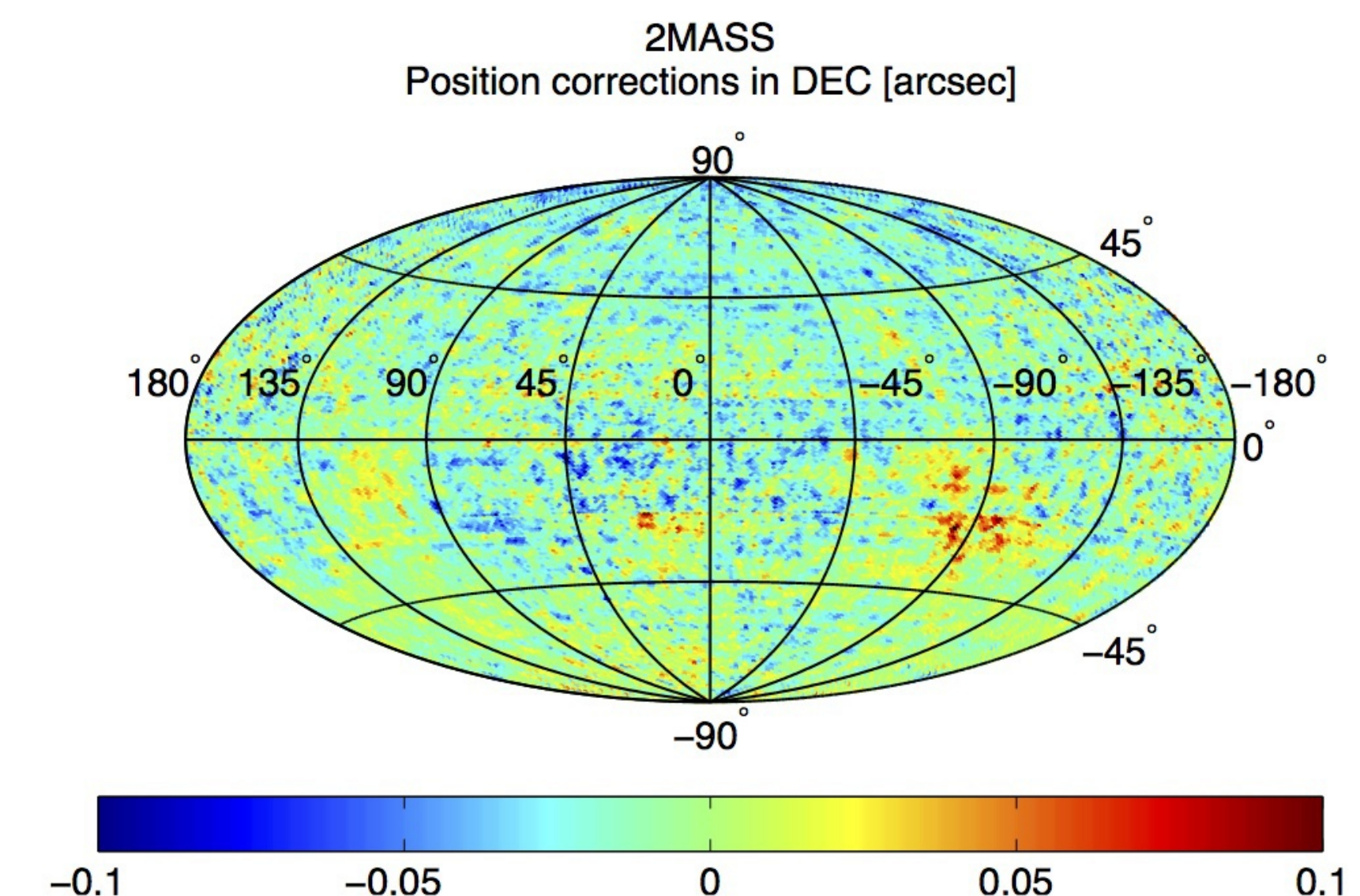}}
\centerline{\includegraphics[width=0.6\textwidth]{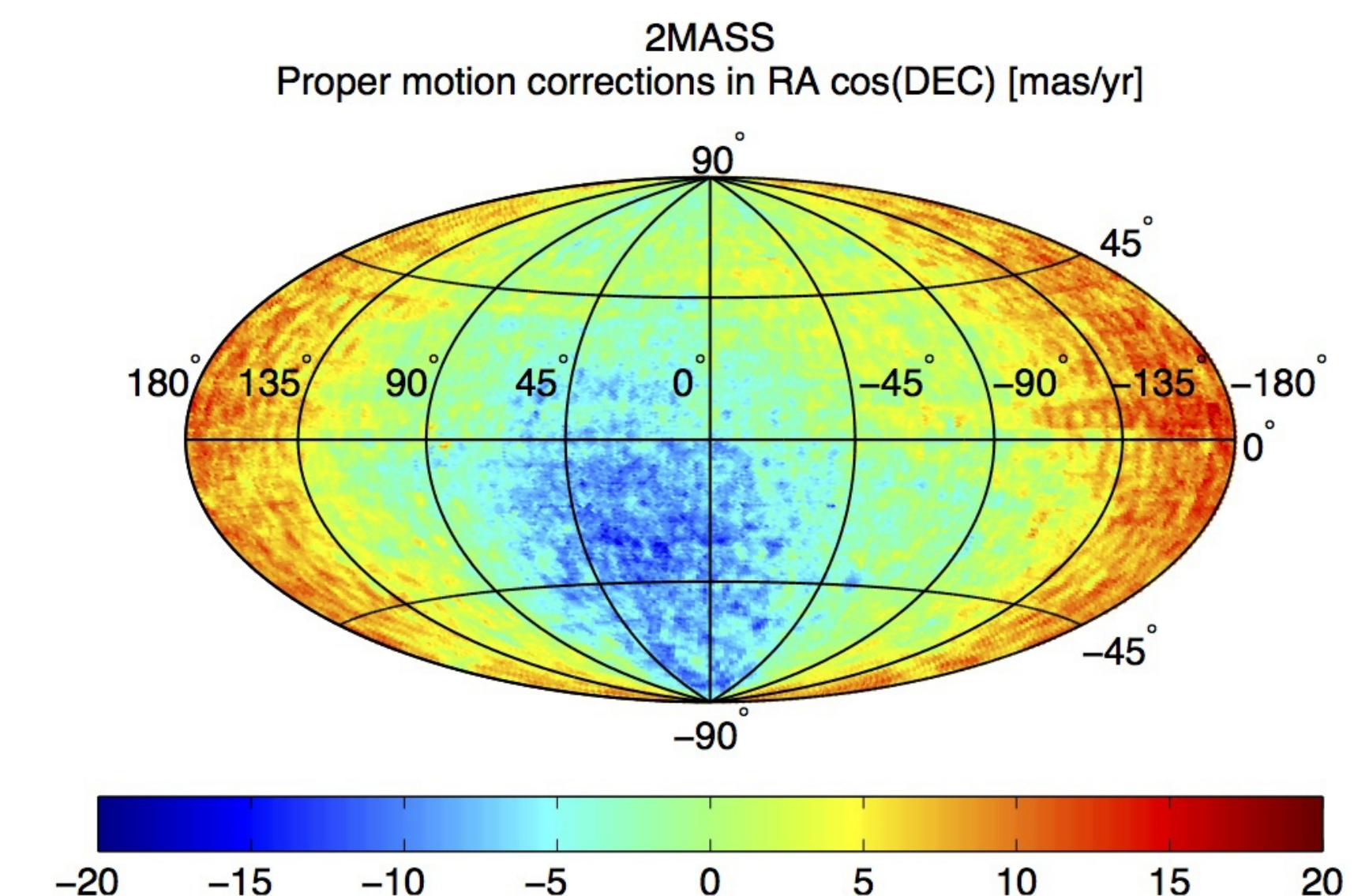}
\includegraphics[width=0.6\textwidth]{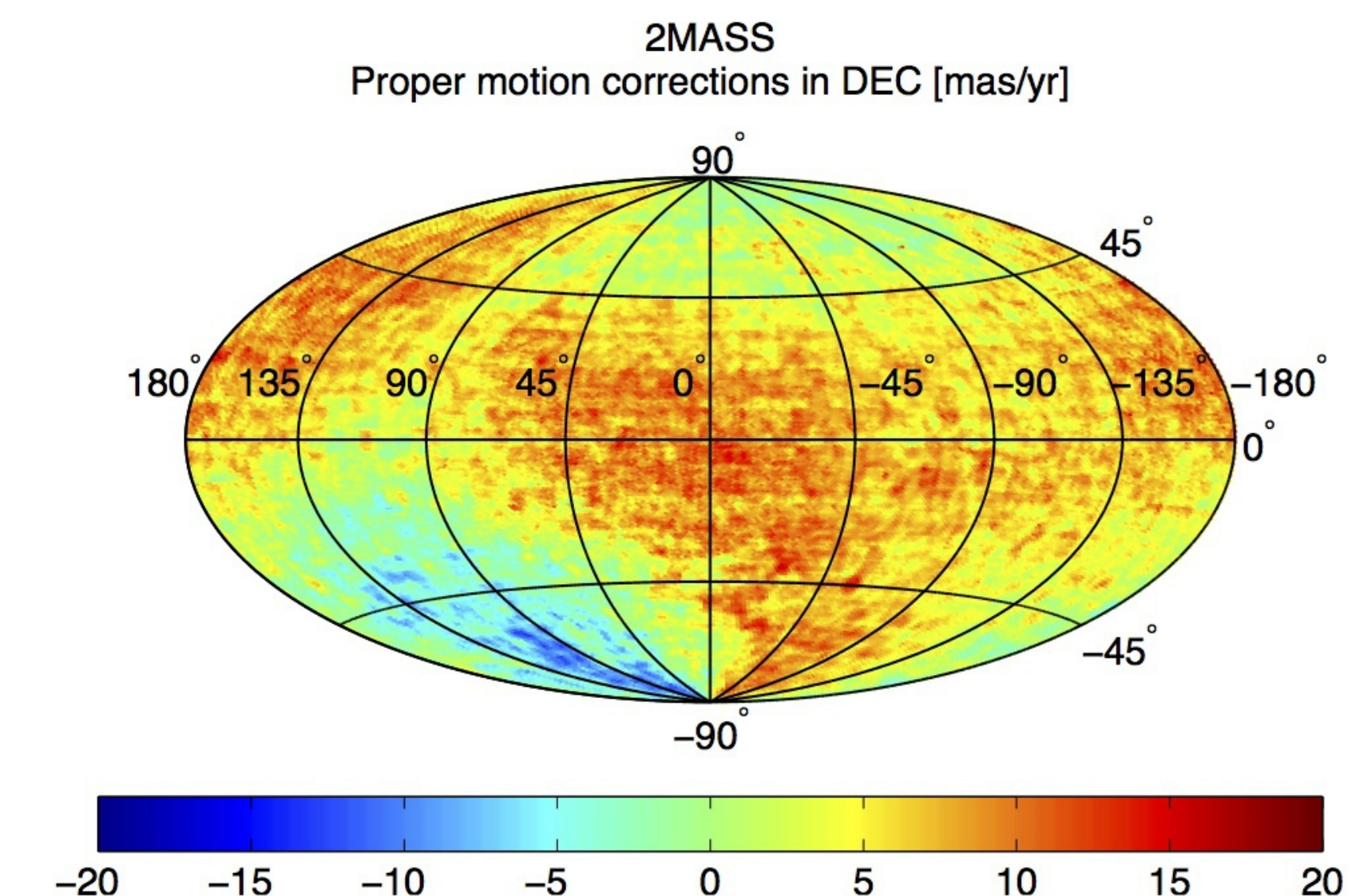}}
\caption{Top: J2000.0 position corrections in right ascension (left)
  and declination (right) for 2MASS. Bottom: proper motion corrections
  in right ascension (left) and declination (right) for
  2MASS.}\label{f:2mass}
\end{figure}

Due to our choice for the reference catalog, we expect small
differences in the 2MASS star positions. As a matter of fact, we have
position differences of the order of 0.01'' -- well consistent
with the 2MASS stated accuracy of $\sim 0.07''$ \citep{2mass}. Though
these corrections are small, the top panels of Fig.~\ref{f:2mass} show
some regional dependence, which may be due to the lack of proper
motion for the time interval in which star positions were integrated.

The bottom panels of Fig.~\ref{f:2mass} show the proper motion
corrections to be applied to 2MASS. Since 2MASS does not have proper
motions, these two panels give the proper motion distribution for
stars that 2MASS and PPMXL have in common. There is an evident
regional dependence and it is clear that the lack of proper motion may
cause significant position errors if the observation epoch is not
close to J2000.0.  Thus, we apply both position and proper motion
corrections to observations reduced with 2MASS.

\subsection{Tycho-2 and ACT}
\begin{figure}
\centerline{\includegraphics[width=0.6\textwidth]{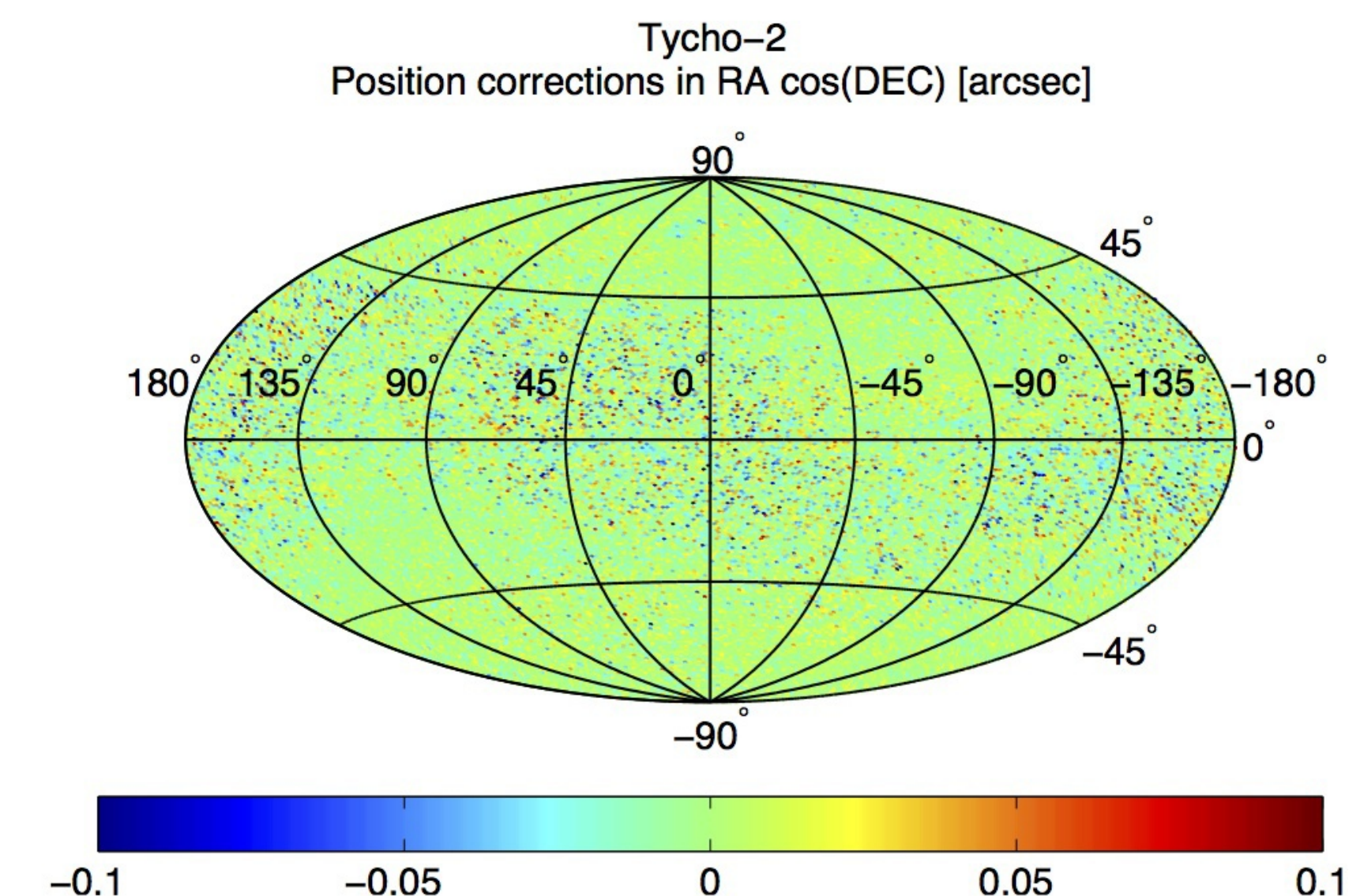}\includegraphics[width=0.6\textwidth]{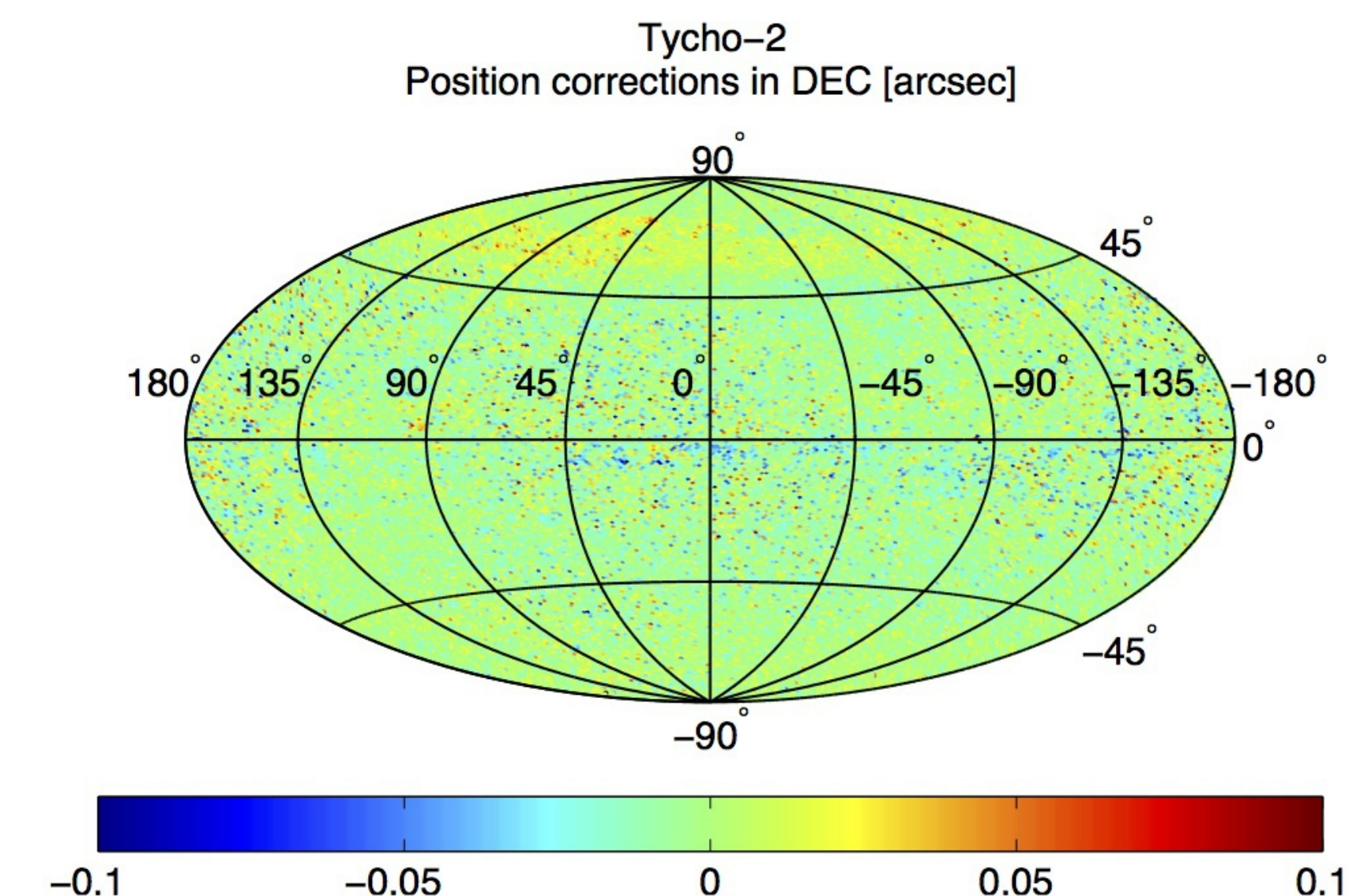}}
\centerline{\includegraphics[width=0.6\textwidth]{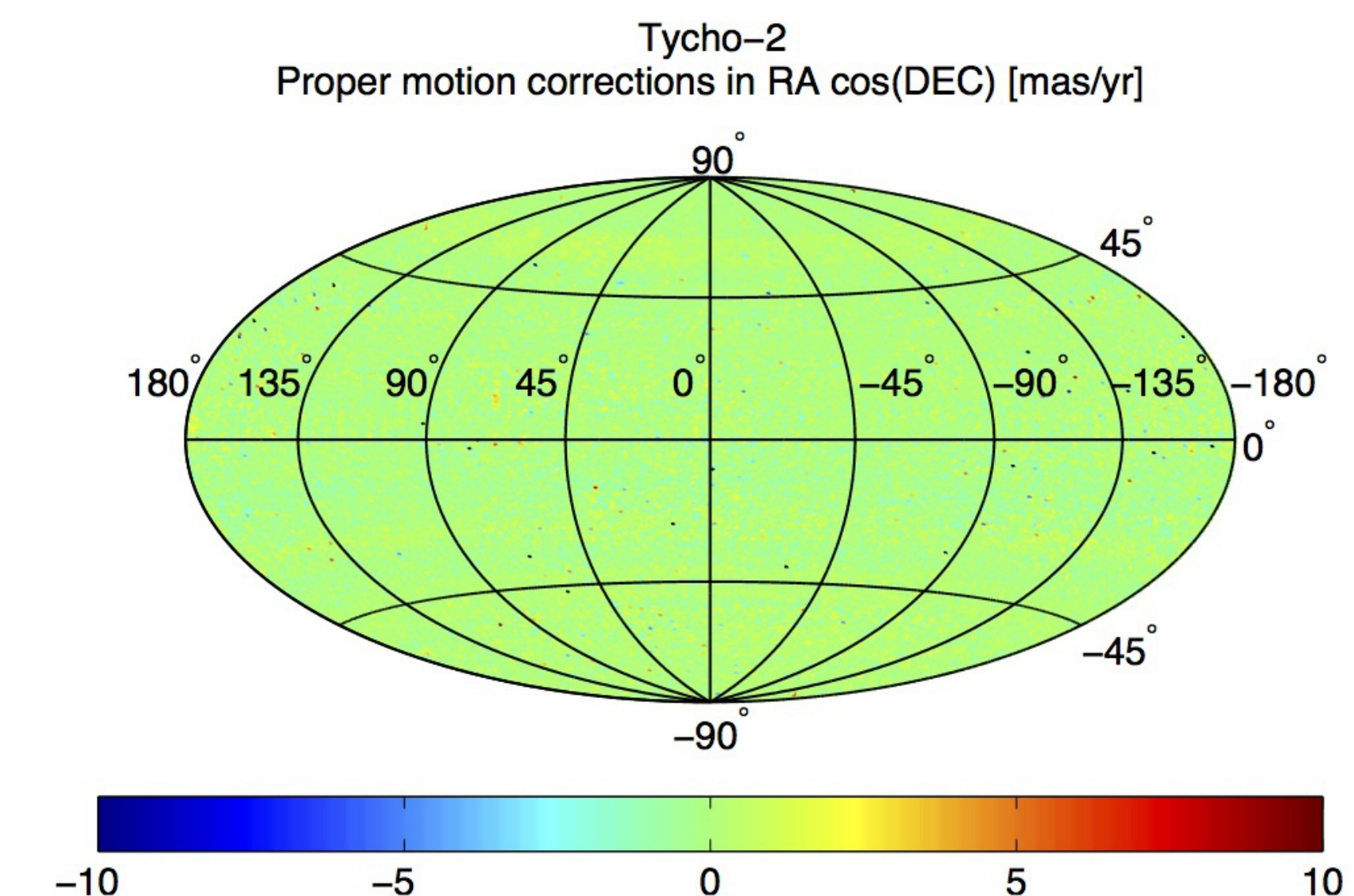}\includegraphics[width=0.6\textwidth]{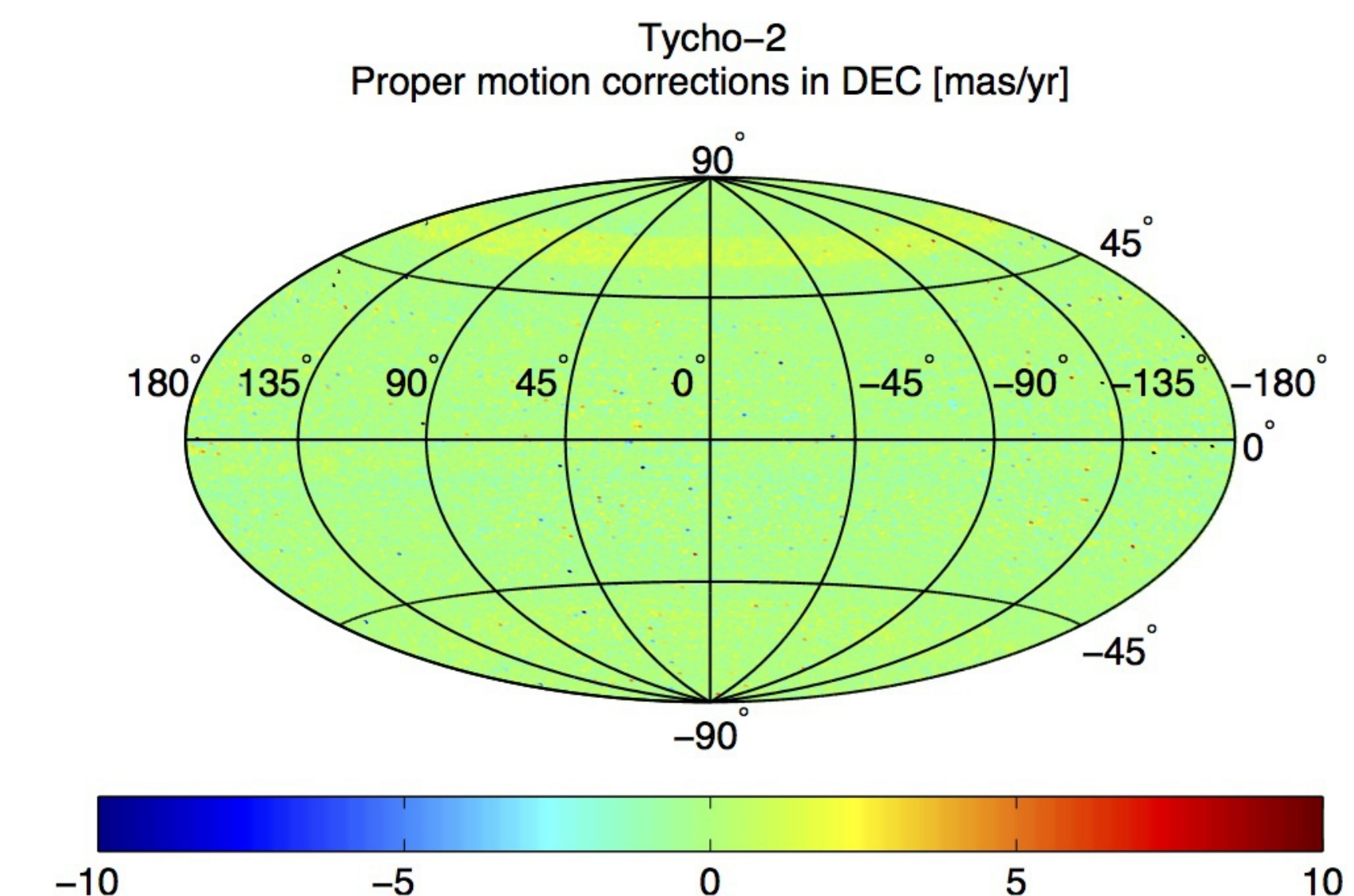}}
\caption{Top: J2000.0 position corrections in right ascension (left)
  and declination (right) for Tycho-2. Bottom: proper motion
  corrections in right ascension (left) and declination (right) for
  Tycho-2.}\label{f:tycho2}
\end{figure}

\begin{figure}
\centerline{\includegraphics[width=0.6\textwidth]{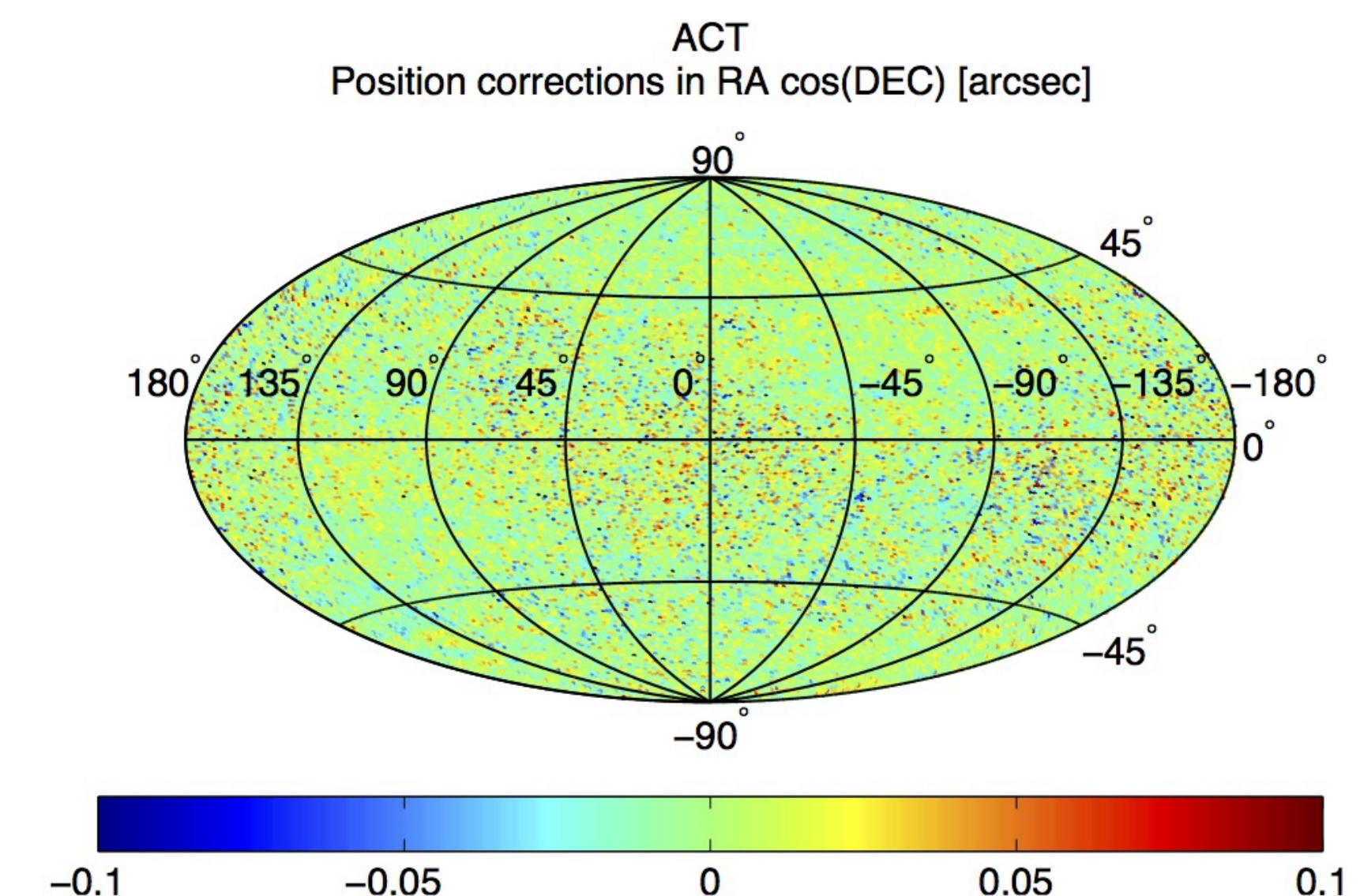}\includegraphics[width=0.6\textwidth]{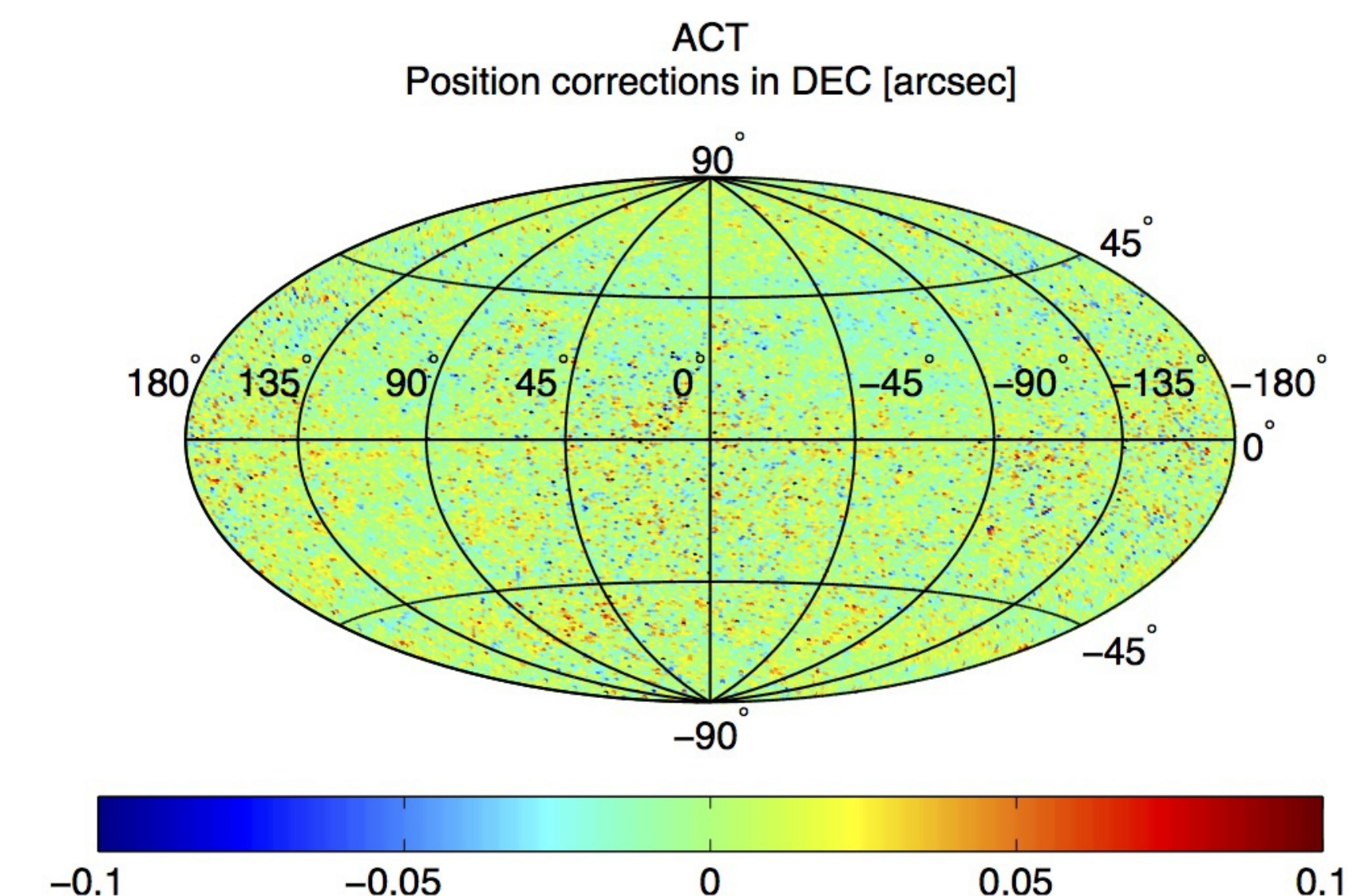}}
\centerline{\includegraphics[width=0.6\textwidth]{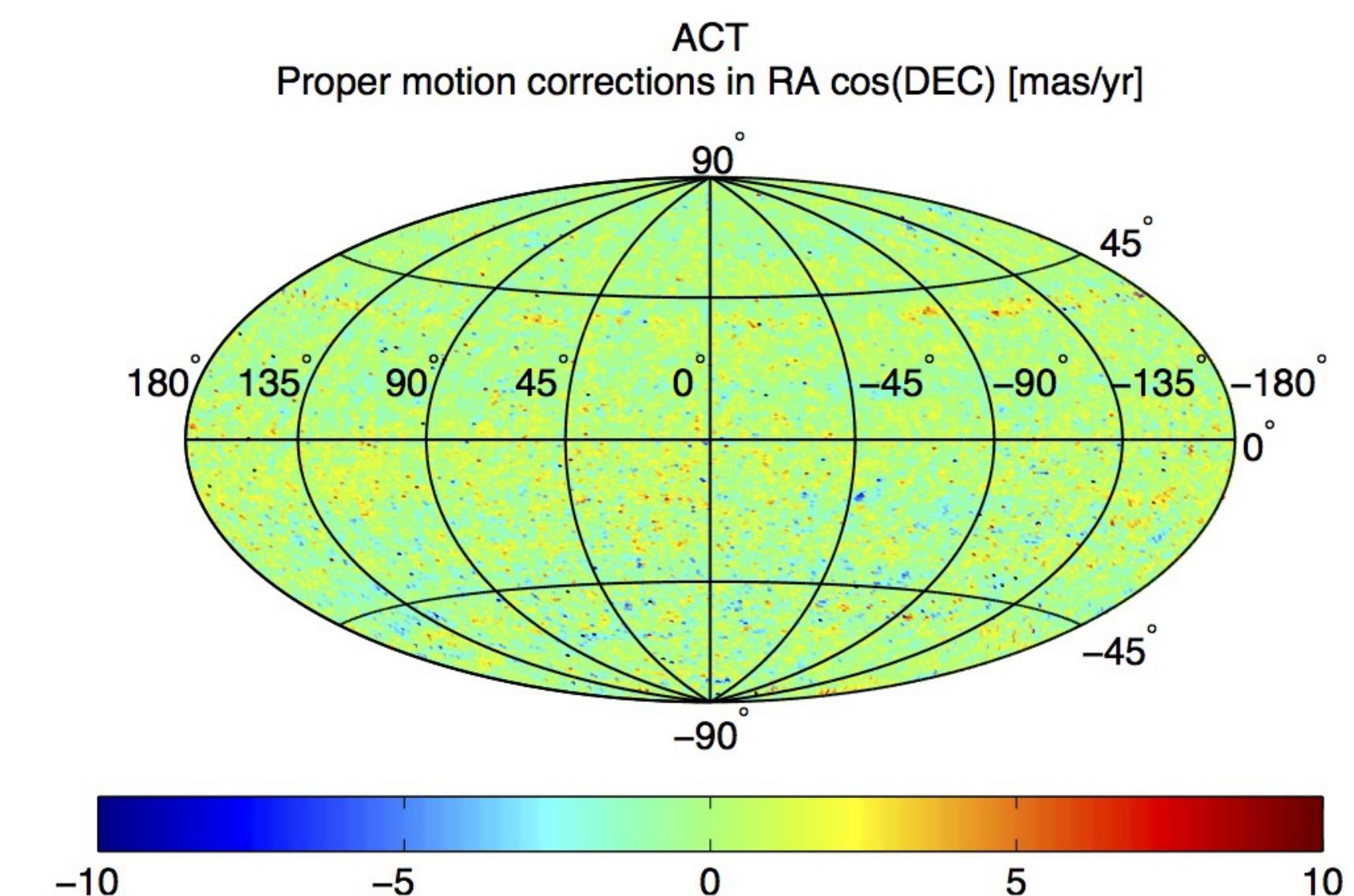}\includegraphics[width=0.6\textwidth]{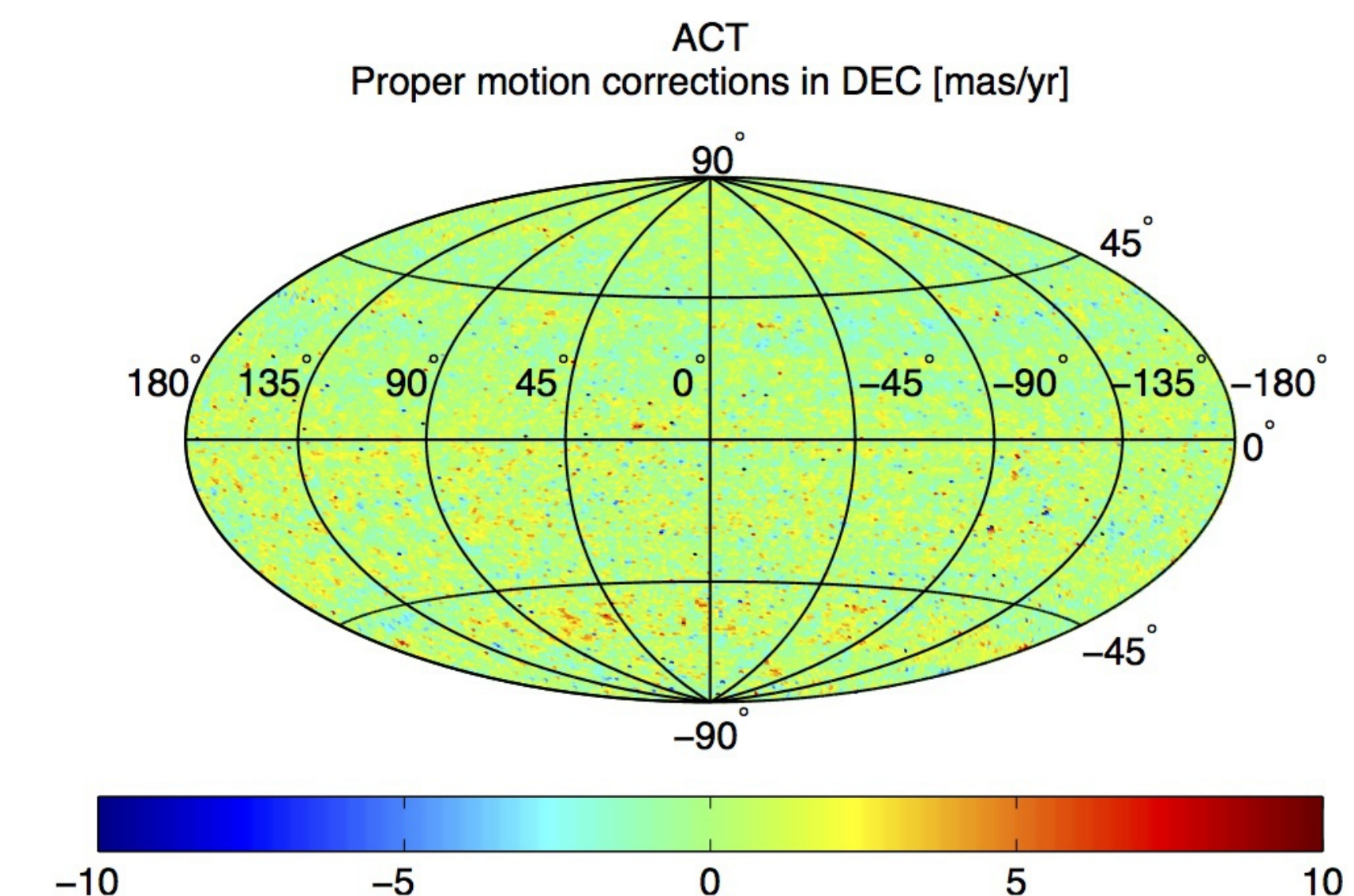}}
\caption{Top: J2000.0 position corrections in right ascension (left)
  and declination (right) for ACT. Bottom: proper motion corrections
  in right ascension (left) and declination (right) for
  ACT.}\label{f:act}
\end{figure}

Tycho-2 (Fig.~\ref{f:tycho2}) and ACT (Fig.~\ref{f:act}) are catalogs
with a relatively low number of stars.  Both positions and proper
motions are close to those of our reference catalog. There is no clear
signature and the differences could simply be noise. Therefore, we
decided to apply no corrections to the Tycho-2 and ACT based
astrometry.  Good agreement between Tycho-2, which is space based, and
our reference catalog gives us some additional confidence that our
reference catalog has good positions and proper motions, at least for
the stars in common.

\subsection{USNO catalogs}
\begin{figure}
\centerline{\includegraphics[width=0.6\textwidth]{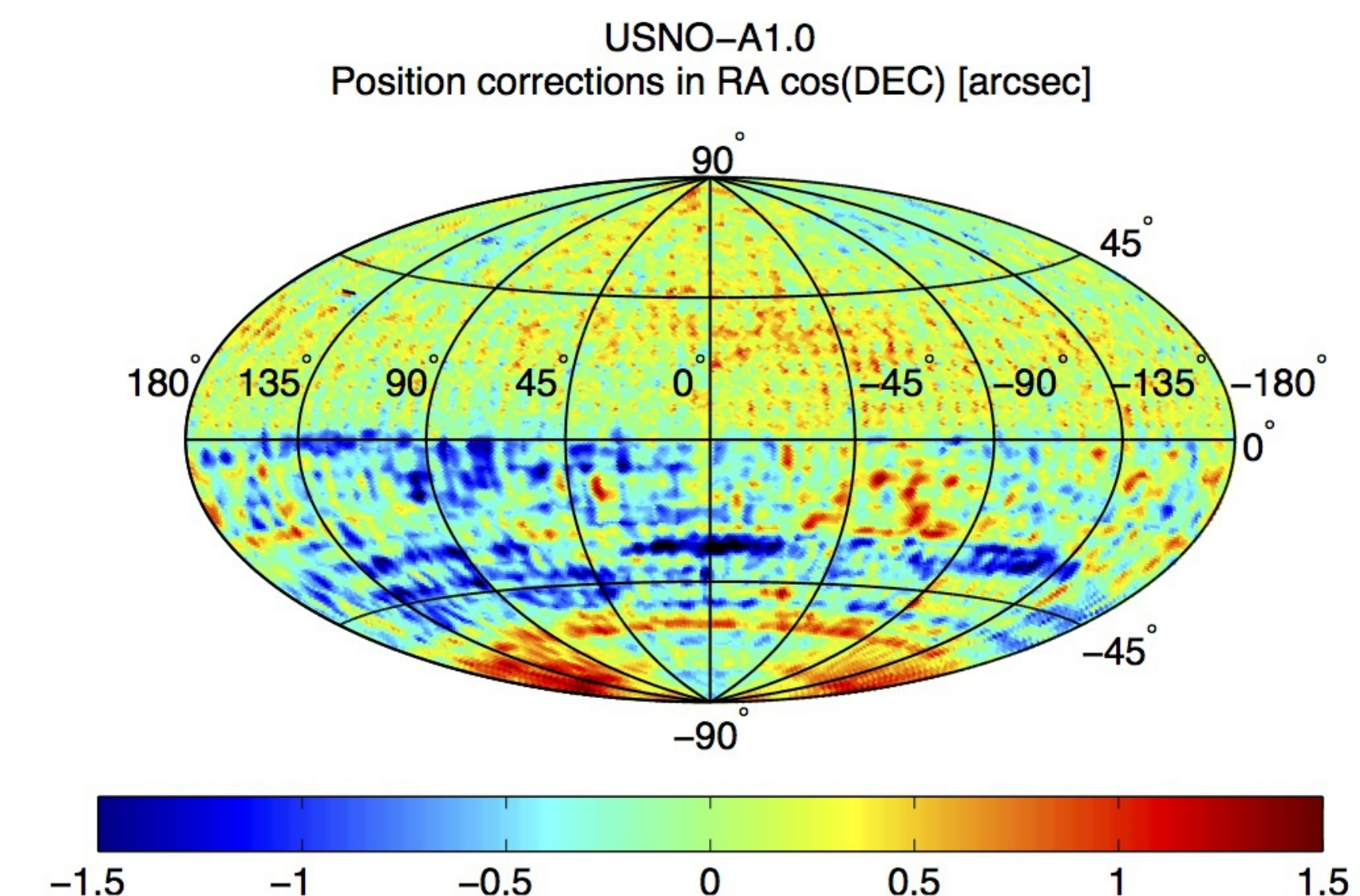}\includegraphics[width=0.6\textwidth]{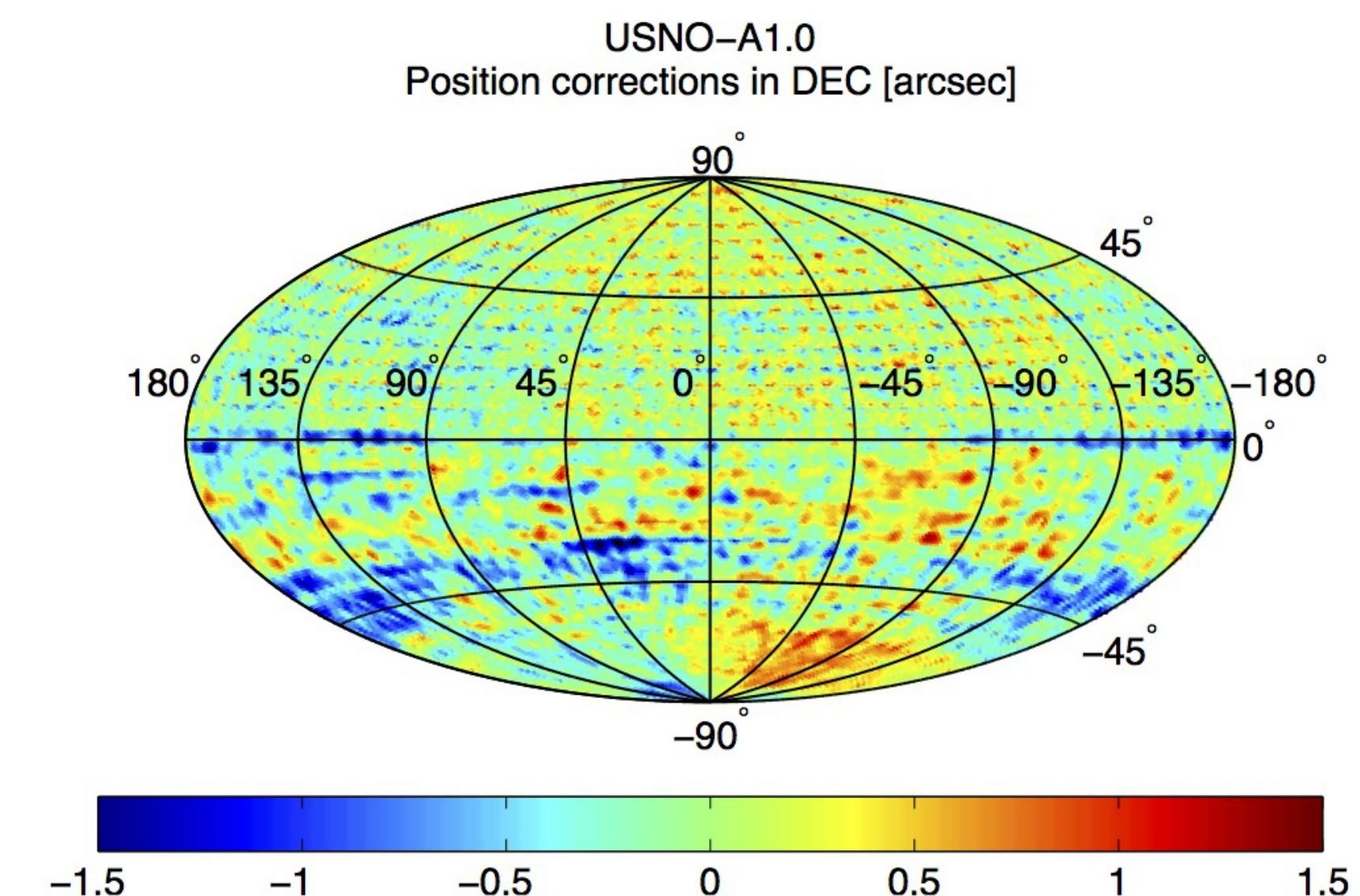}}
\centerline{\includegraphics[width=0.6\textwidth]{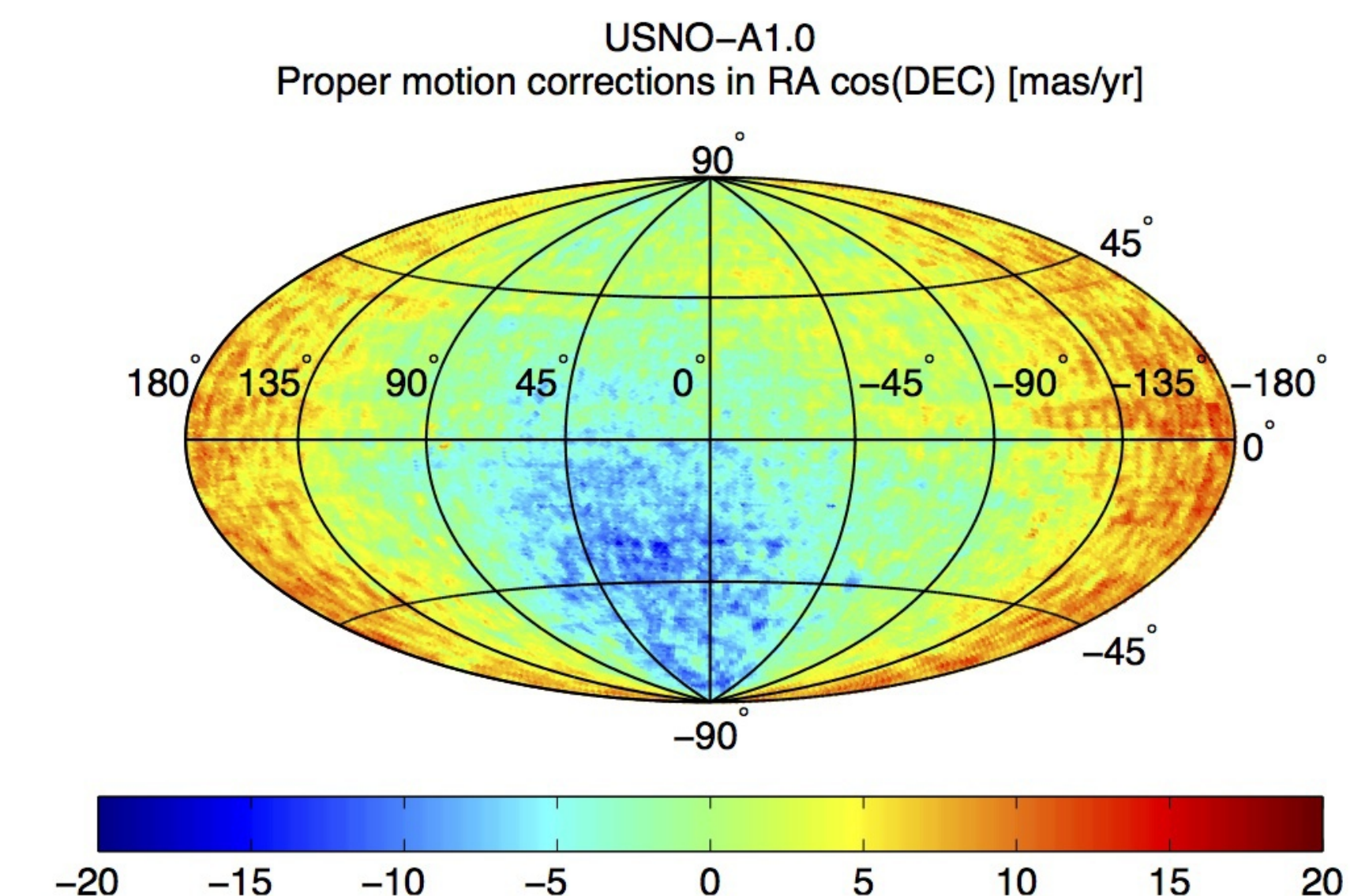}\includegraphics[width=0.6\textwidth]{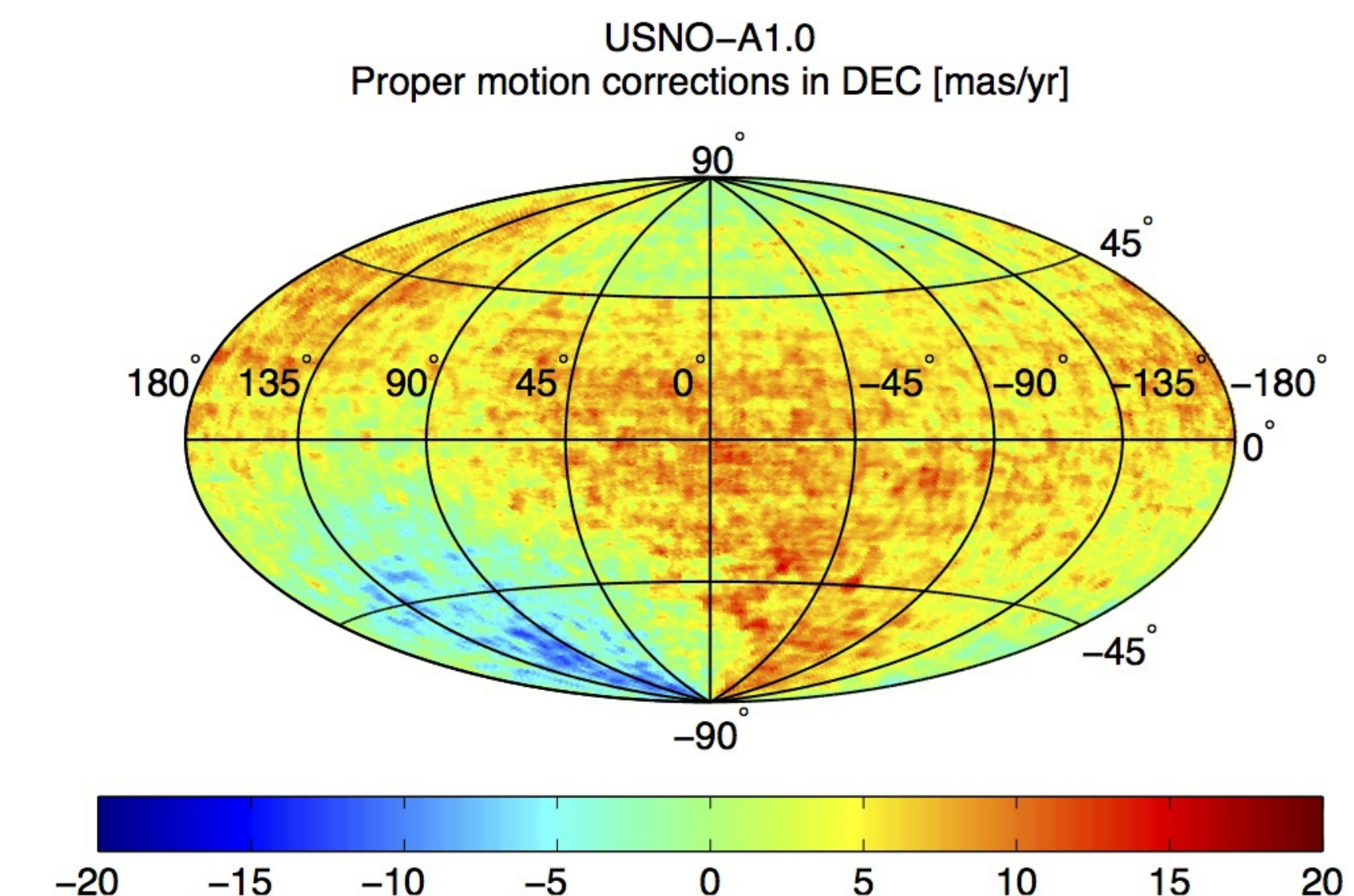}}
\caption{Top: J2000.0 position corrections in right ascension (left)
  and declination (right) for USNO-A1.0. Bottom: proper motion
  corrections in right ascension (left) and declination (right) for
  USNO-A1.0.}\label{f:usno_a1}
\end{figure}

\begin{figure}
\centerline{\includegraphics[width=0.6\textwidth]{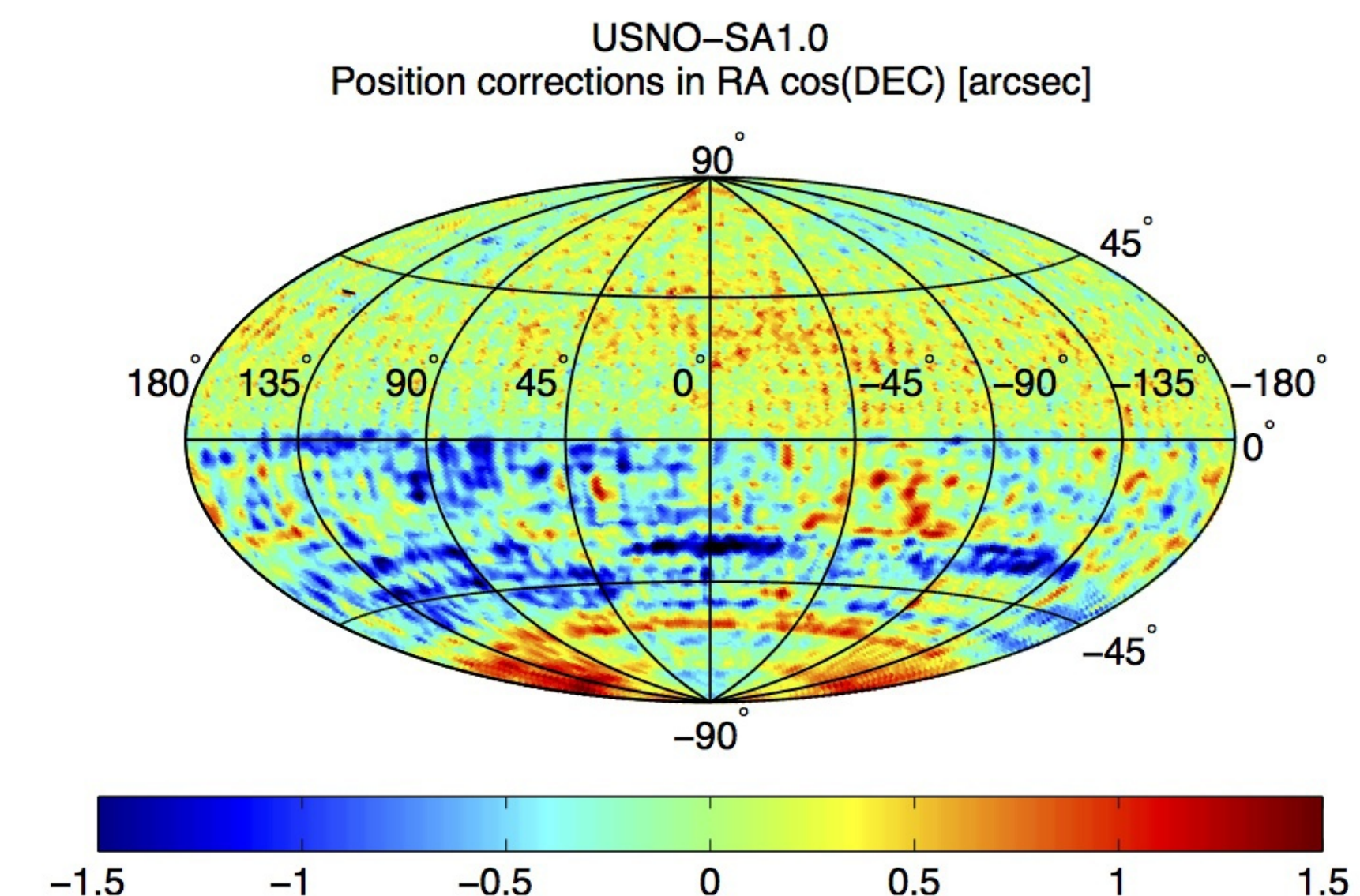}\includegraphics[width=0.6\textwidth]{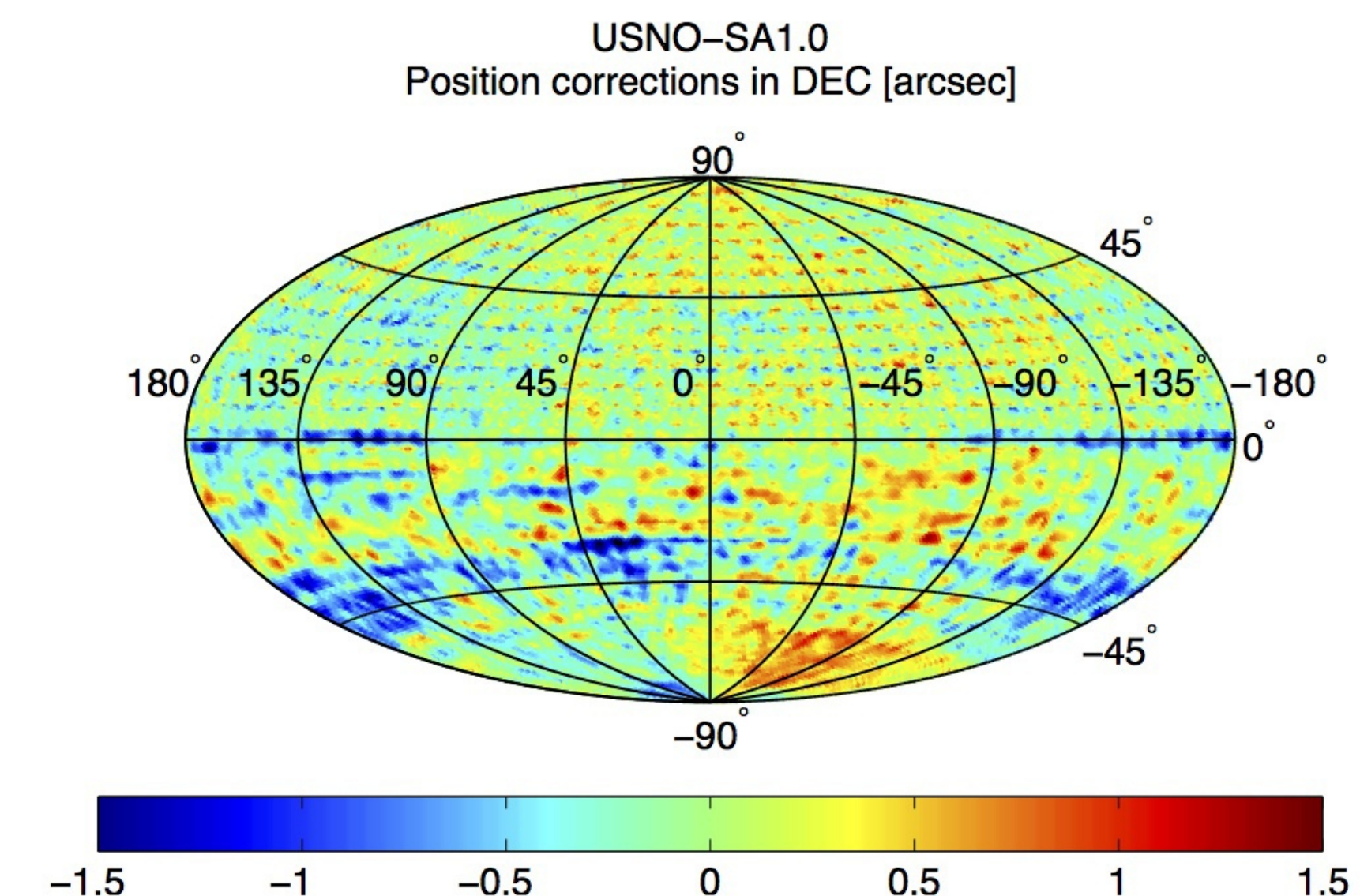}}
\centerline{\includegraphics[width=0.6\textwidth]{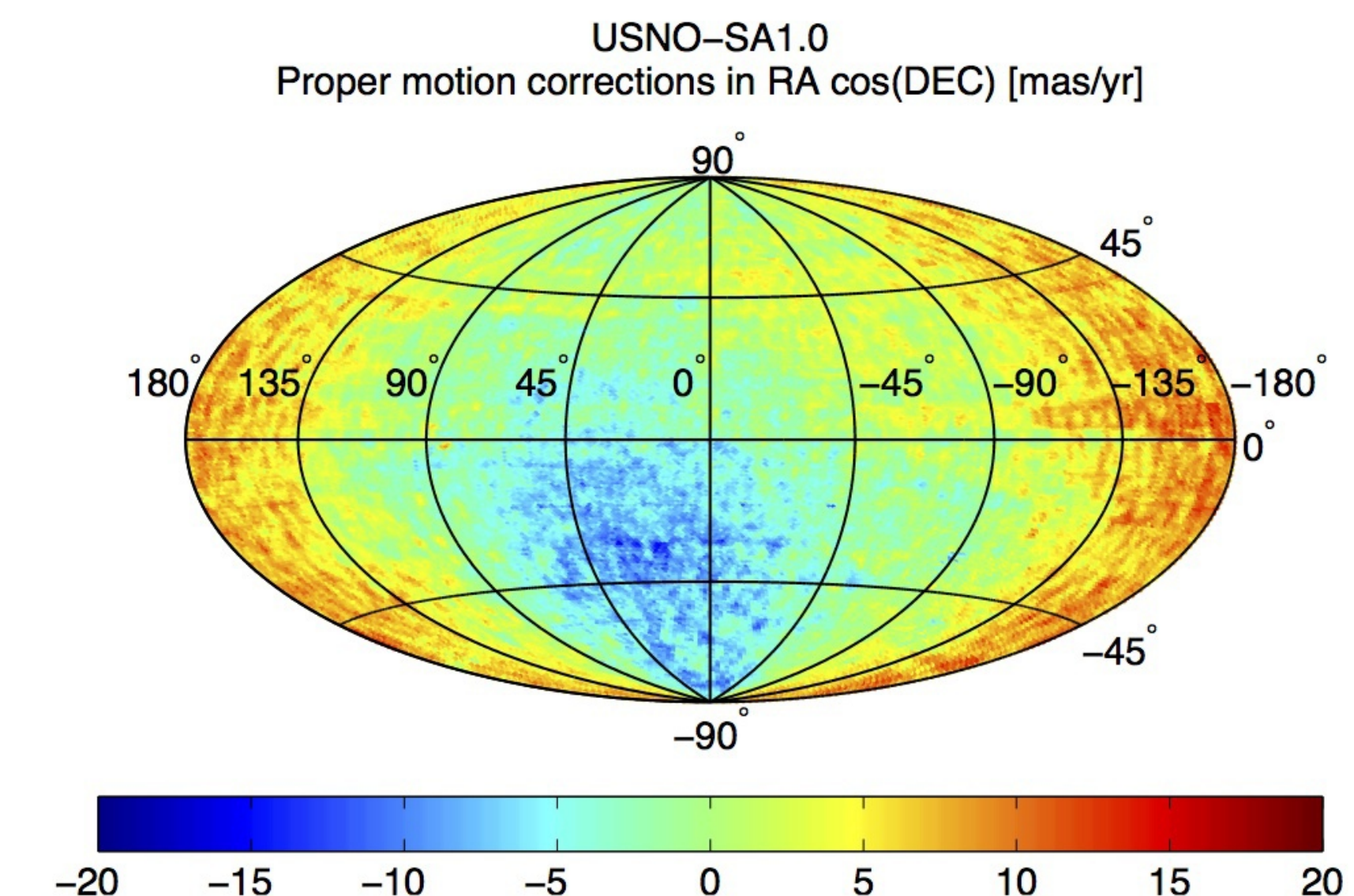}\includegraphics[width=0.6\textwidth]{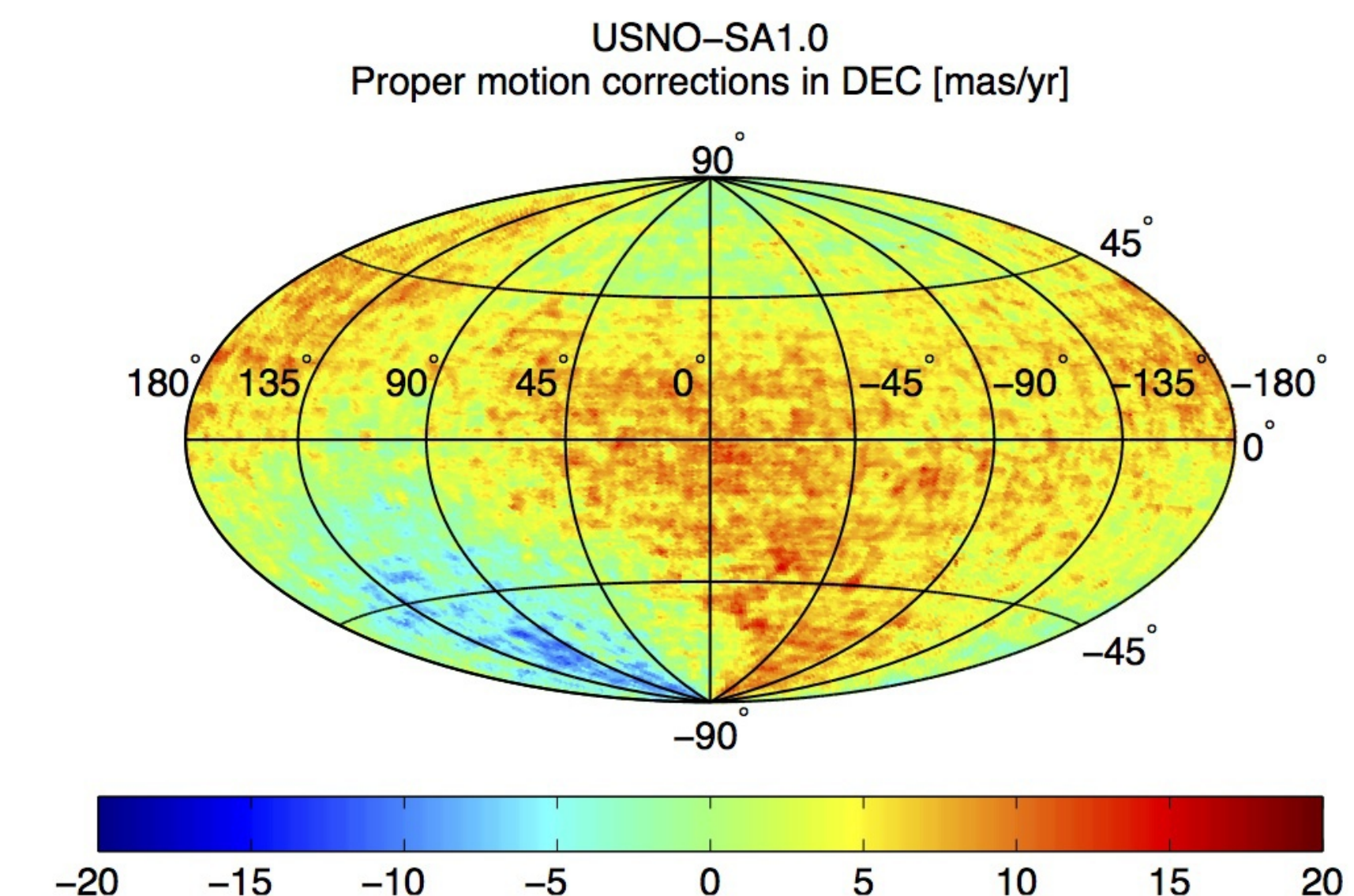}}
\caption{Top: J2000.0 position corrections in right ascension (left)
  and declination (right) for USNO-SA1.0. Bottom: proper motion corrections
  in right ascension (left) and declination (right) for USNO-SA1.0.}\label{f:usno_sa1}
\end{figure}

\begin{figure}
\centerline{\includegraphics[width=0.6\textwidth]{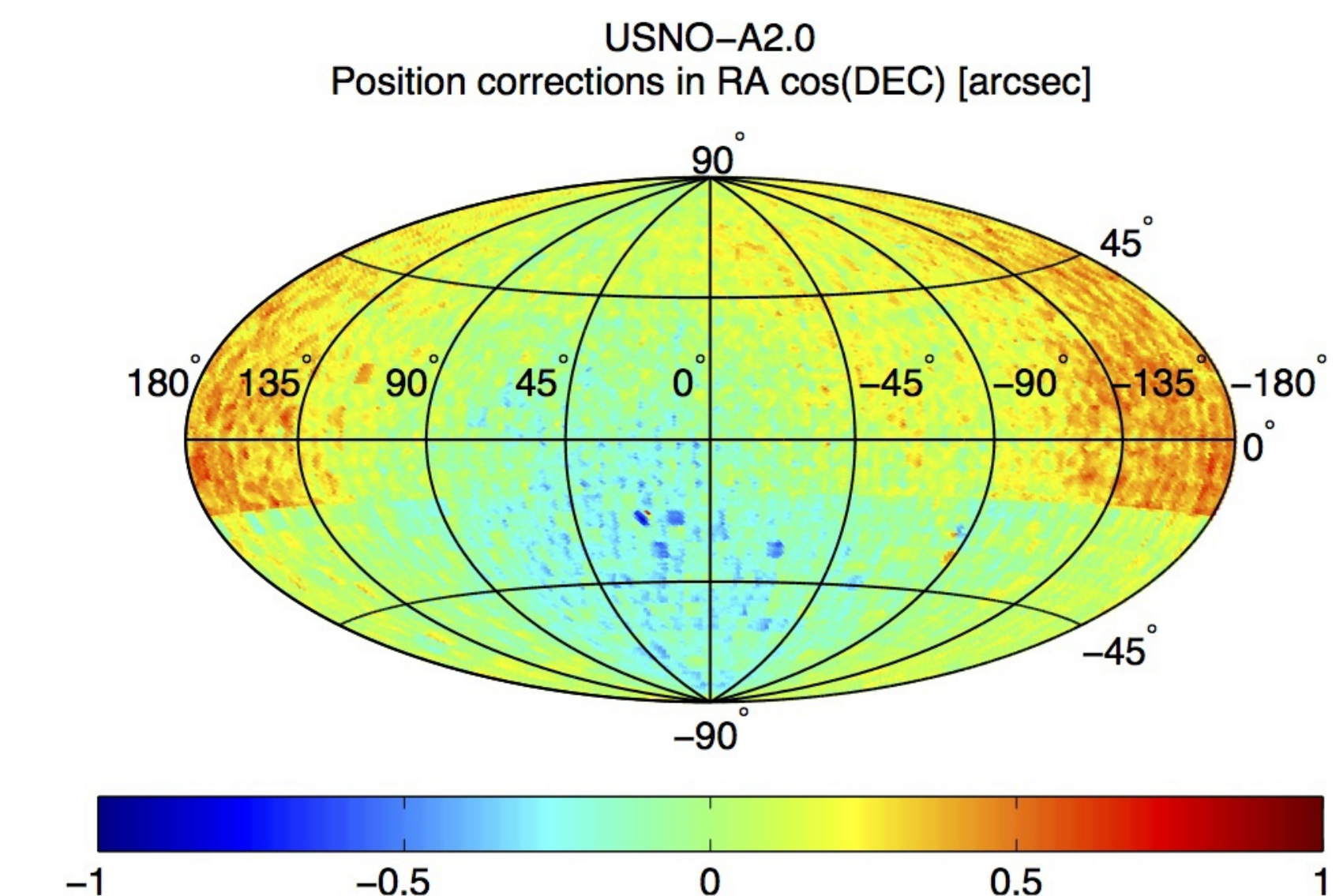}\includegraphics[width=0.6\textwidth]{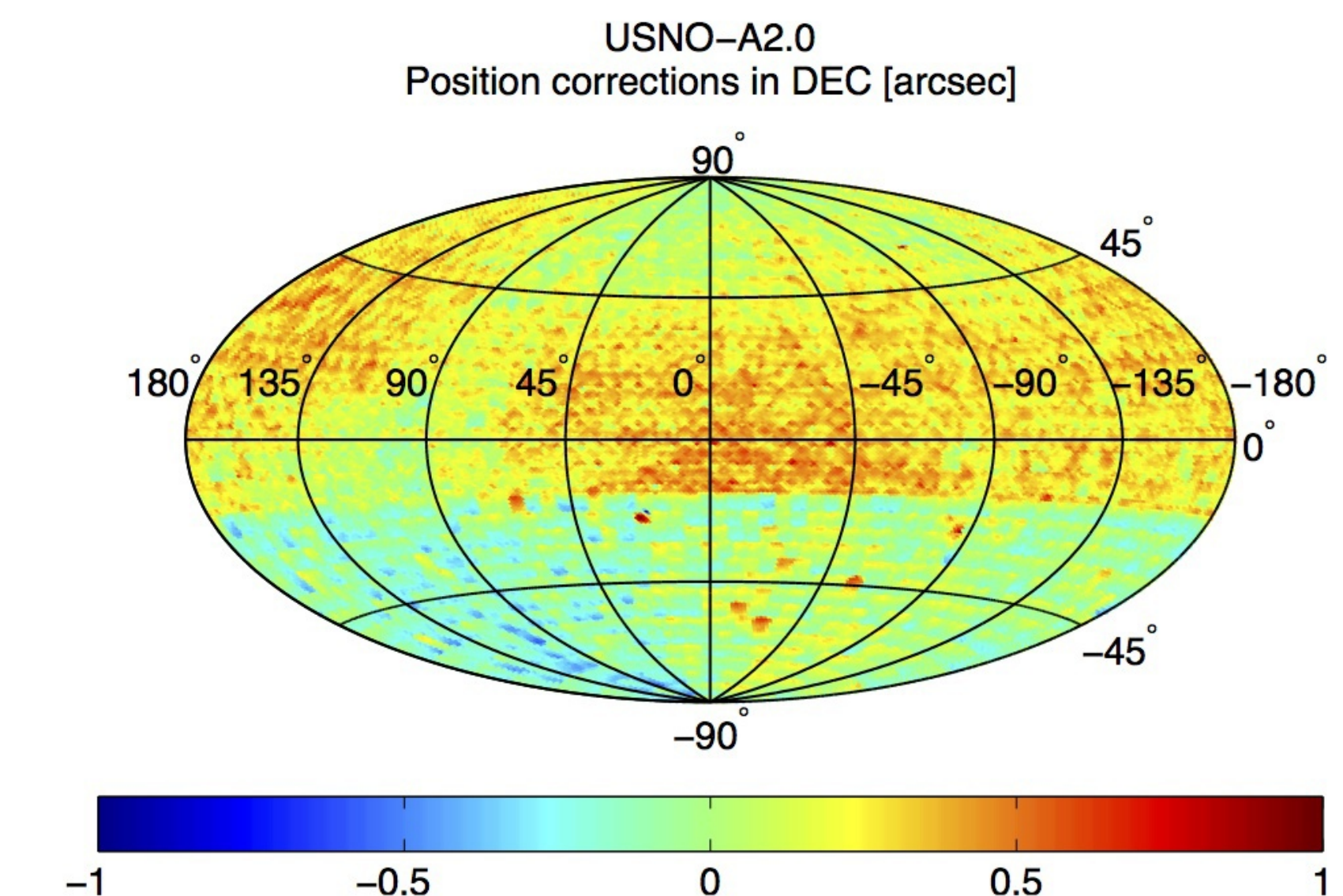}}
\centerline{\includegraphics[width=0.6\textwidth]{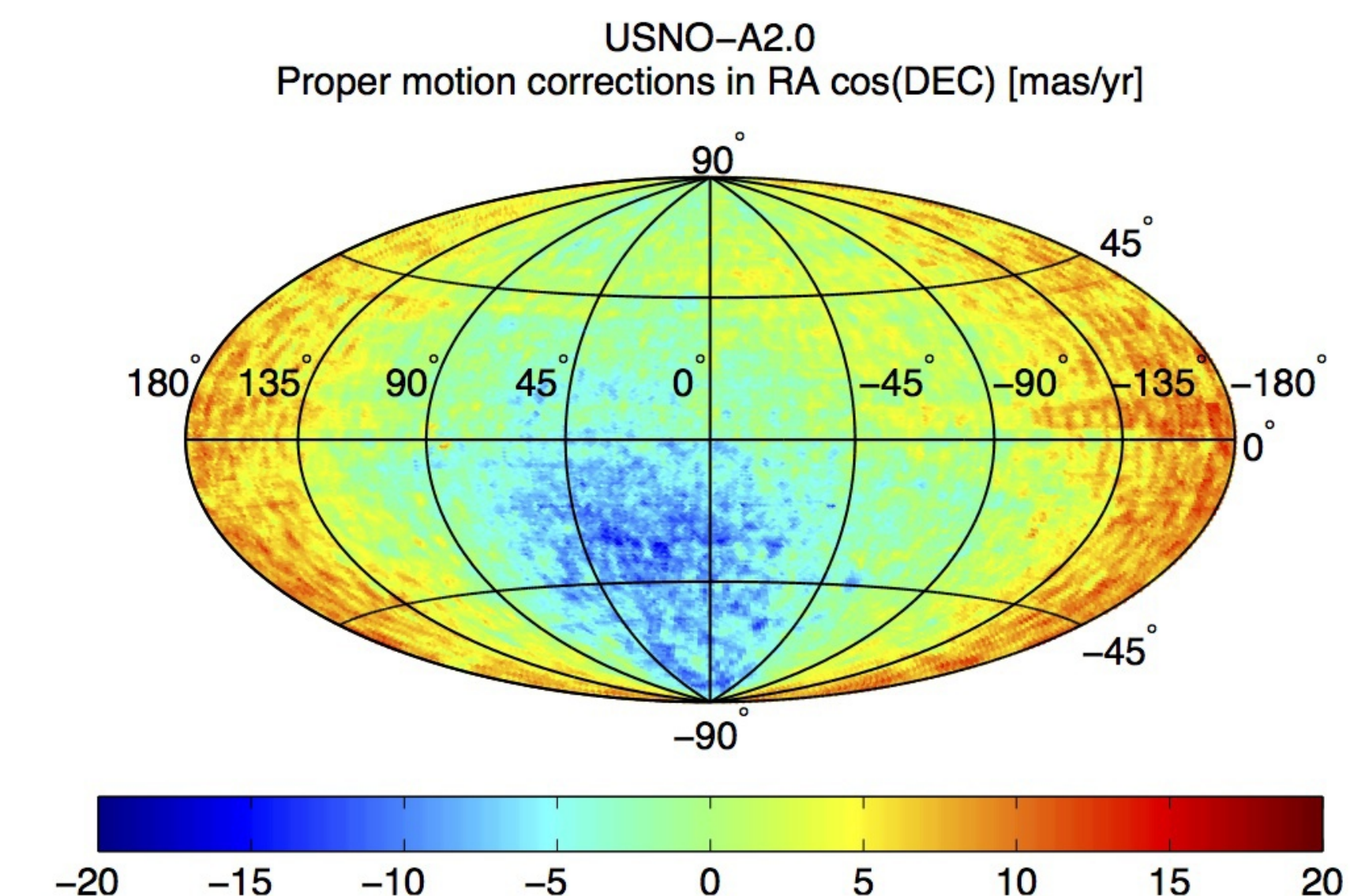}\includegraphics[width=0.6\textwidth]{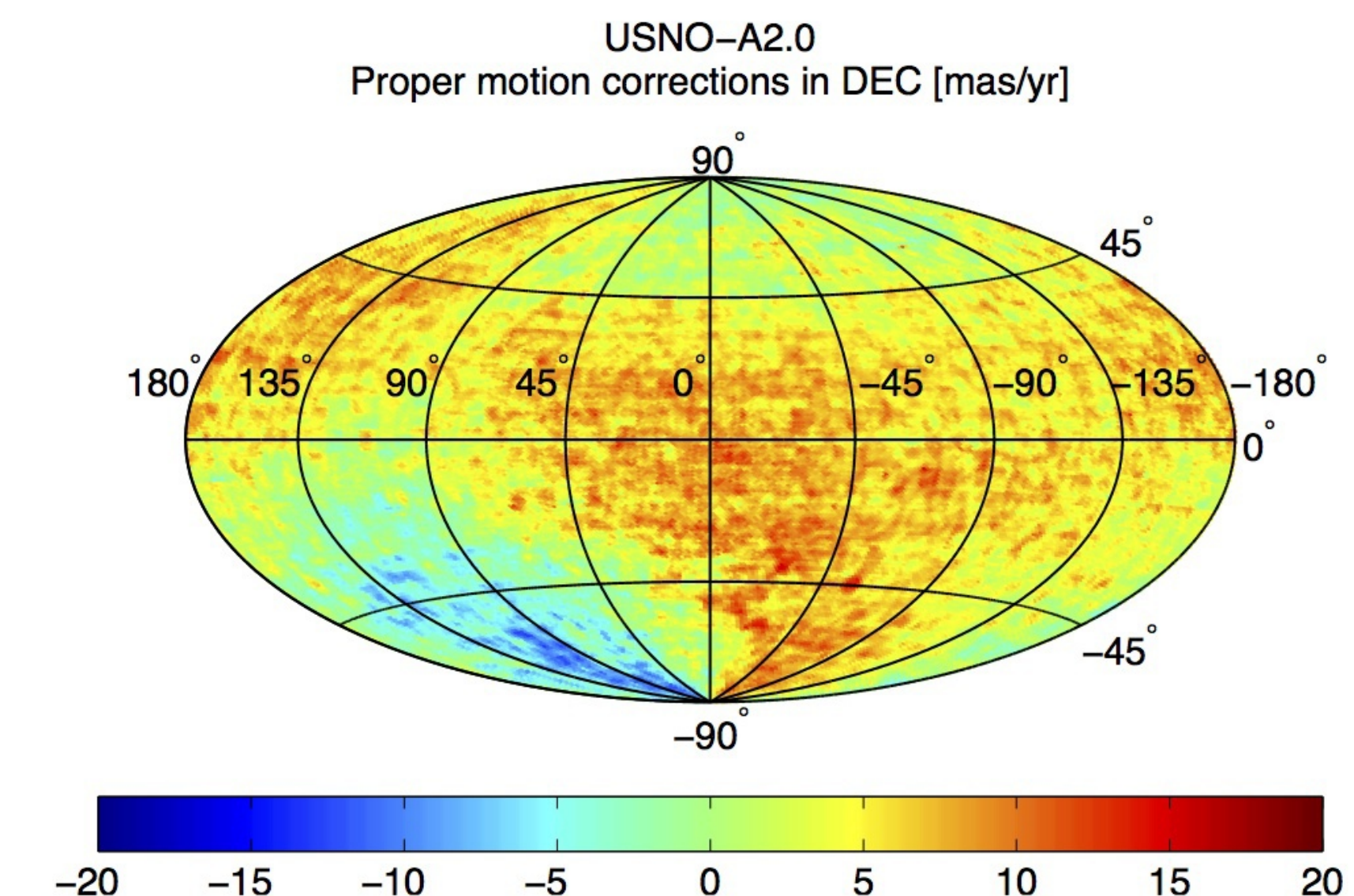}}
\caption{Top: J2000.0 position corrections in right ascension (left)
  and declination (right) for USNO-A2.0. Bottom: proper motion corrections
  in right ascension (left) and declination (right) for USNO-A2.0.}\label{f:usno_a2}
\end{figure}

\begin{figure}
\centerline{\includegraphics[width=0.6\textwidth]{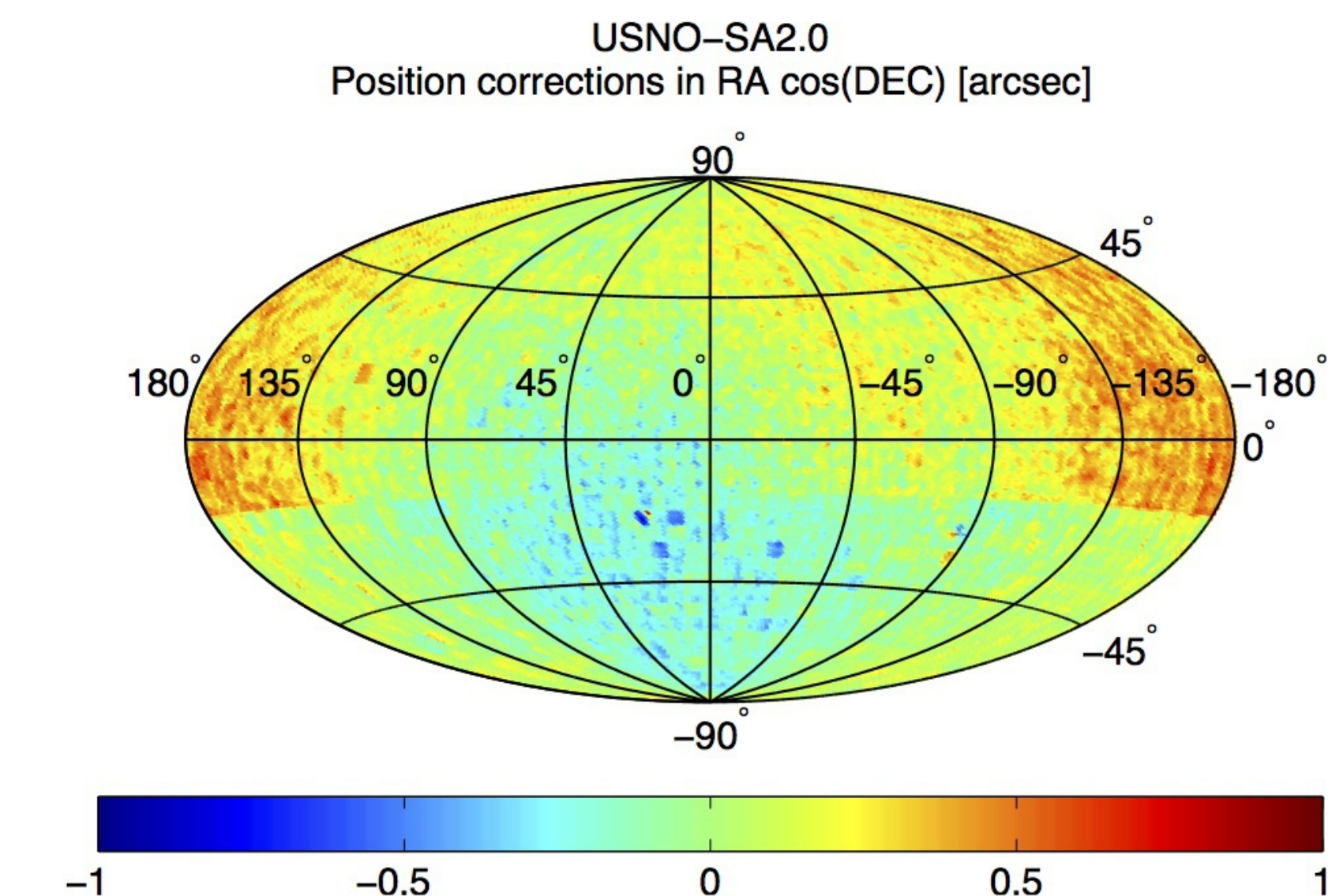}\includegraphics[width=0.6\textwidth]{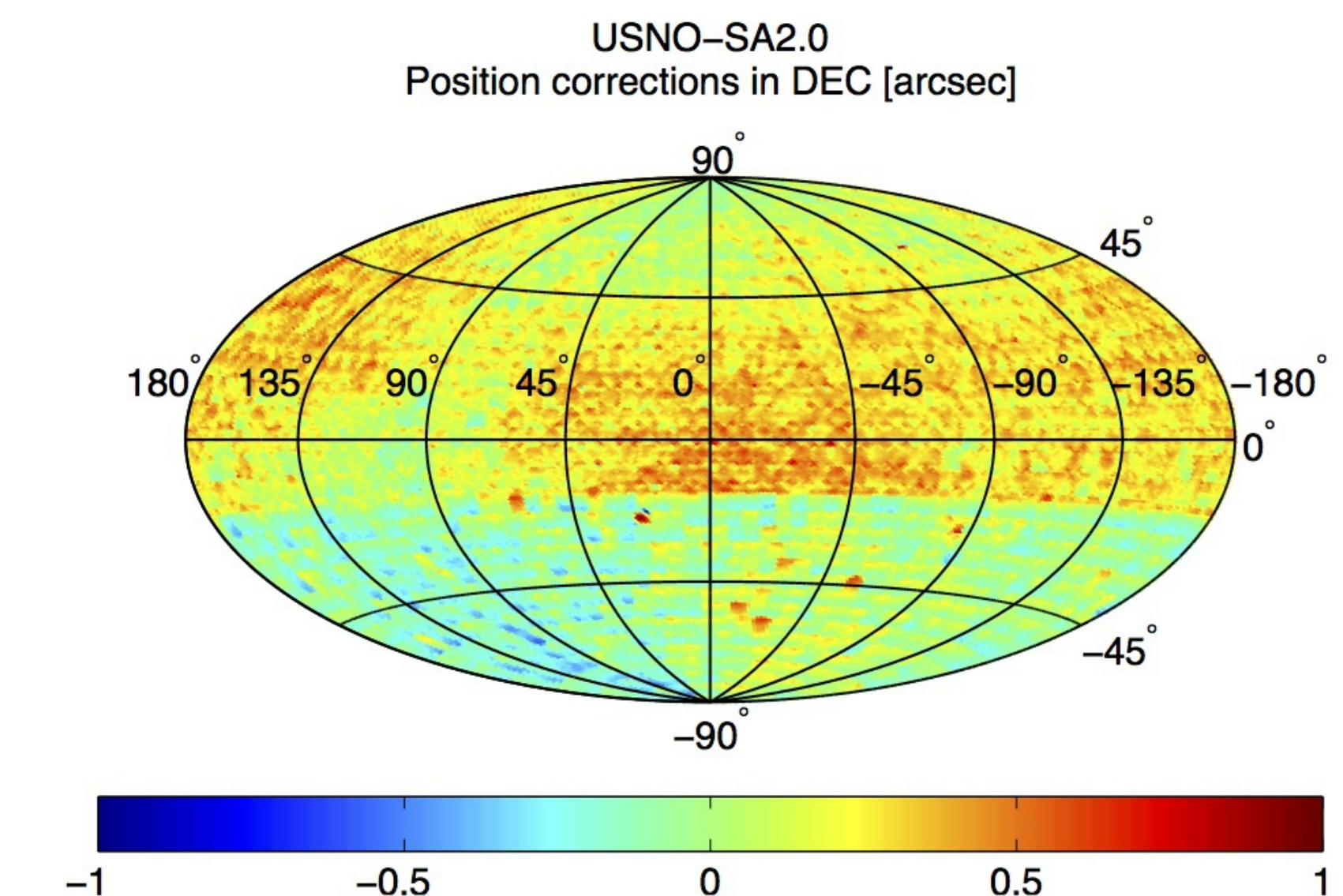}}
\centerline{\includegraphics[width=0.6\textwidth]{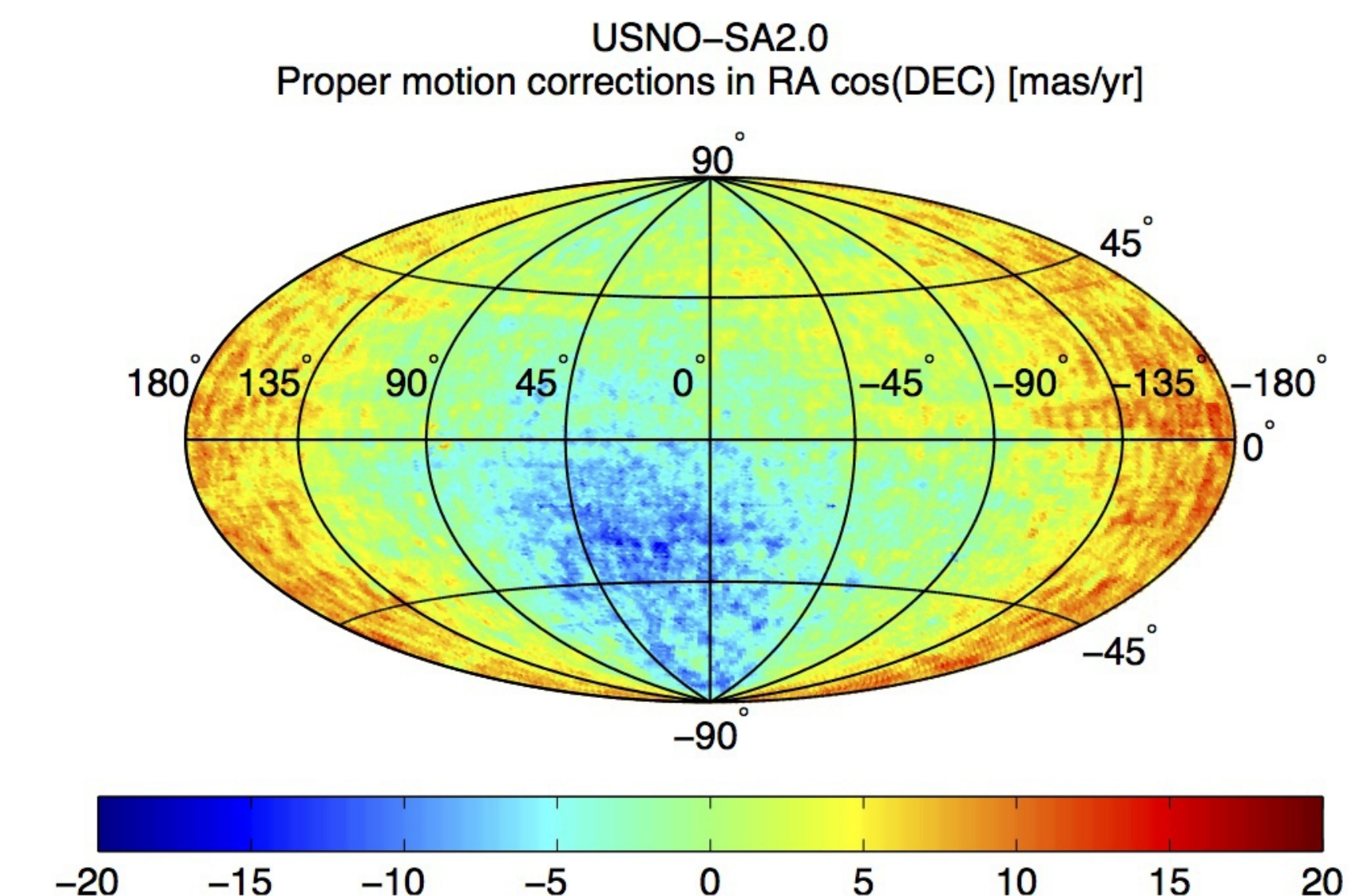}\includegraphics[width=0.6\textwidth]{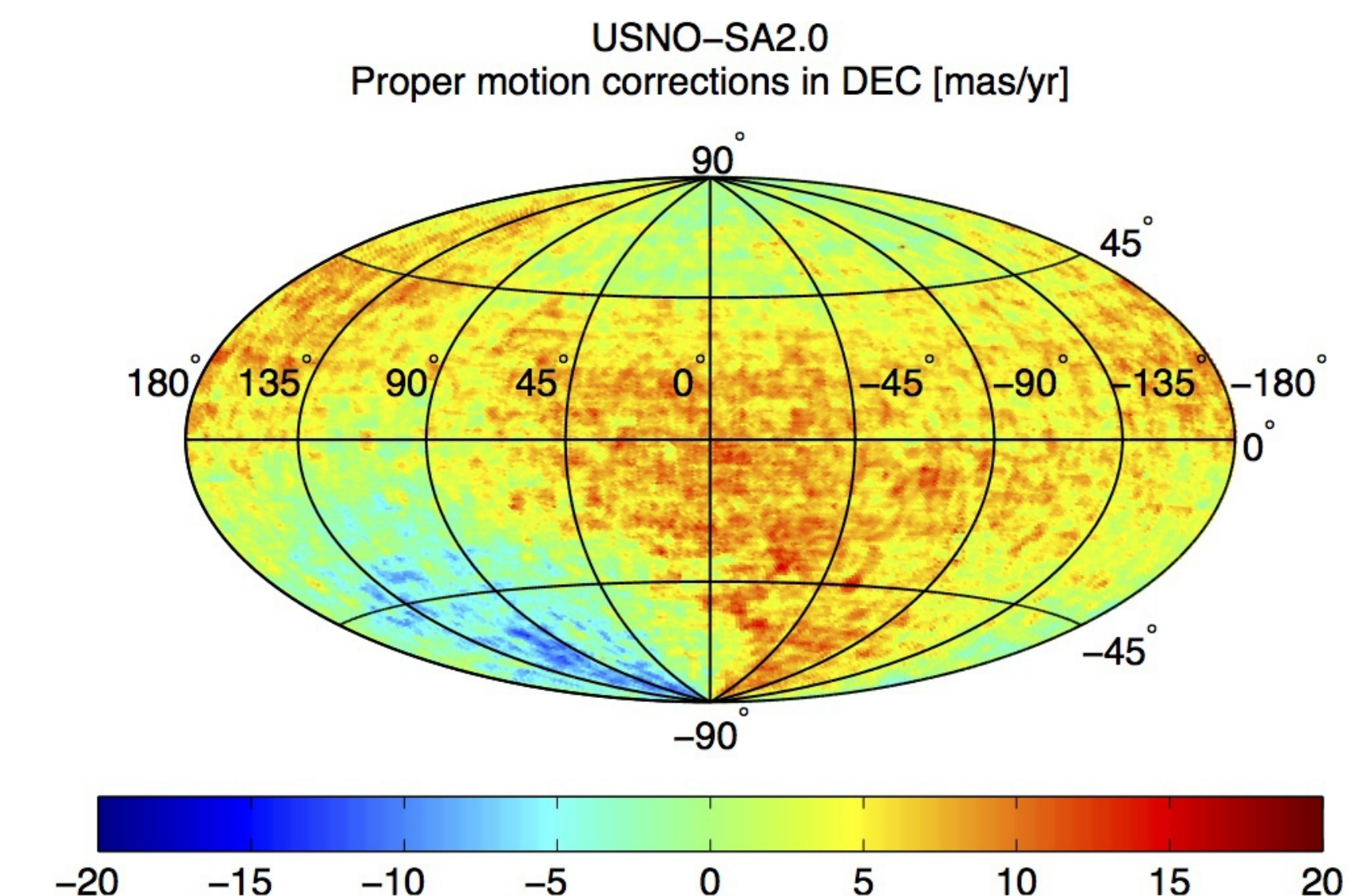}}
\caption{Top: J2000.0 position corrections in right ascension (left)
  and declination (right) for USNO-SA2.0. Bottom: proper motion corrections
  in right ascension (left) and declination (right) for USNO-SA2.0.}\label{f:usno_sa2}
\end{figure}

\begin{figure}
\centerline{\includegraphics[width=0.6\textwidth]{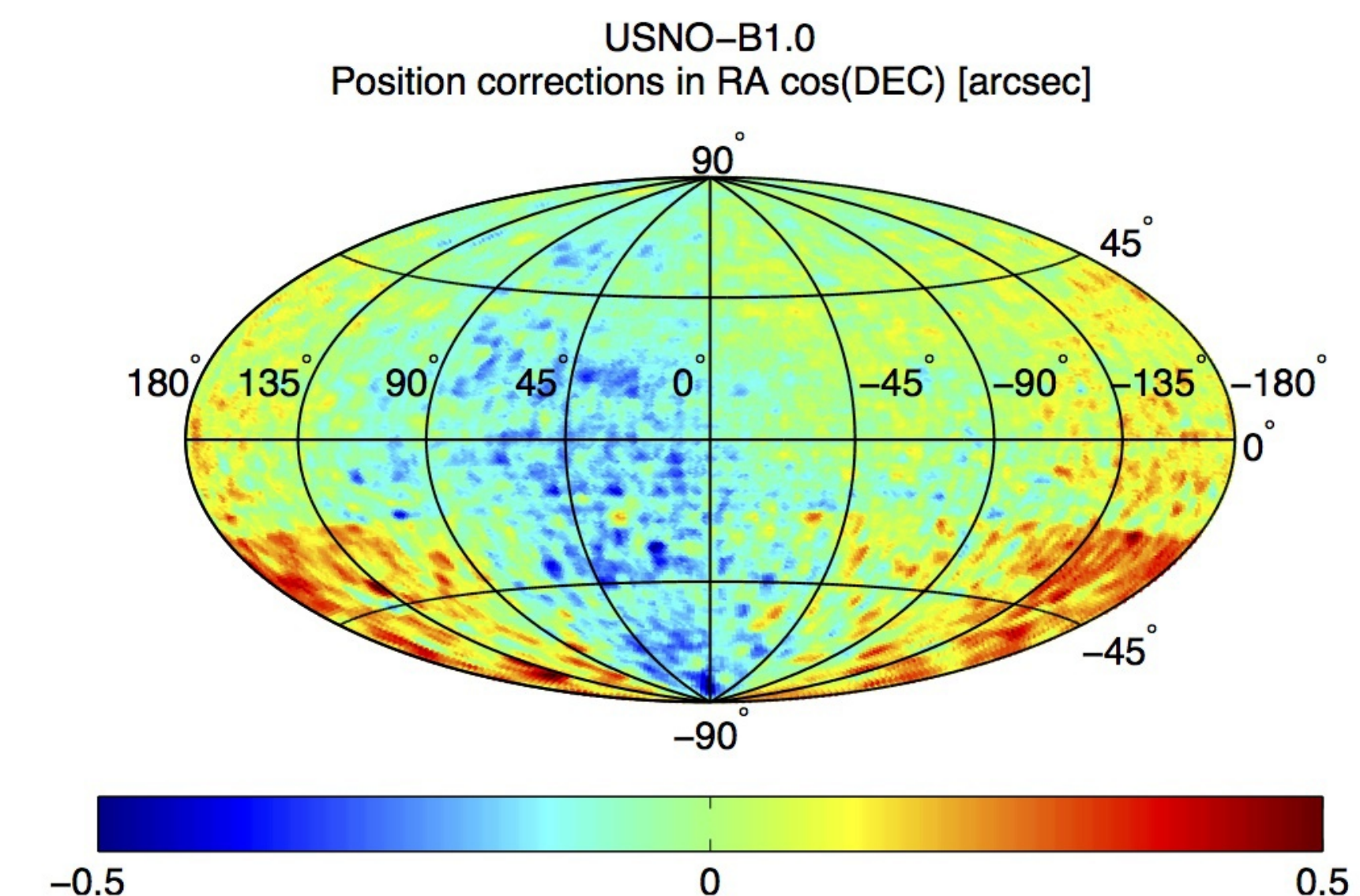}\includegraphics[width=0.6\textwidth]{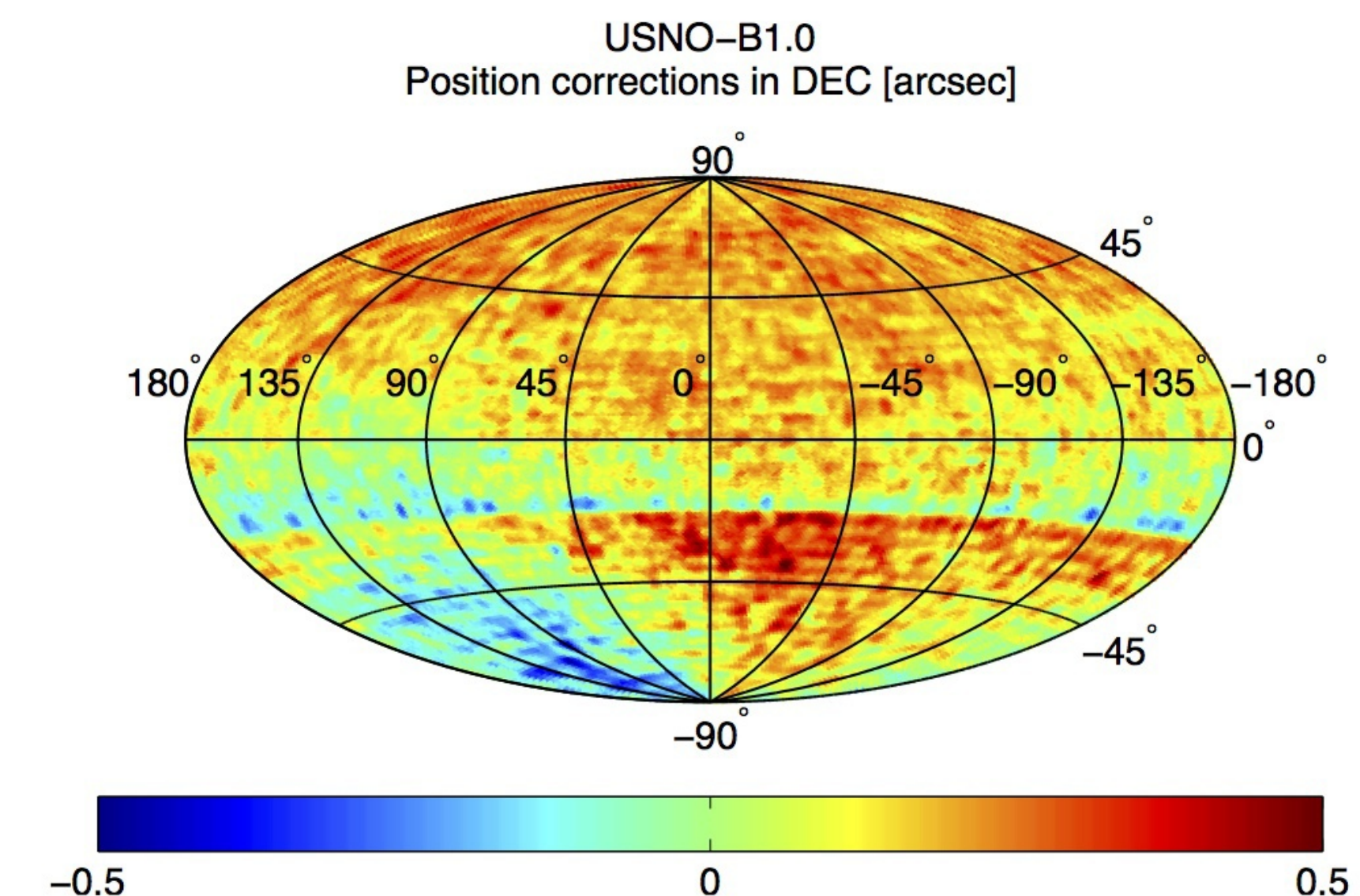}}
\centerline{\includegraphics[width=0.6\textwidth]{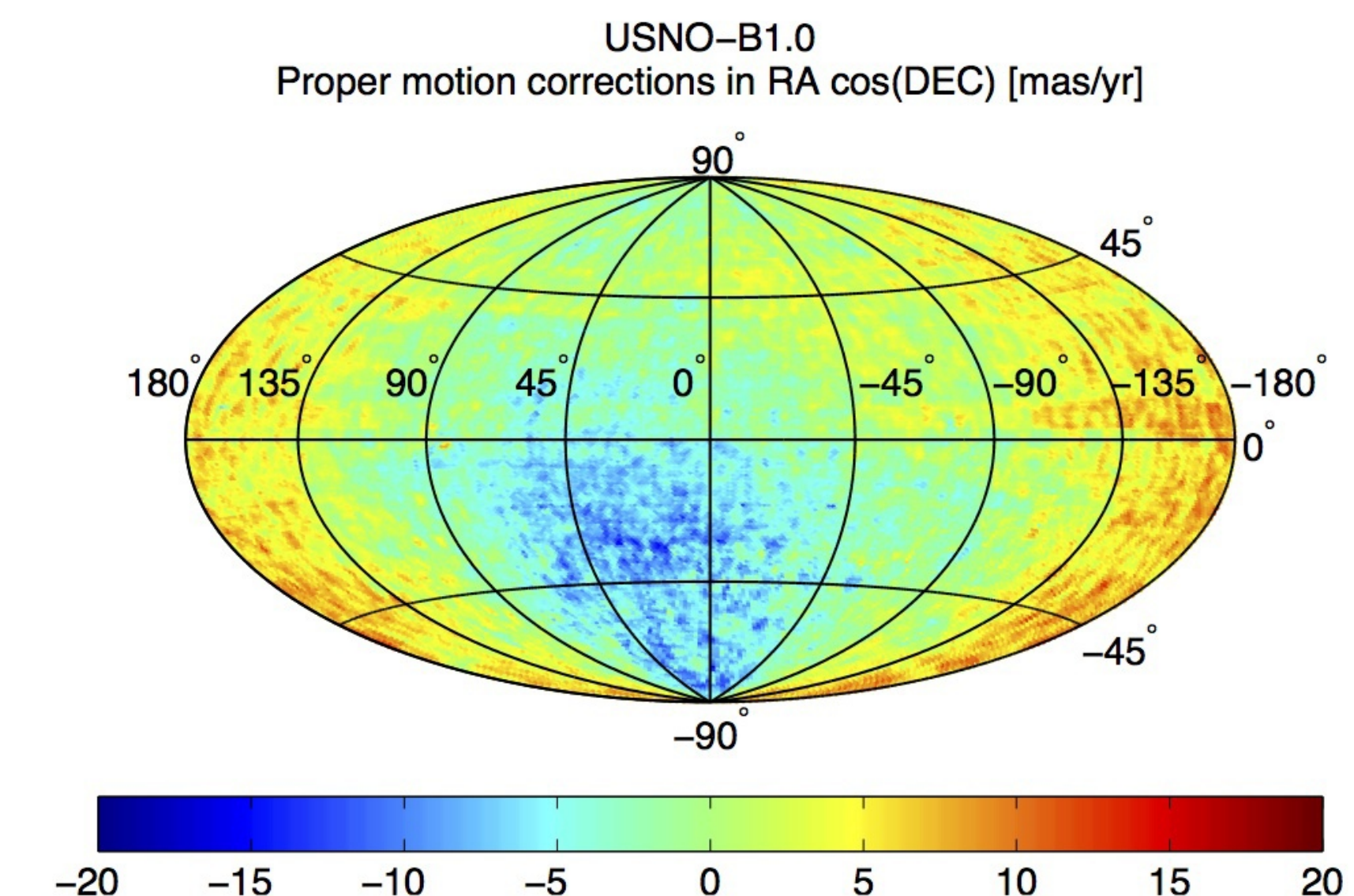}\includegraphics[width=0.6\textwidth]{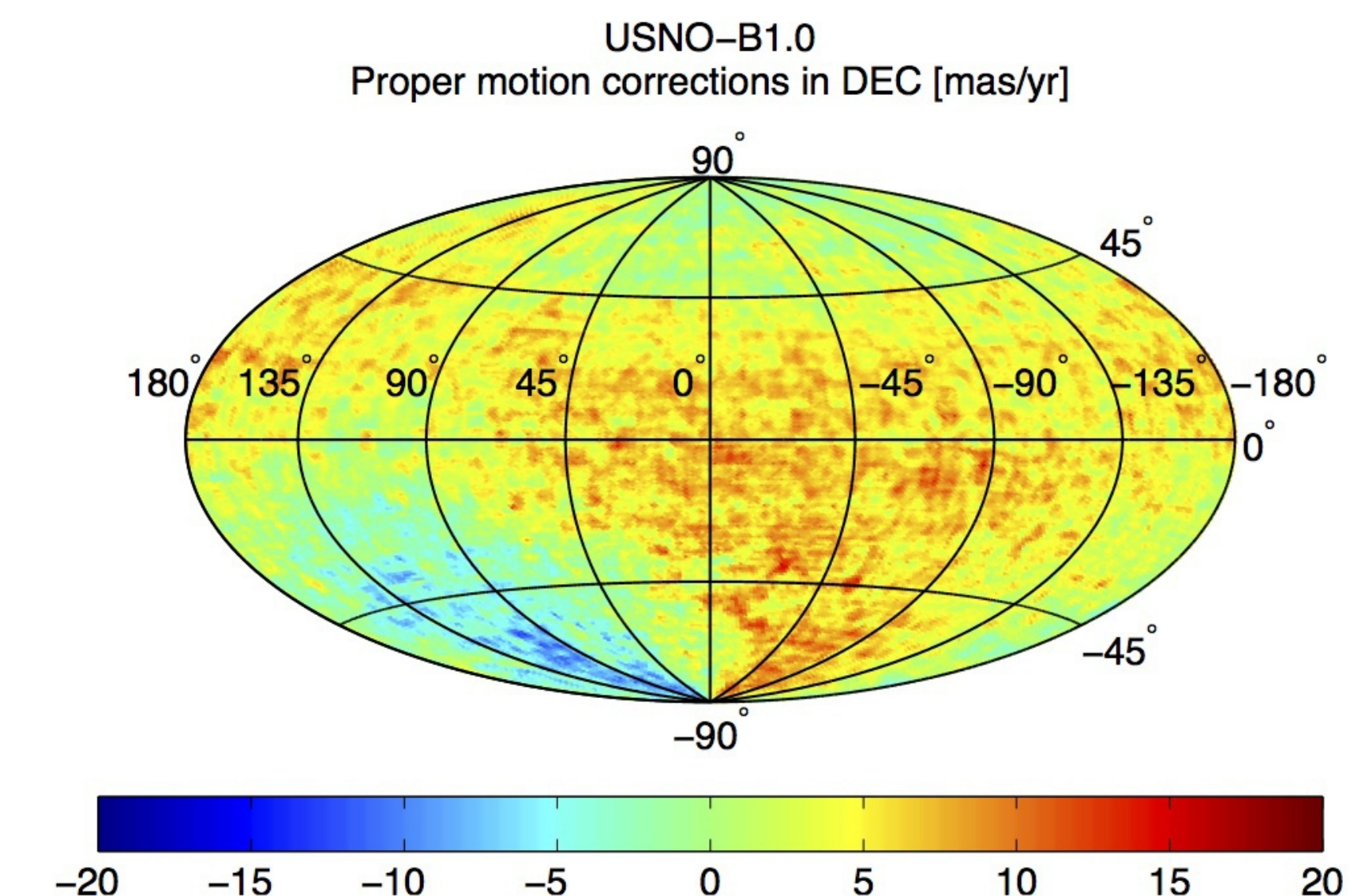}}
\caption{Top: J2000.0 position corrections in right ascension (left)
  and declination (right) for USNO-B1.0. Bottom: proper motion corrections
  in right ascension (left) and declination (right) for USNO-B1.0.}\label{f:usno_b1}
\end{figure}

\citet{cbm10} showed that the USNO catalogs
(Fig.~\ref{f:usno_a1} -- Fig.~\ref{f:usno_b1}) present significant
position biases. Moreover, the USNO-A catalogs do not account for
proper motions.  Although USNO-B1.0 does have proper motions for some
of its stars, the proper motion differences with our reference are of
the same order of the USNO-A catalogs thus indicating that USNO-B1.0
proper motions are not generally accurate enough. We therefore correct
all the astrometry based on the USNO catalogs for both position and
proper motion errors.

\subsection{UCAC catalogs}
\begin{figure}
\centerline{\includegraphics[width=0.6\textwidth]{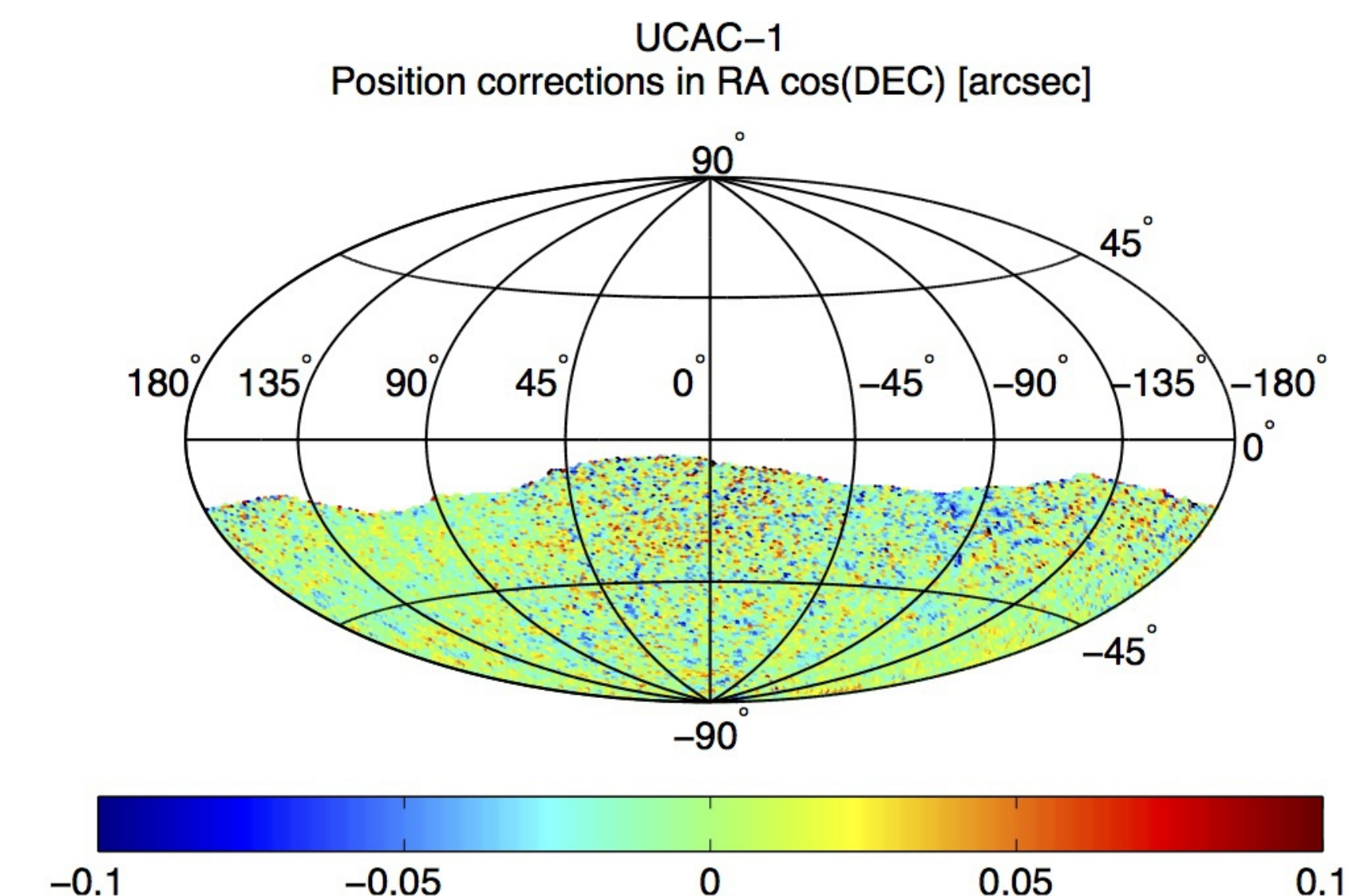}\includegraphics[width=0.6\textwidth]{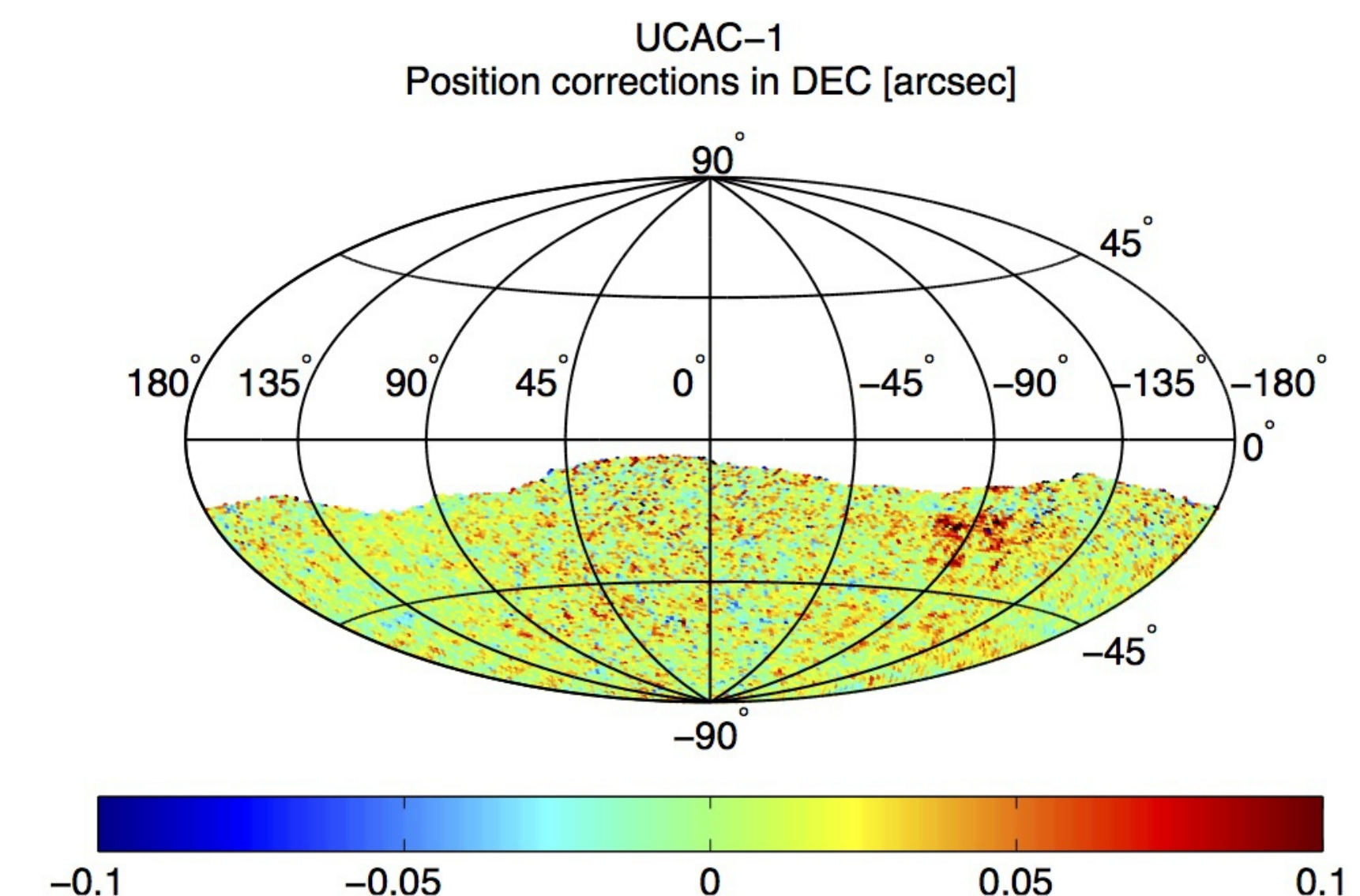}}
\centerline{\includegraphics[width=0.6\textwidth]{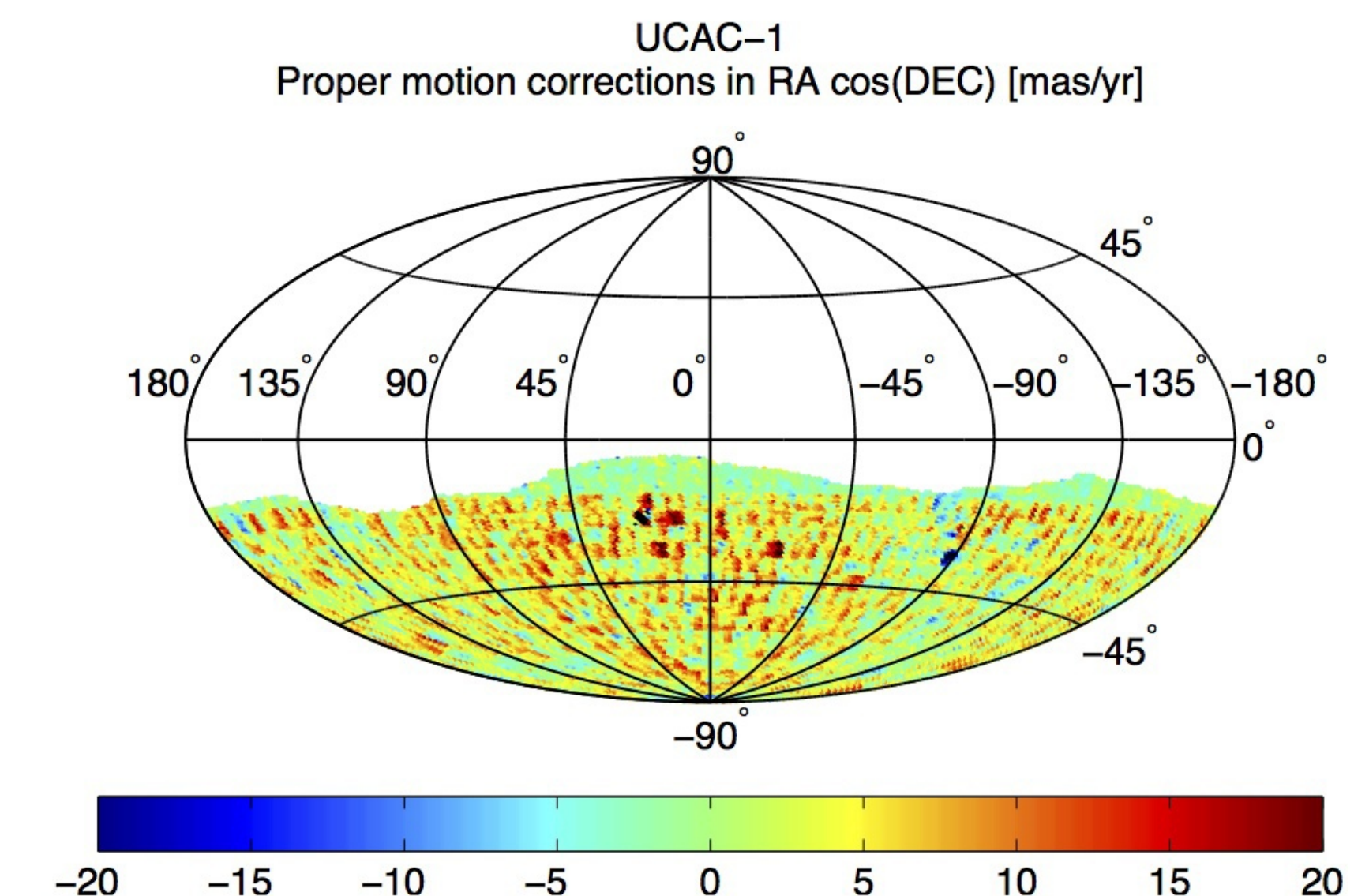}\includegraphics[width=0.6\textwidth]{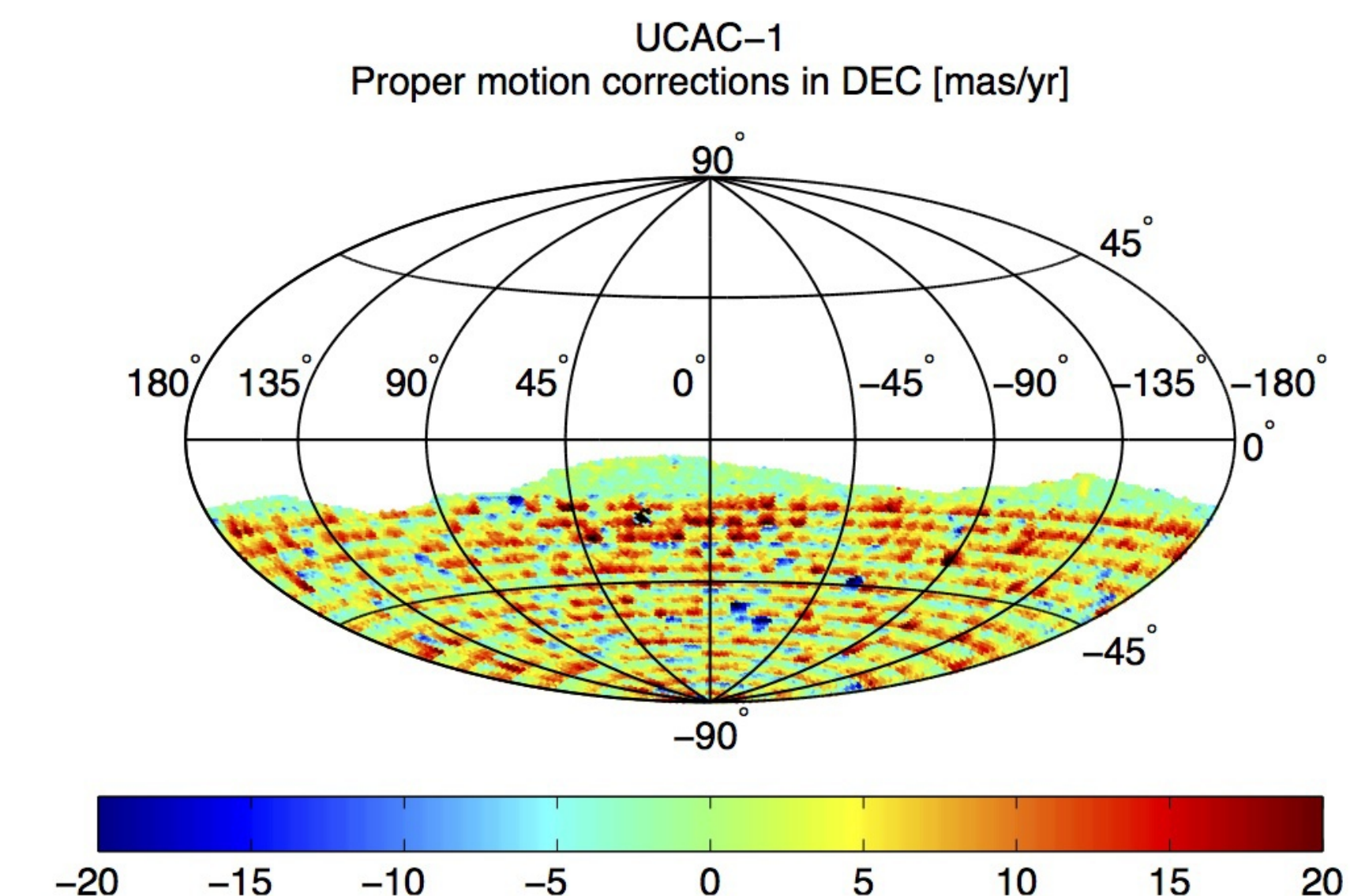}}
\caption{Top: J2000.0 position corrections in right ascension (left)
  and declination (right) for UCAC-1. Bottom: proper motion corrections
  in right ascension (left) and declination (right) for UCAC-1.}\label{f:ucac1}
\end{figure}

\begin{figure}
\centerline{\includegraphics[width=0.6\textwidth]{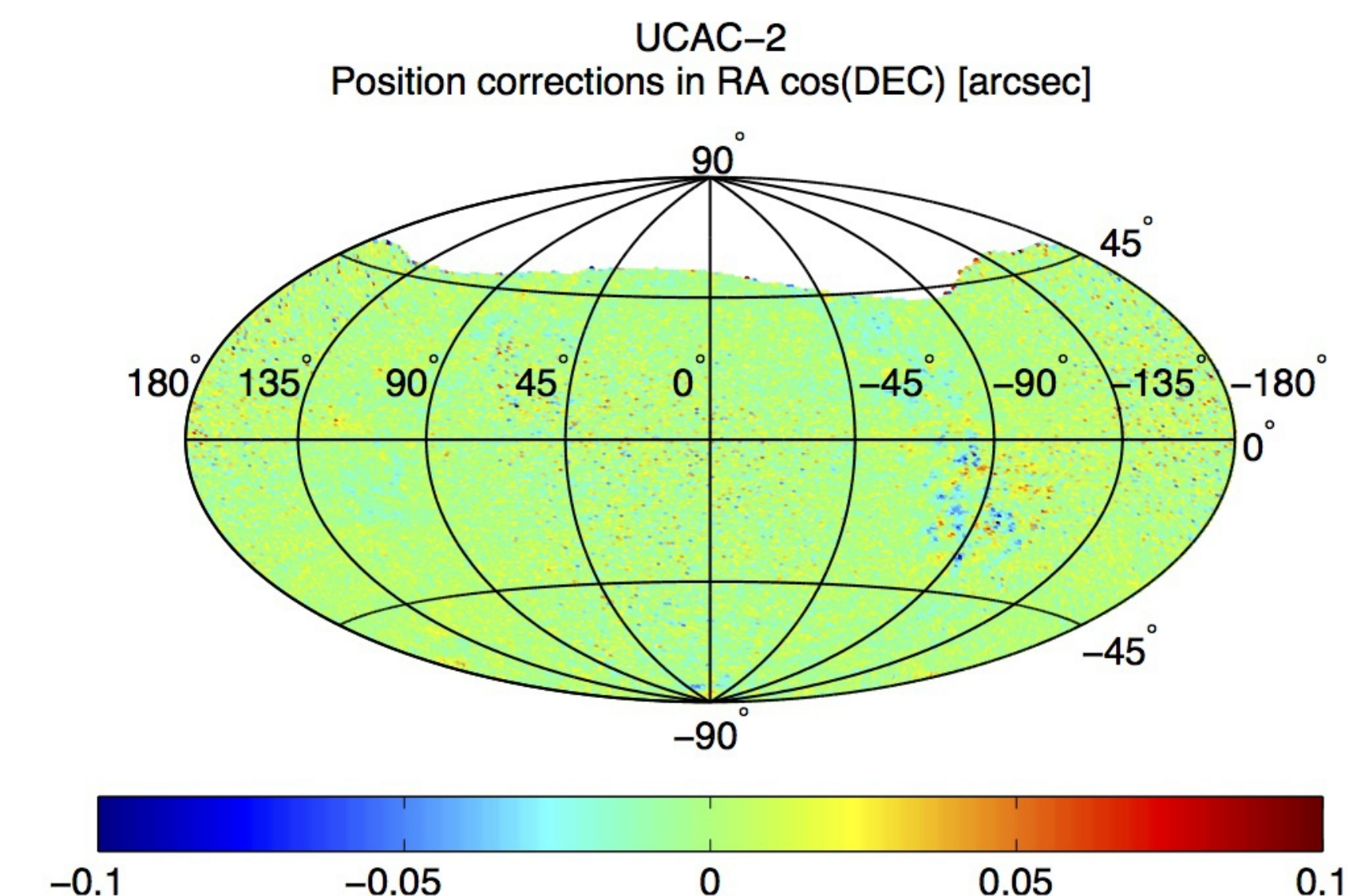}\includegraphics[width=0.6\textwidth]{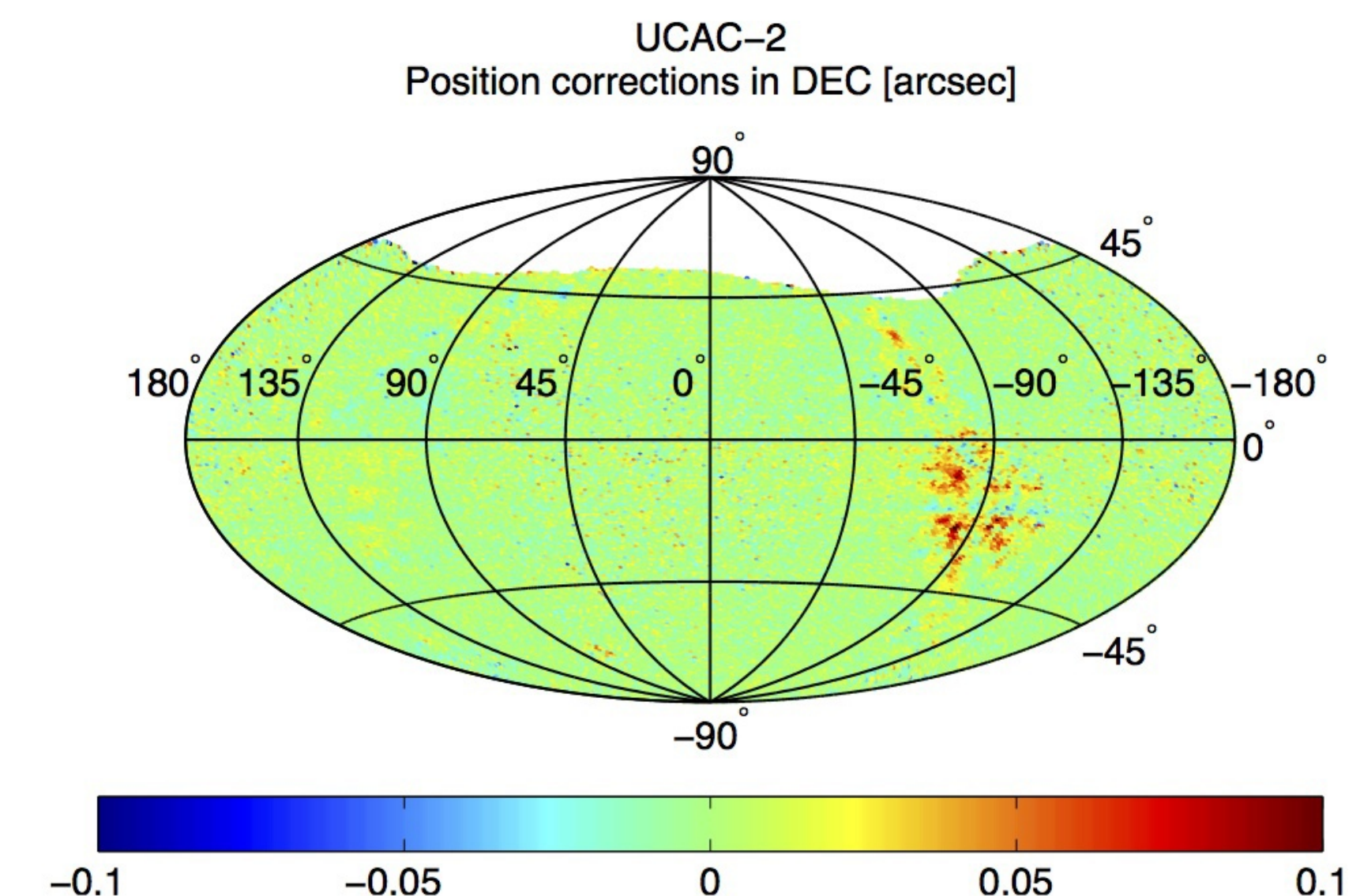}}
\centerline{\includegraphics[width=0.6\textwidth]{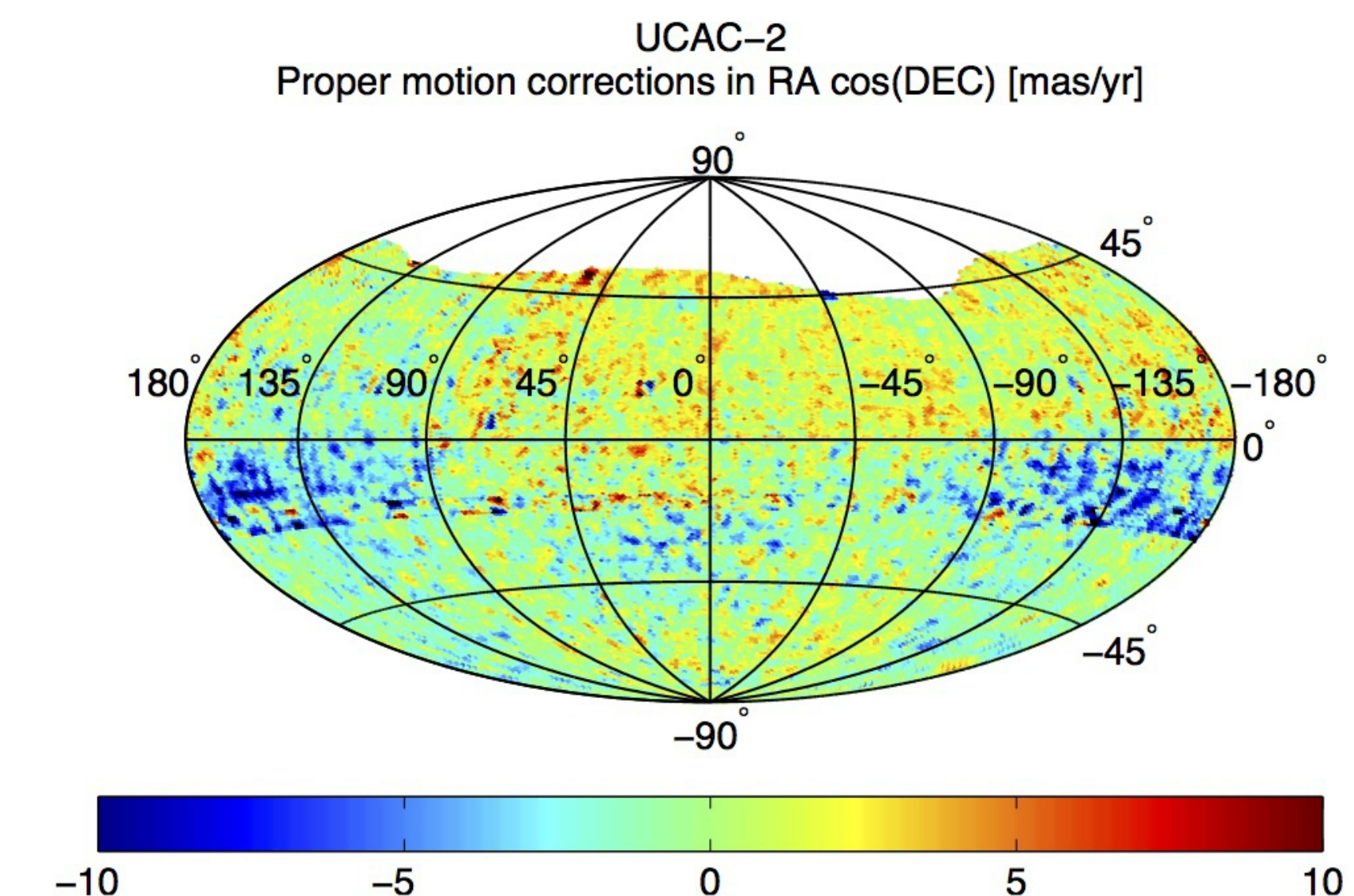}\includegraphics[width=0.6\textwidth]{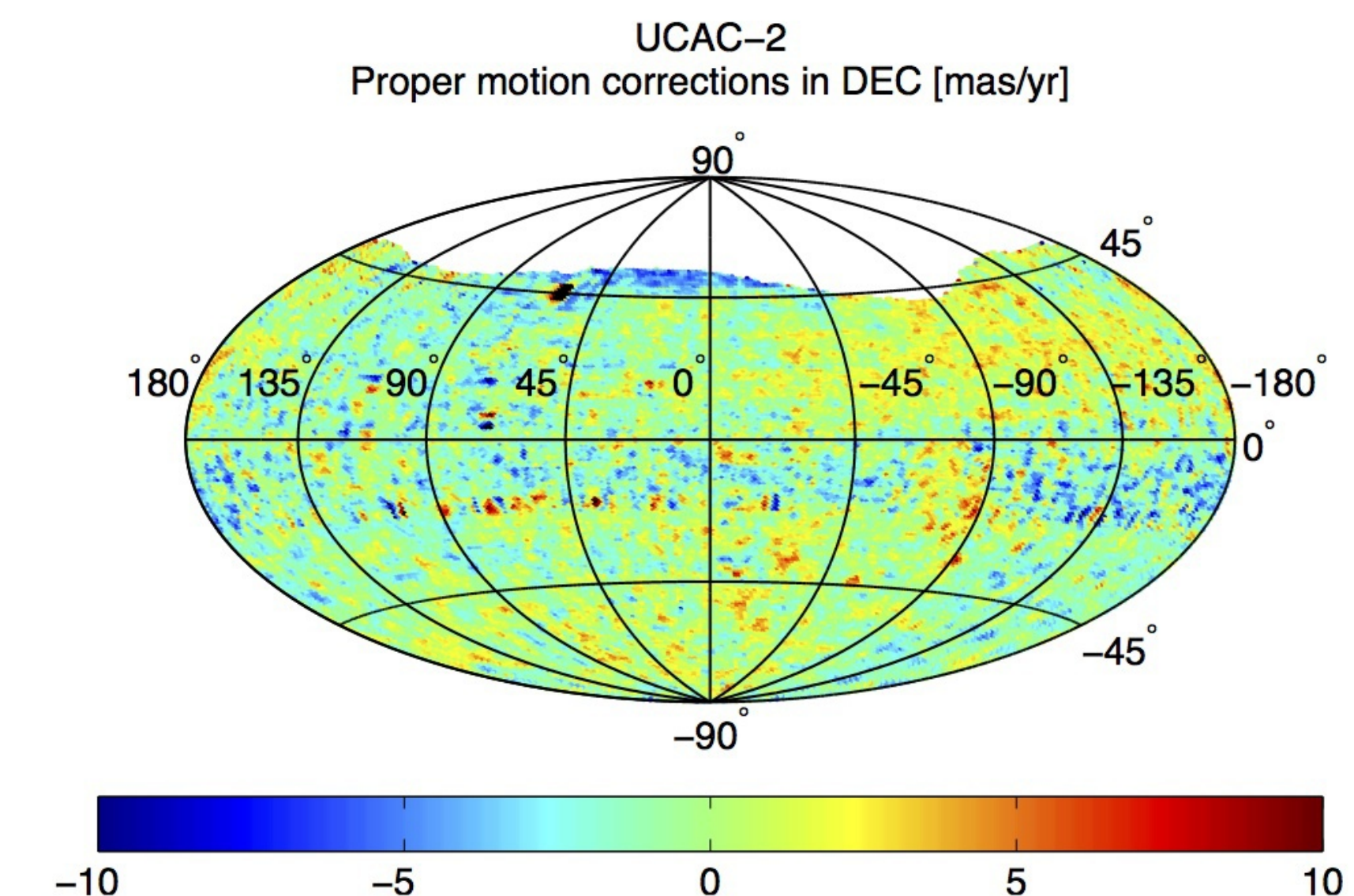}}
\caption{Top: J2000.0 position corrections in right ascension (left)
  and declination (right) for UCAC-2. Bottom: proper motion corrections
  in right ascension (left) and declination (right) for UCAC-2.}\label{f:ucac2}
\end{figure}

\begin{figure}
\centerline{\includegraphics[width=0.6\textwidth]{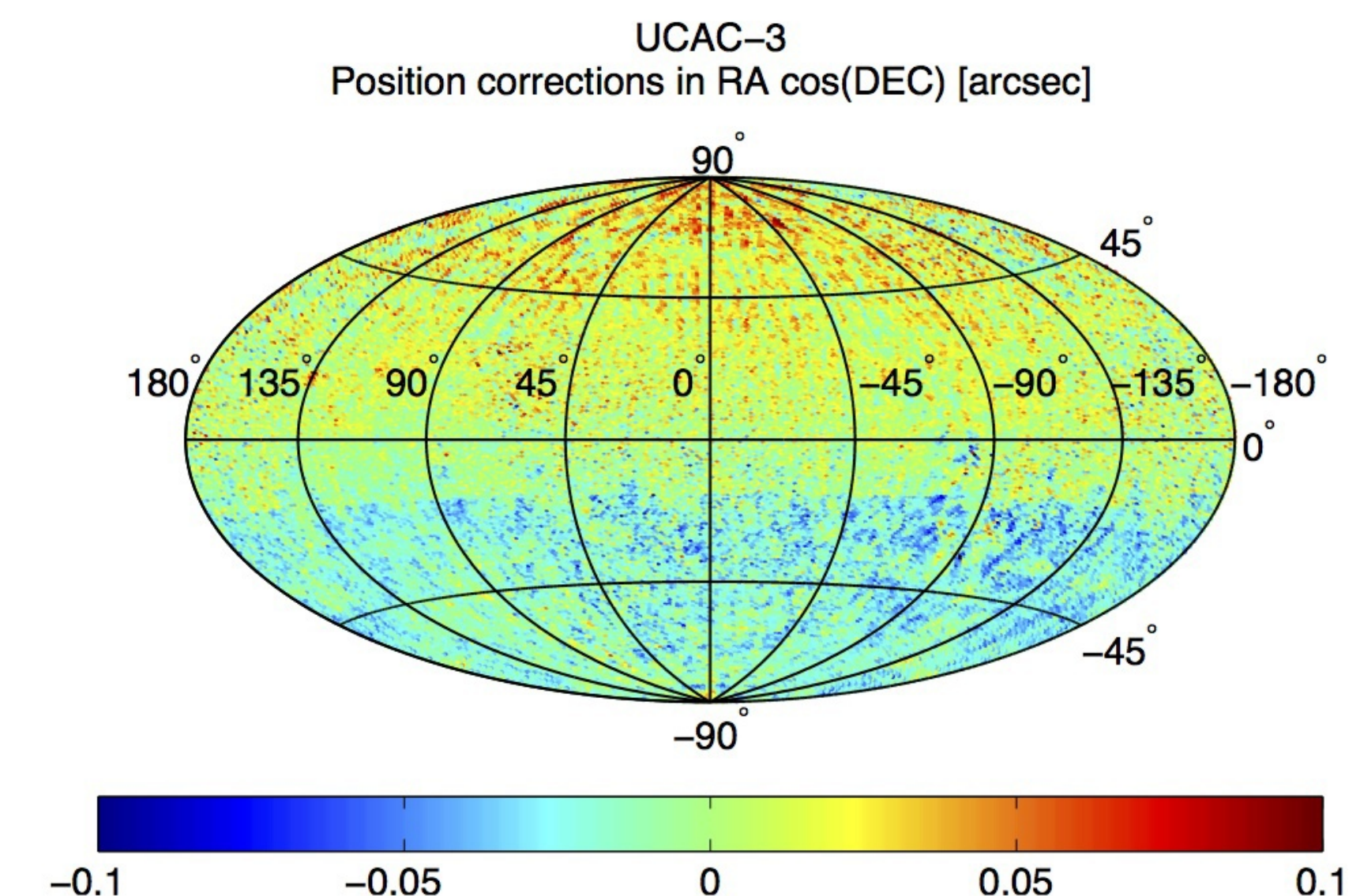}\includegraphics[width=0.6\textwidth]{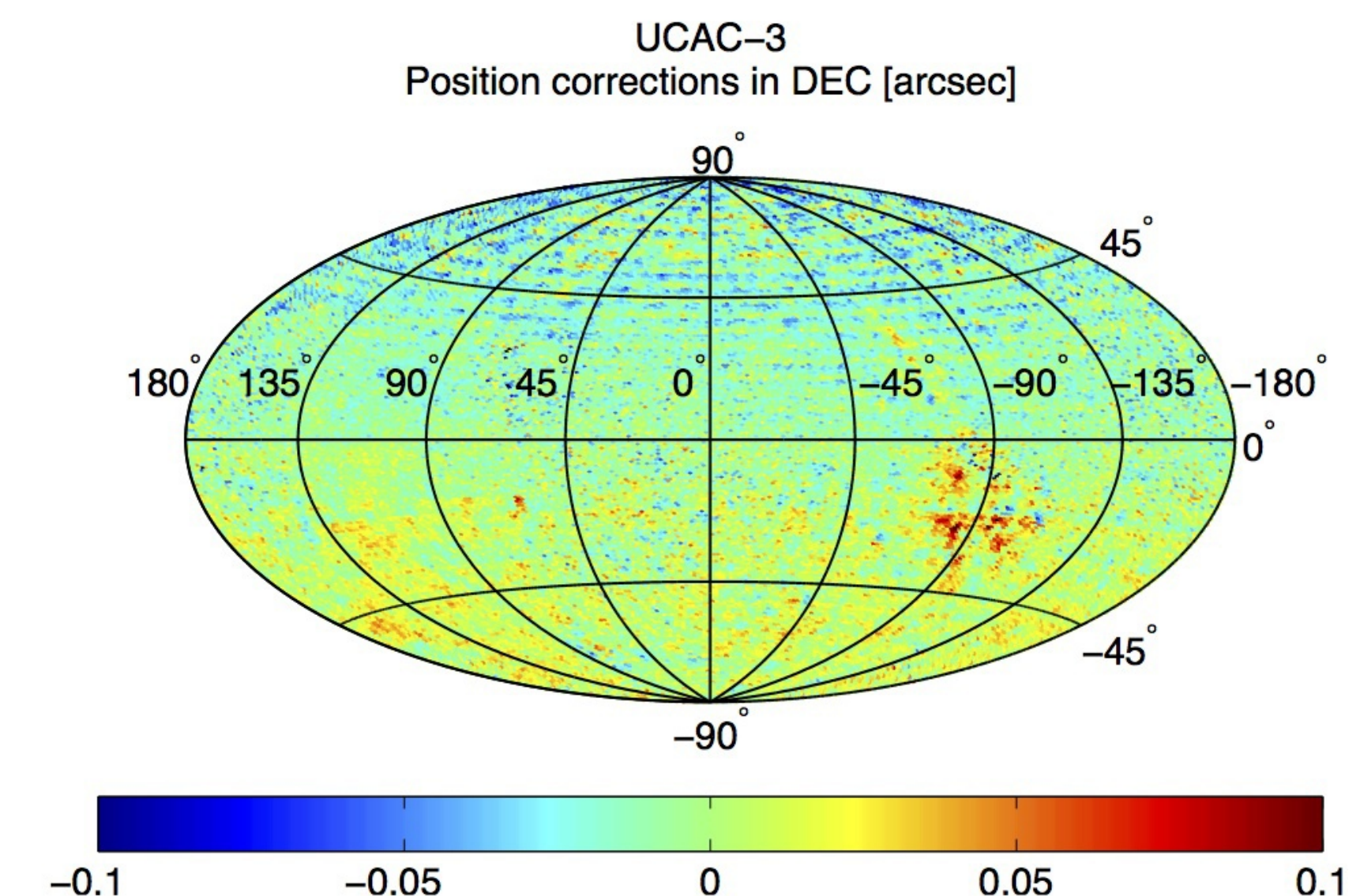}}
\centerline{\includegraphics[width=0.6\textwidth]{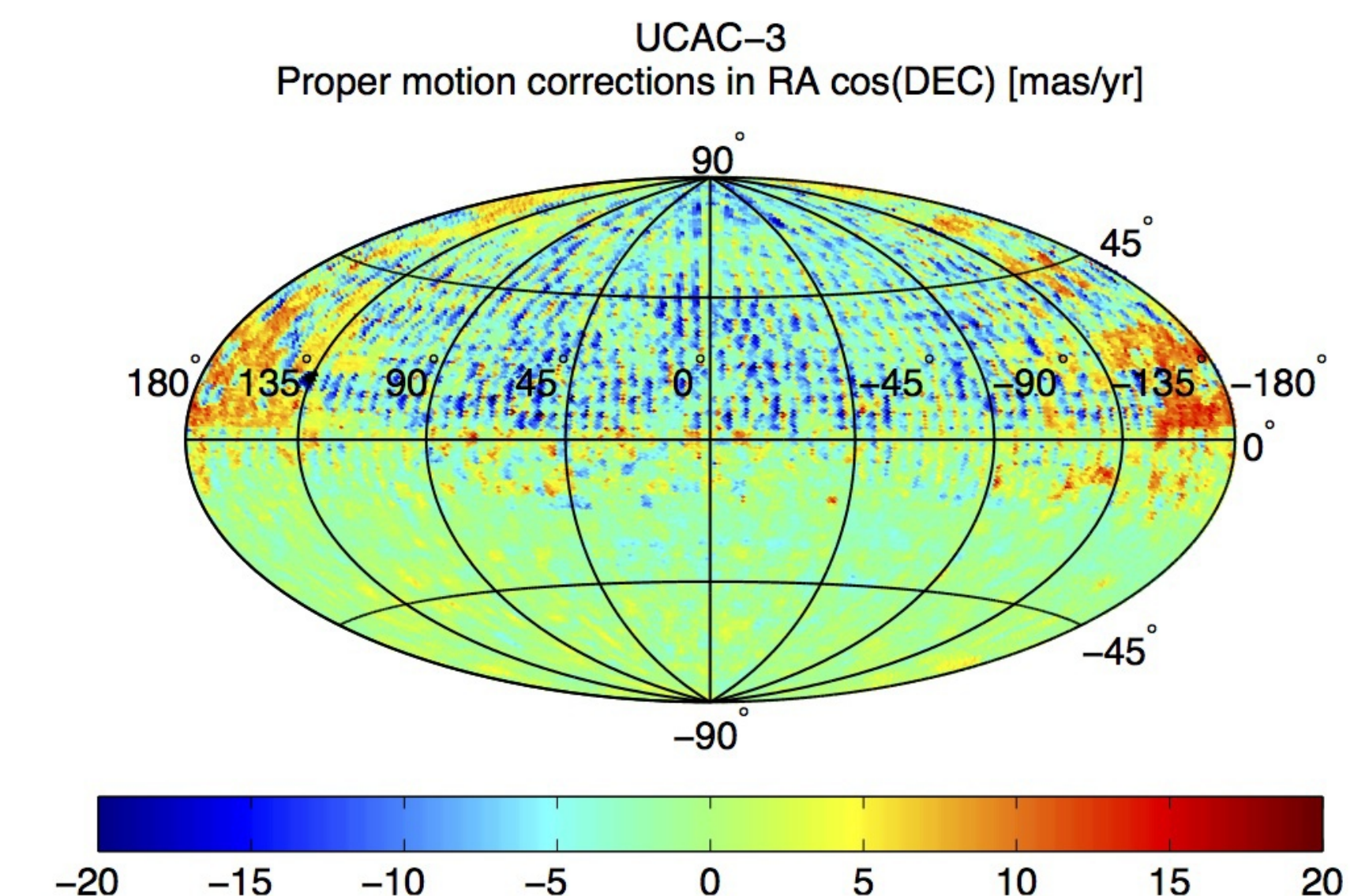}\includegraphics[width=0.6\textwidth]{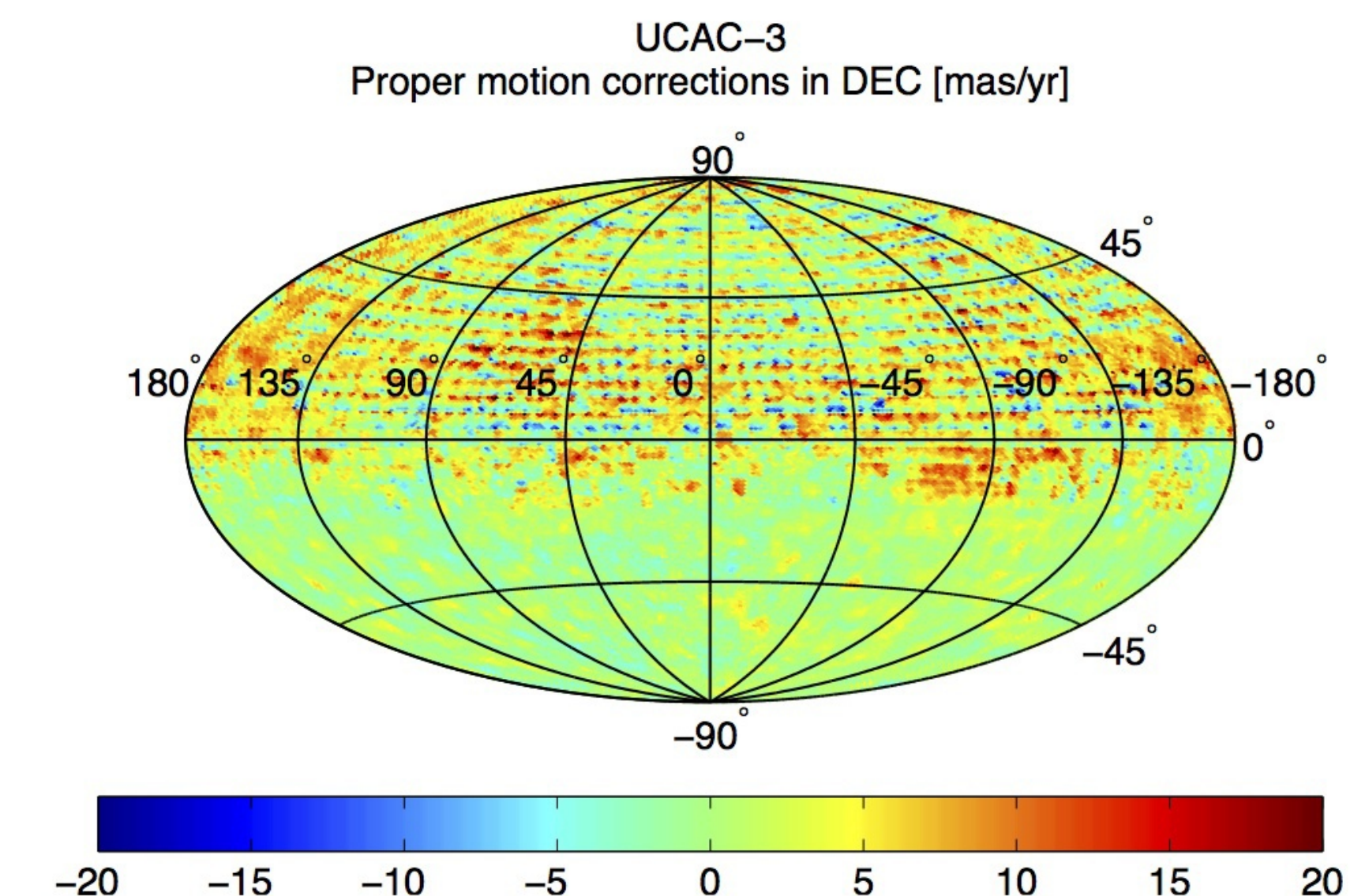}}
\caption{Top: J2000.0 position corrections in right ascension (left)
  and declination (right) for UCAC-3. Bottom: proper motion corrections
  in right ascension (left) and declination (right) for UCAC-3.}\label{f:ucac3}
\end{figure}

\begin{figure}
\centerline{\includegraphics[width=0.6\textwidth]{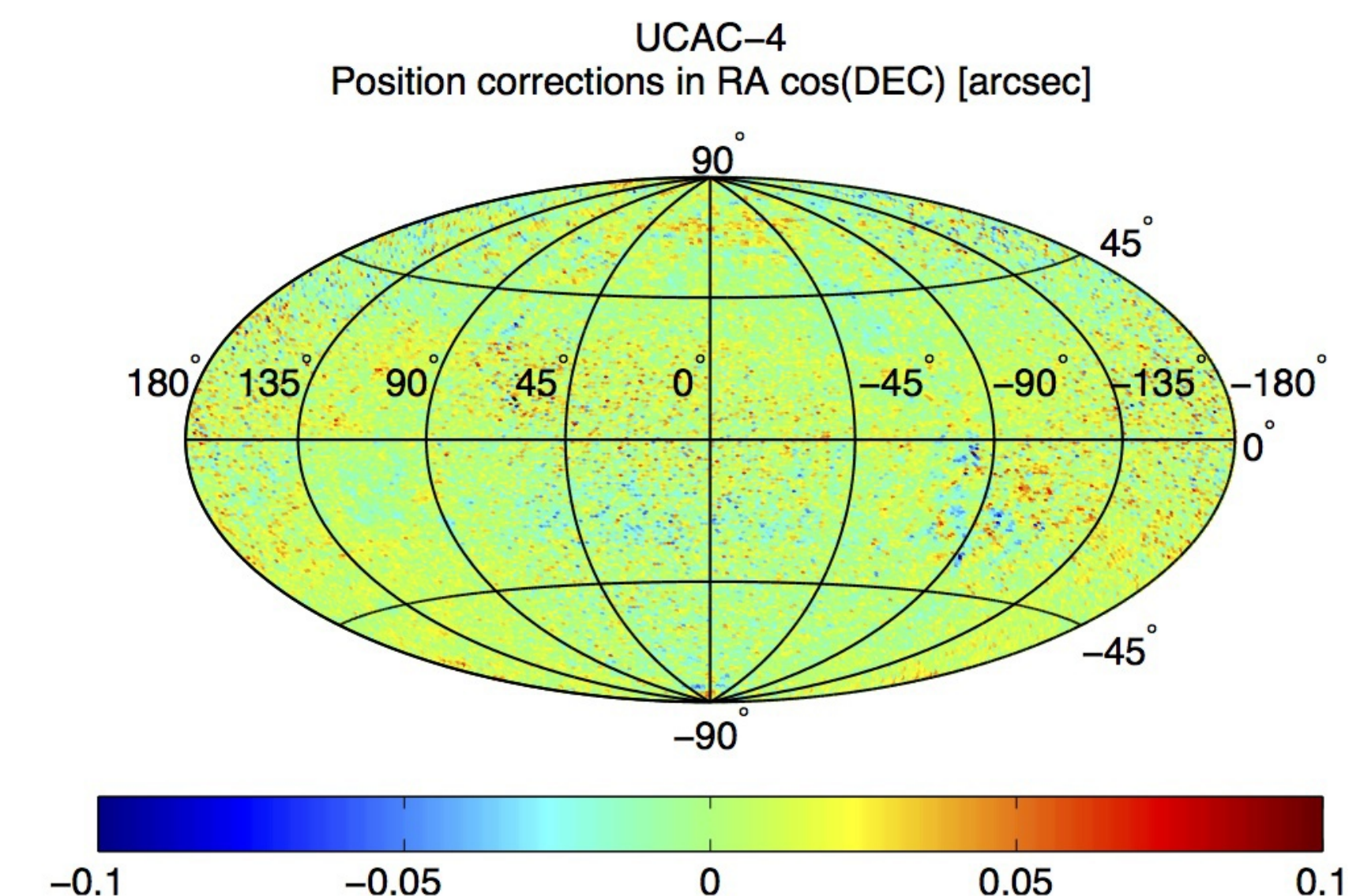}\includegraphics[width=0.6\textwidth]{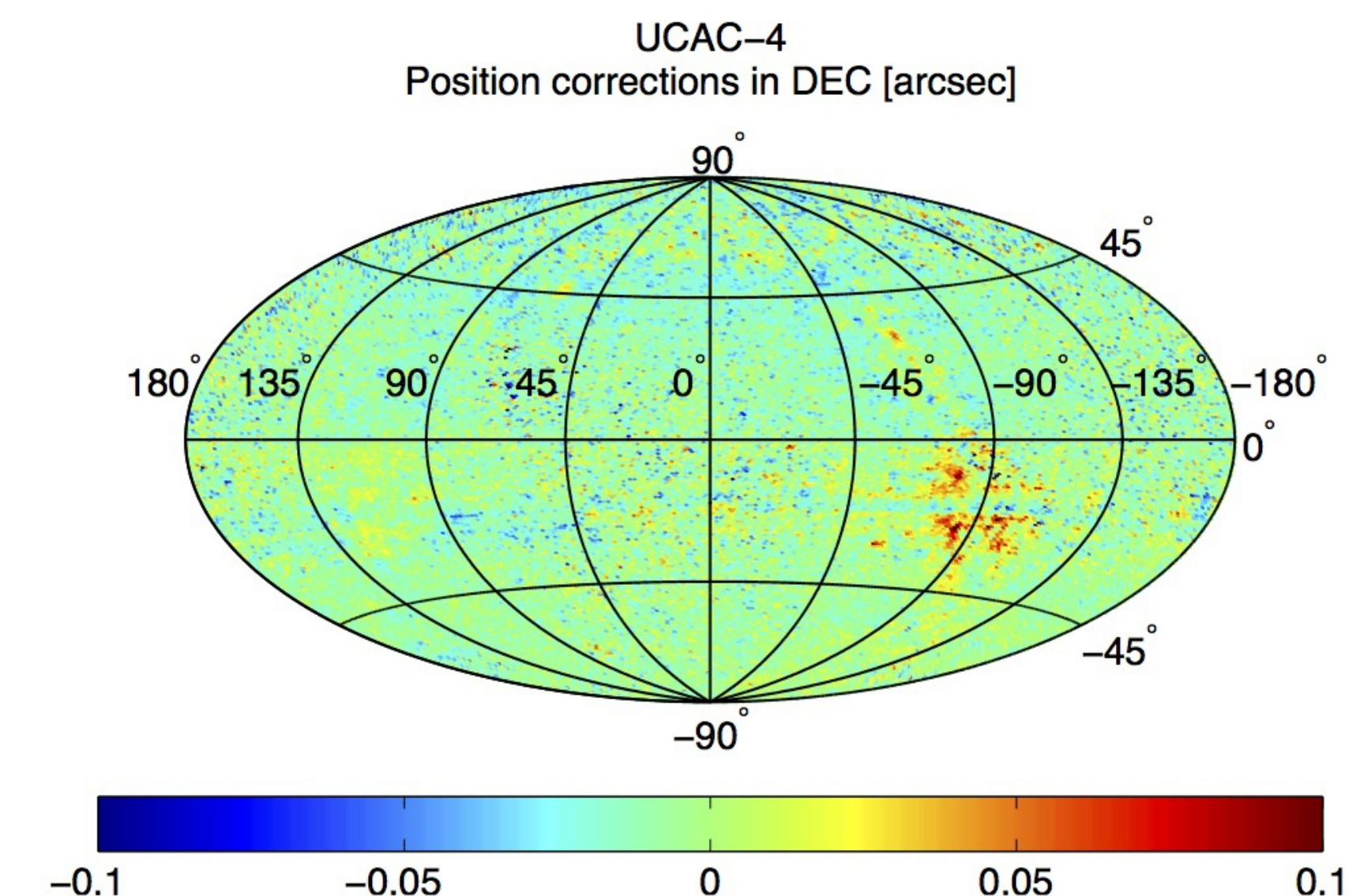}}
\centerline{\includegraphics[width=0.6\textwidth]{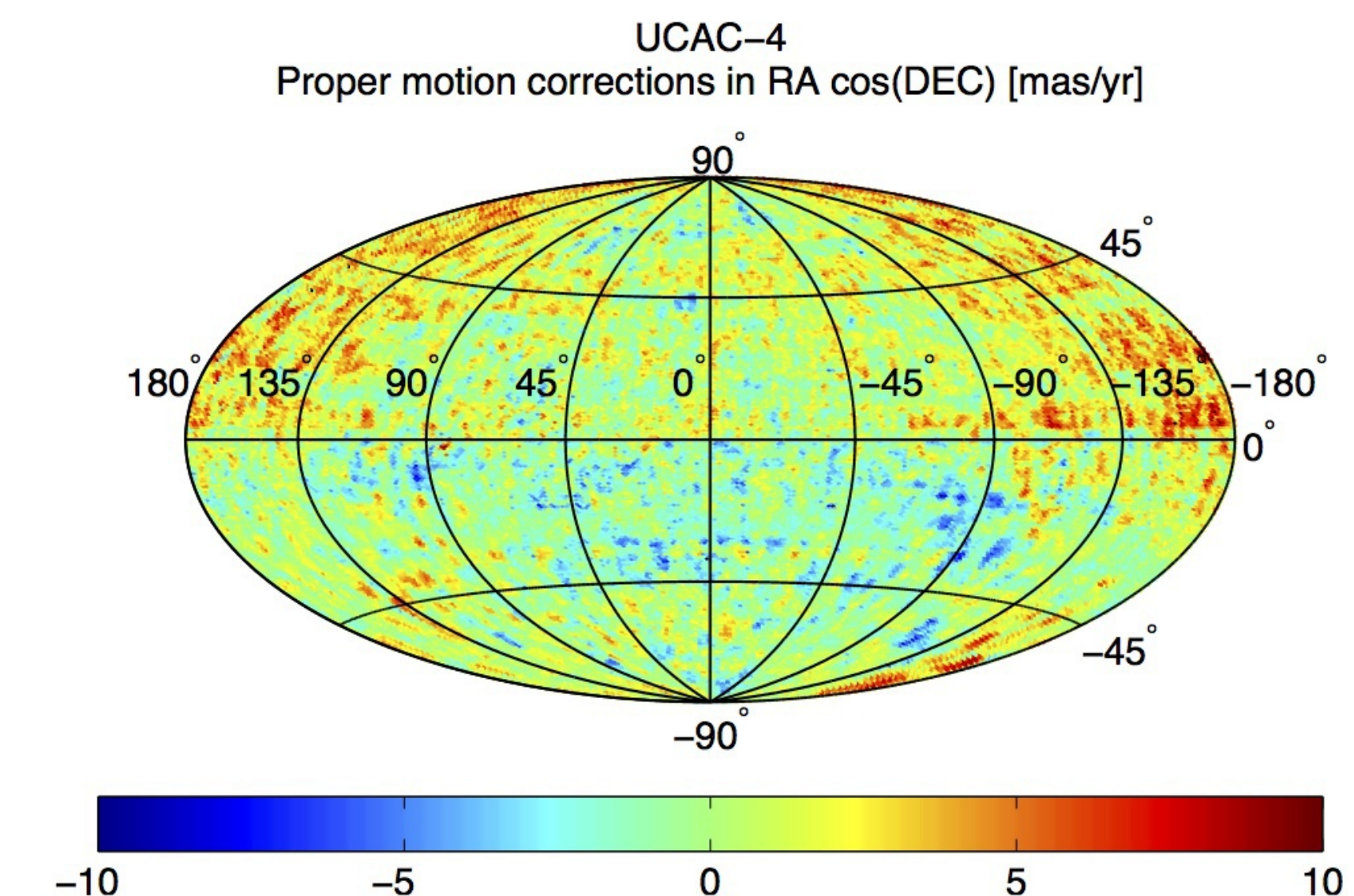}\includegraphics[width=0.6\textwidth]{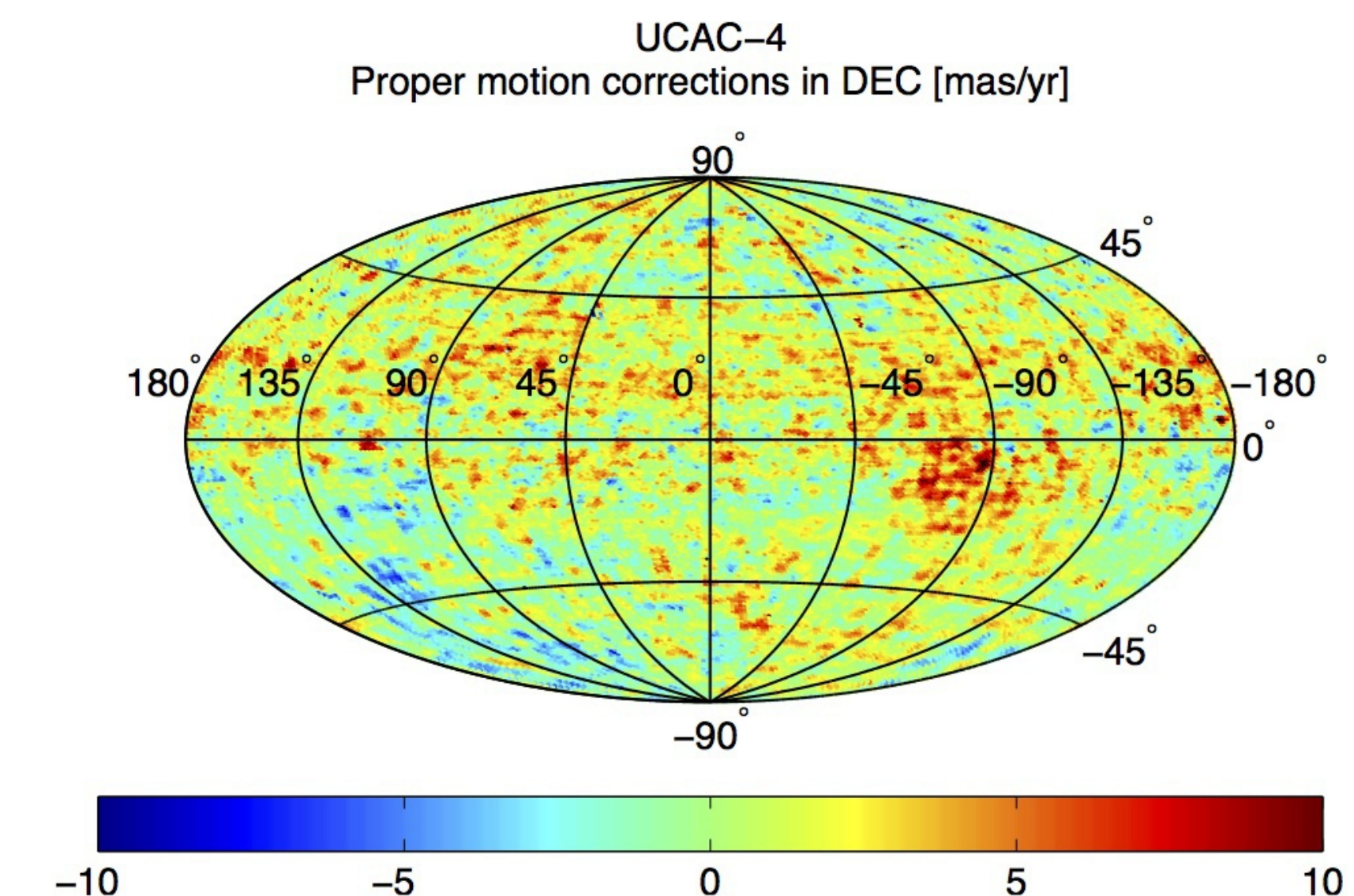}}
\caption{Top: J2000.0 position corrections in right ascension (left)
  and declination (right) for UCAC-4. Bottom: proper motion corrections
  in right ascension (left) and declination (right) for UCAC-4.}\label{f:ucac4}
\end{figure}

The UCAC catalogs (Fig.~\ref{f:ucac1} -- Fig.~\ref{f:ucac4}) provide
extremely good star positions, very close to those of our
reference. However, proper motions look problematic: UCAC-1 and UCAC-3
have significant corrections of the order of 5 mas/yr, while UCAC-2
and UCAC-4 seem to have better proper motions. \citet{ppmxl} report
problems in UCAC-3 proper motions, in particular for declinations
greater than $-20^\circ$.

Despite being the final product of the UCAC series, UCAC-4 has some
regional dependence of the proper motions suggesting that there are
still unresolved issues with proper motions.  Moreover, the comparison
between UCAC-4 and Tycho-2 provides average proper motion corrections
of $\sim$ 2 mas/yr, while the comparison between the subset of PPMXL
that we are using as reference and Tycho-2 gives average proper motion
differences $< 1$ mas/yr. Due to the high quality of Tycho-2, these
differences further suggest that UCAC-4 proper motions have
correctable errors.

These indications suggest that proper motion in the UCAC catalogs should be
corrected, but corrections in positions are small. For consistency,
we correct both positions and proper motions.

\subsection{GSC catalogs}
\begin{figure}
\centerline{\includegraphics[width=0.6\textwidth]{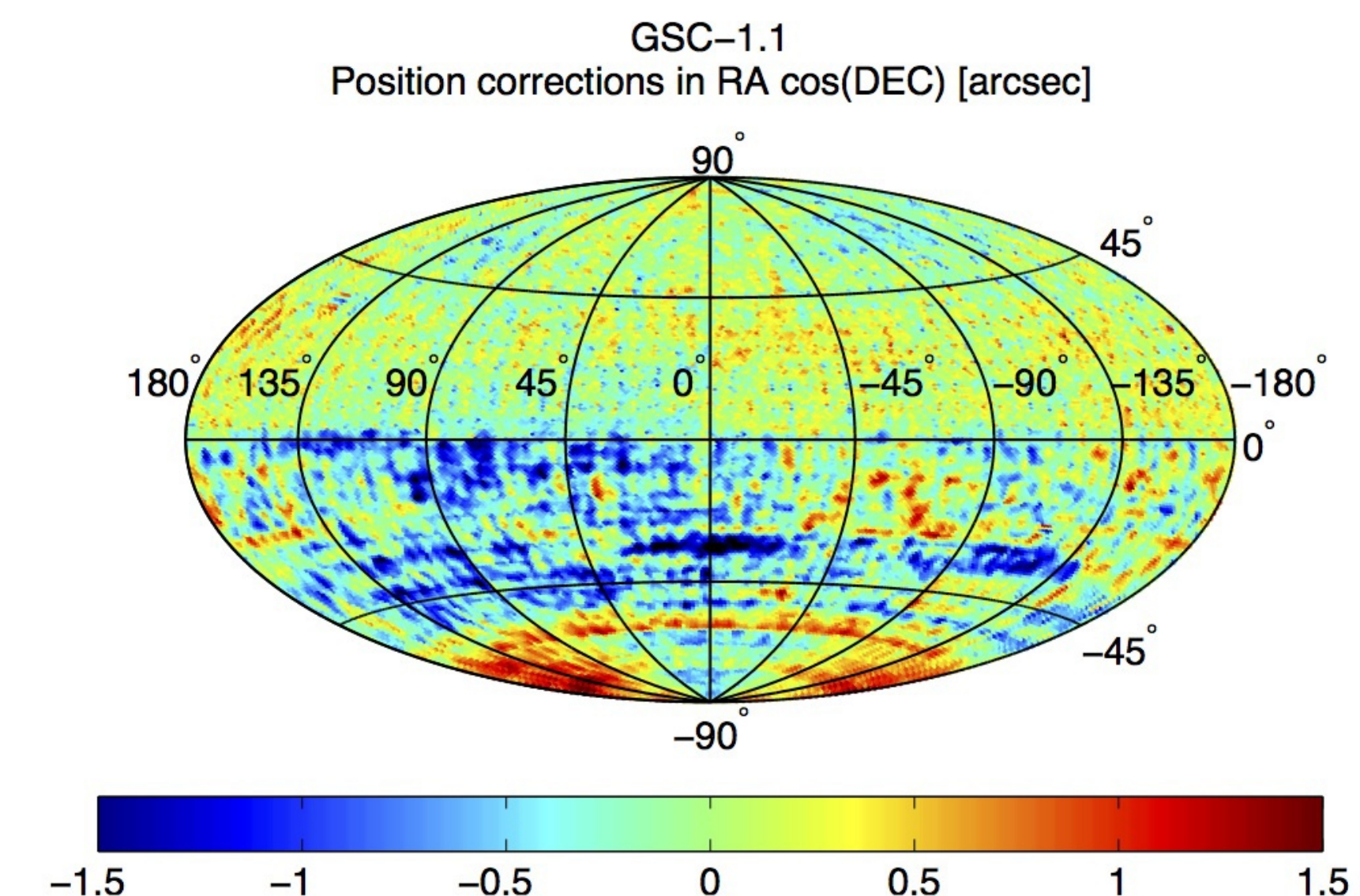}\includegraphics[width=0.6\textwidth]{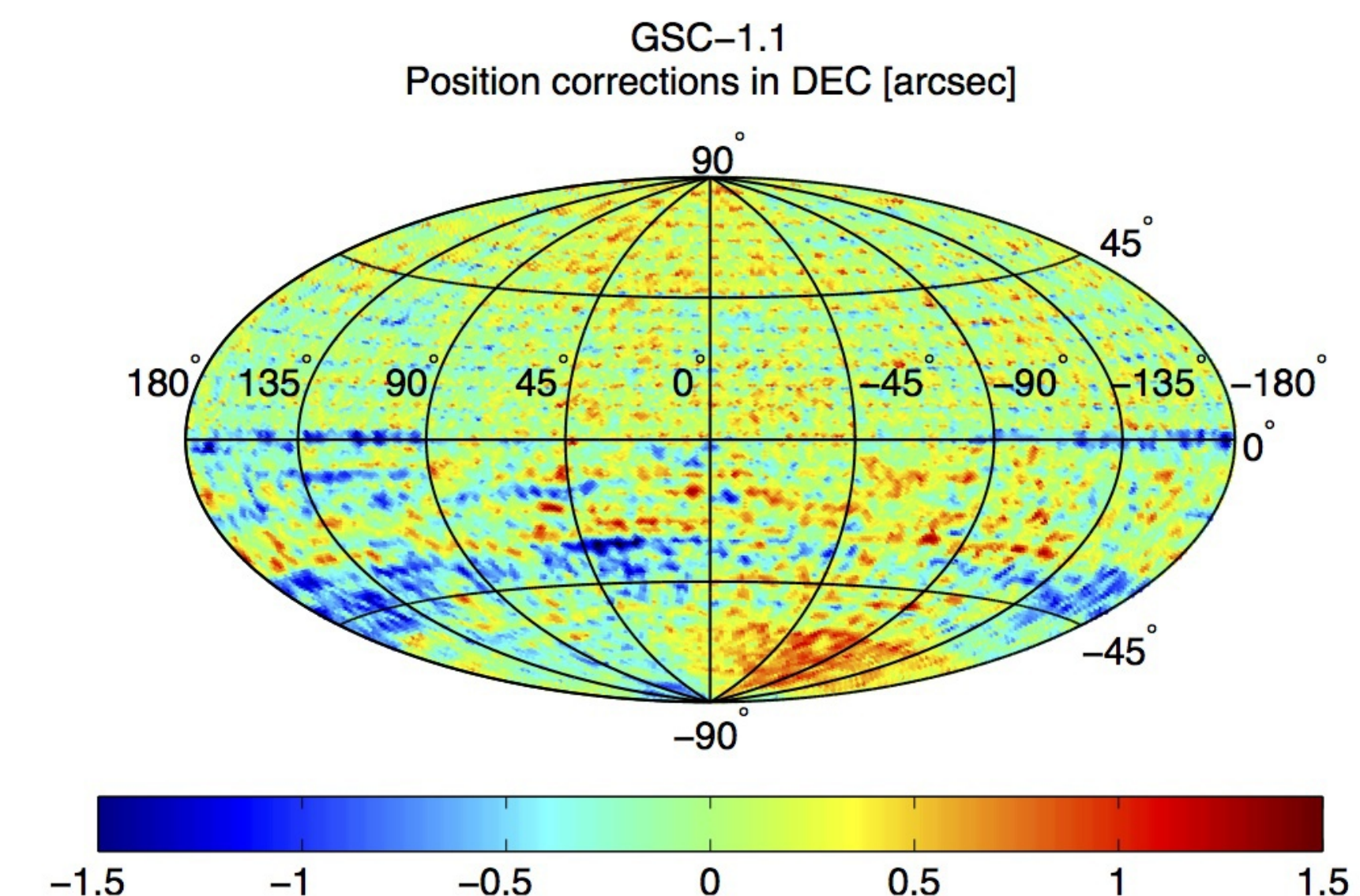}}
\centerline{\includegraphics[width=0.6\textwidth]{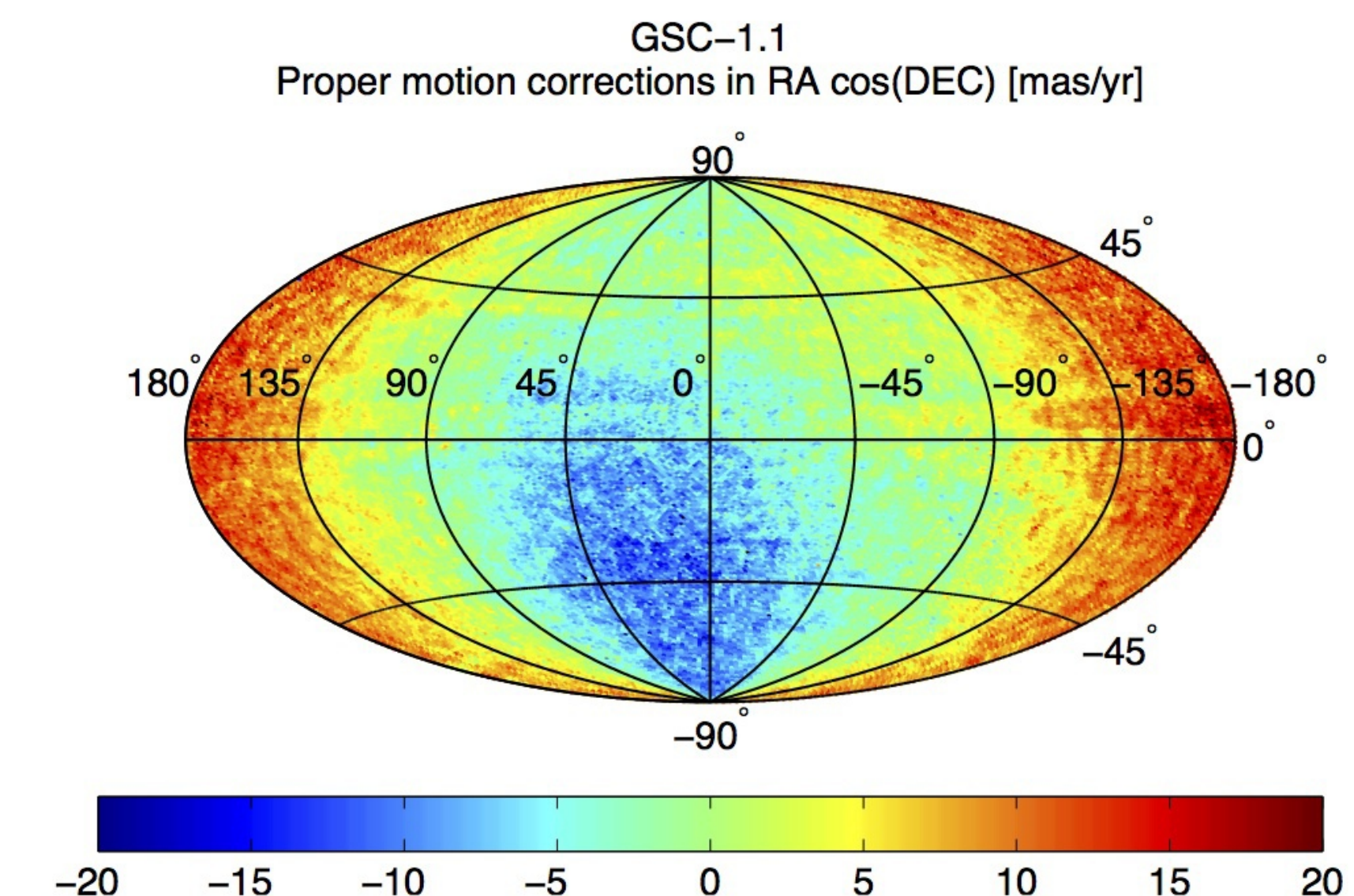}\includegraphics[width=0.6\textwidth]{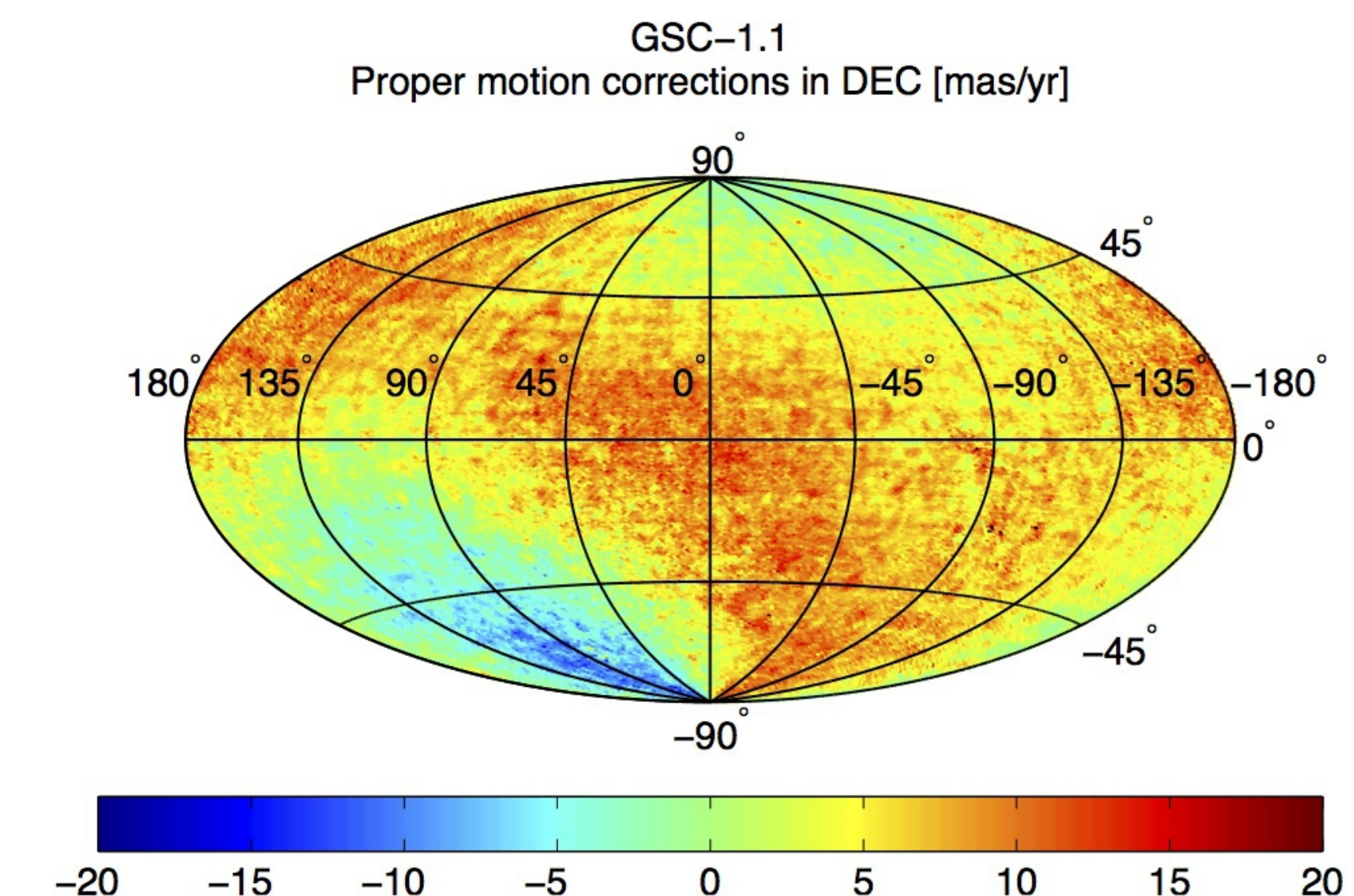}}
\caption{Top: J2000.0 position corrections in right ascension (left)
  and declination (right) for GSC-1.1. Bottom: proper motion corrections
  in right ascension (left) and declination (right) for GSC-1.1.}\label{f:gsc_1.1}
\end{figure}

\begin{figure}
\centerline{\includegraphics[width=0.6\textwidth]{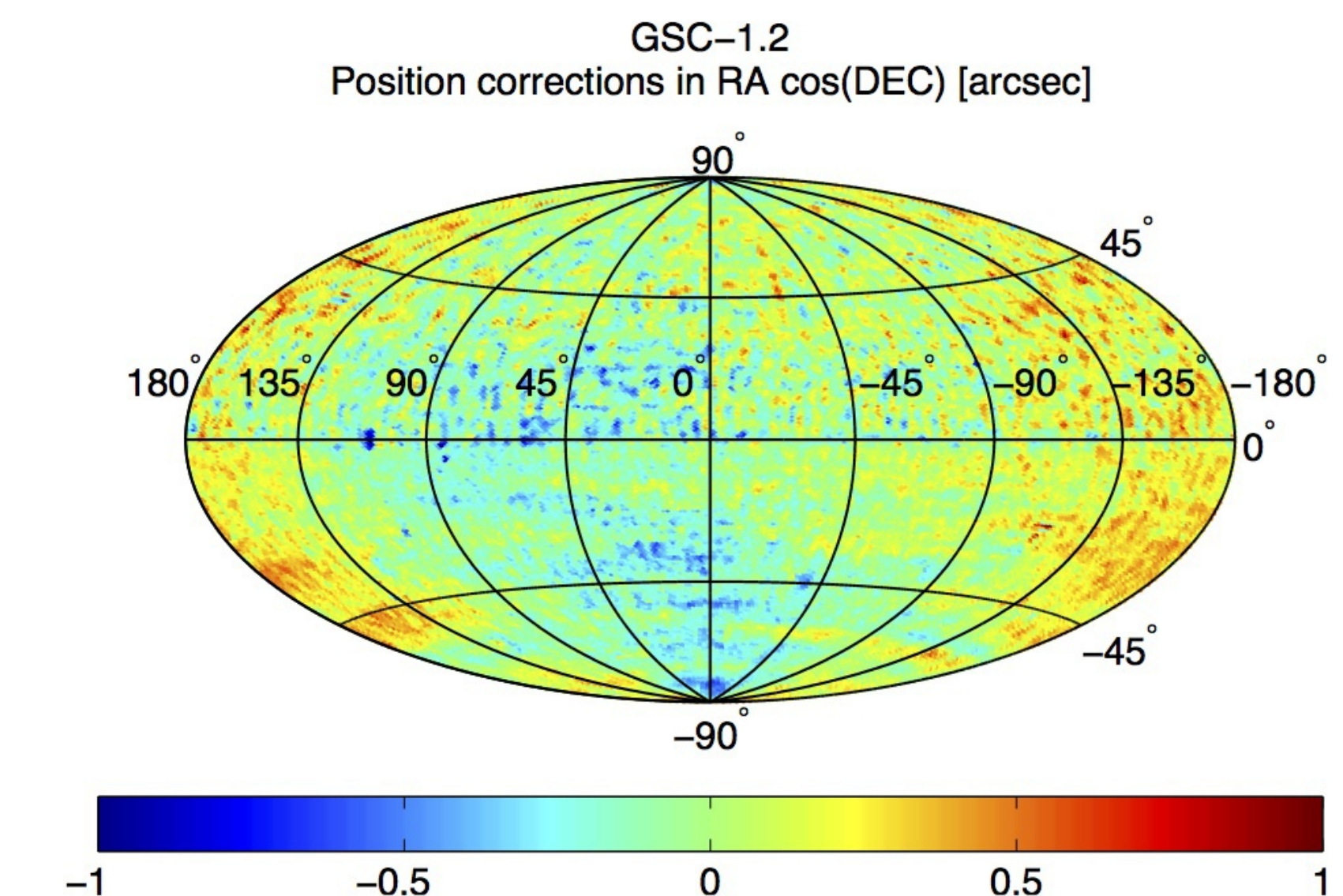}\includegraphics[width=0.6\textwidth]{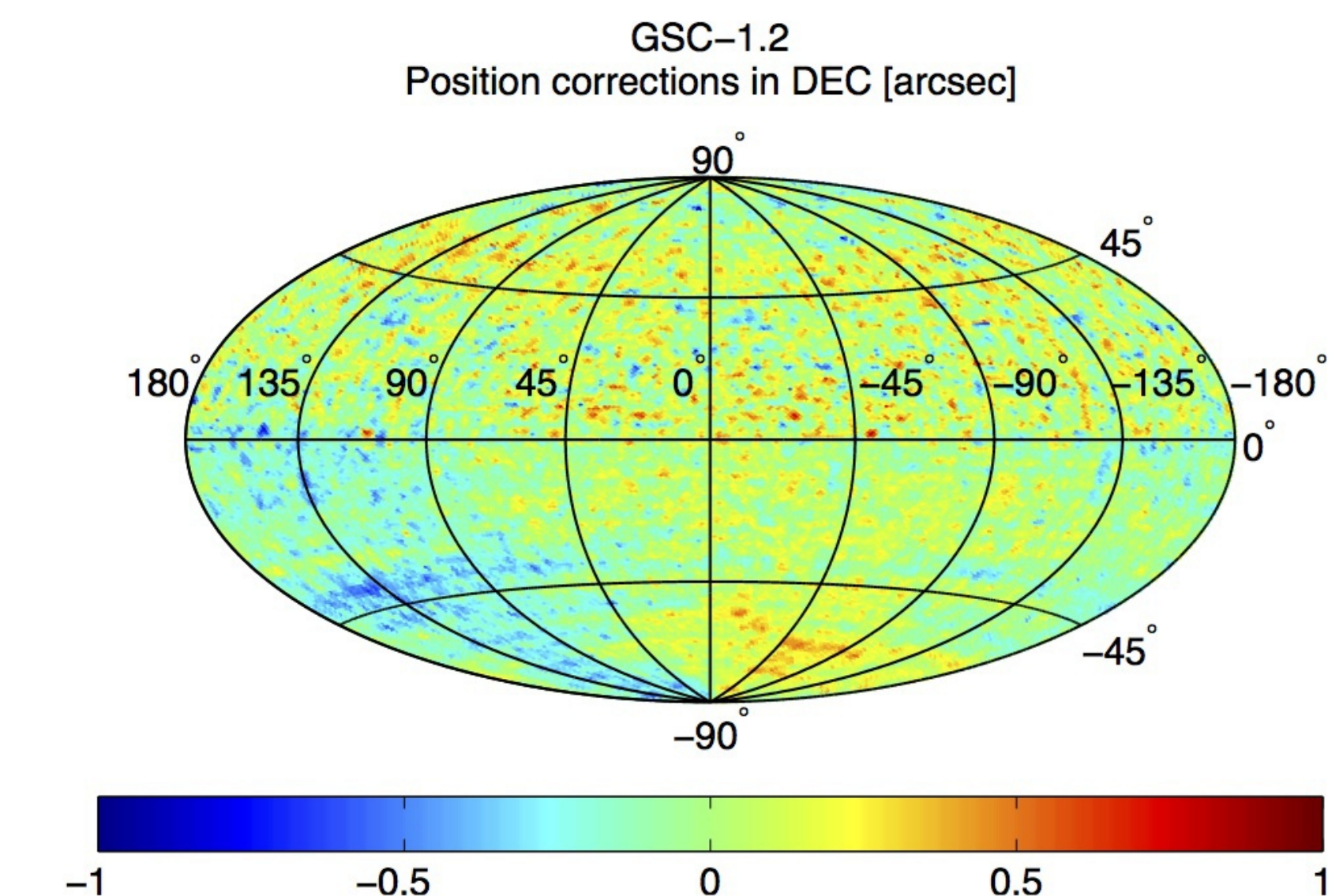}}
\centerline{\includegraphics[width=0.6\textwidth]{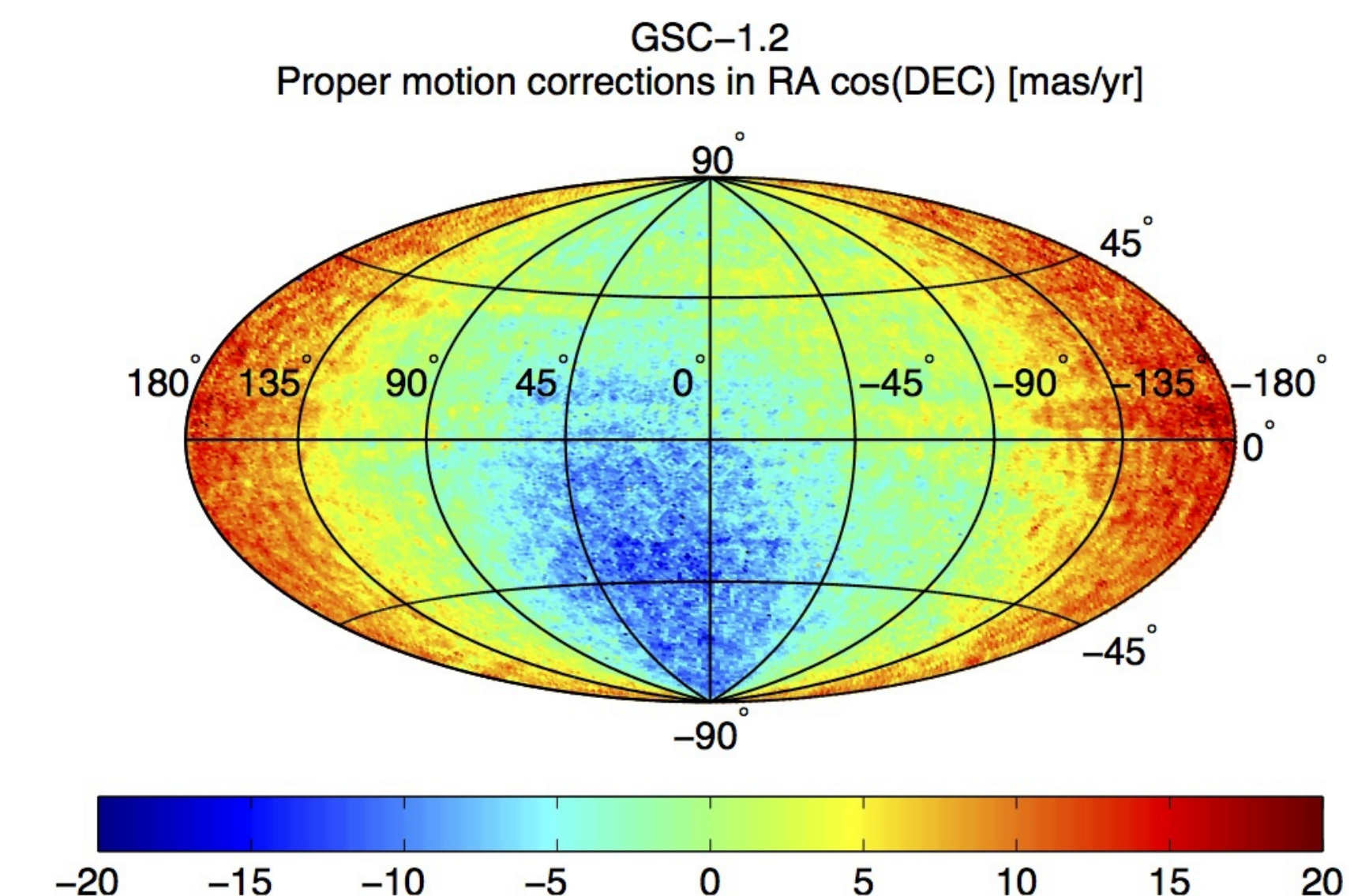}\includegraphics[width=0.6\textwidth]{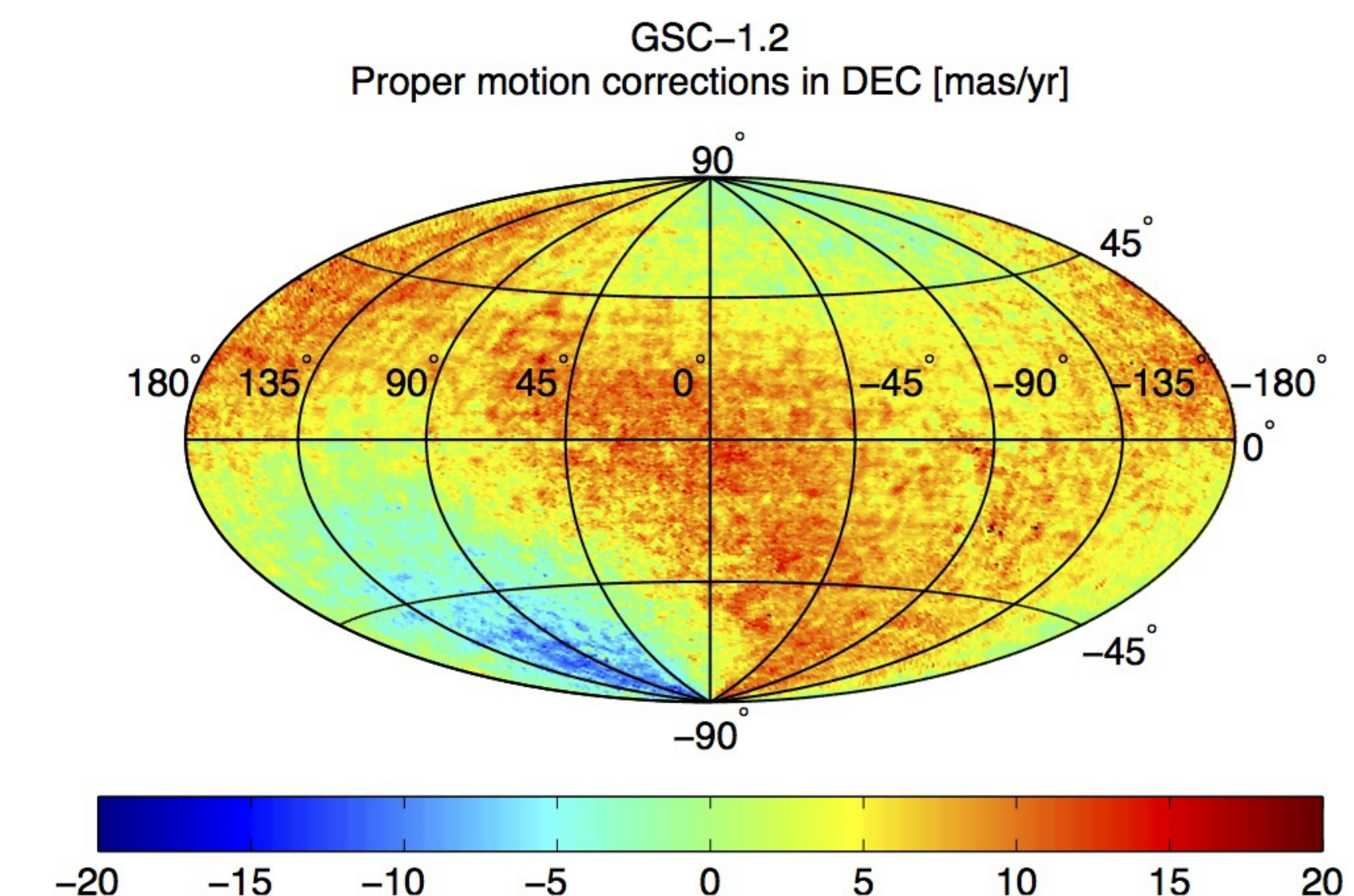}}
\caption{Top: J2000.0 position corrections in right ascension (left)
  and declination (right) for GSC-1.2. Bottom: proper motion corrections
  in right ascension (left) and declination (right) for GSC-1.2.}\label{f:gsc_1.2}
\end{figure}

\begin{figure}
\centerline{\includegraphics[width=0.6\textwidth]{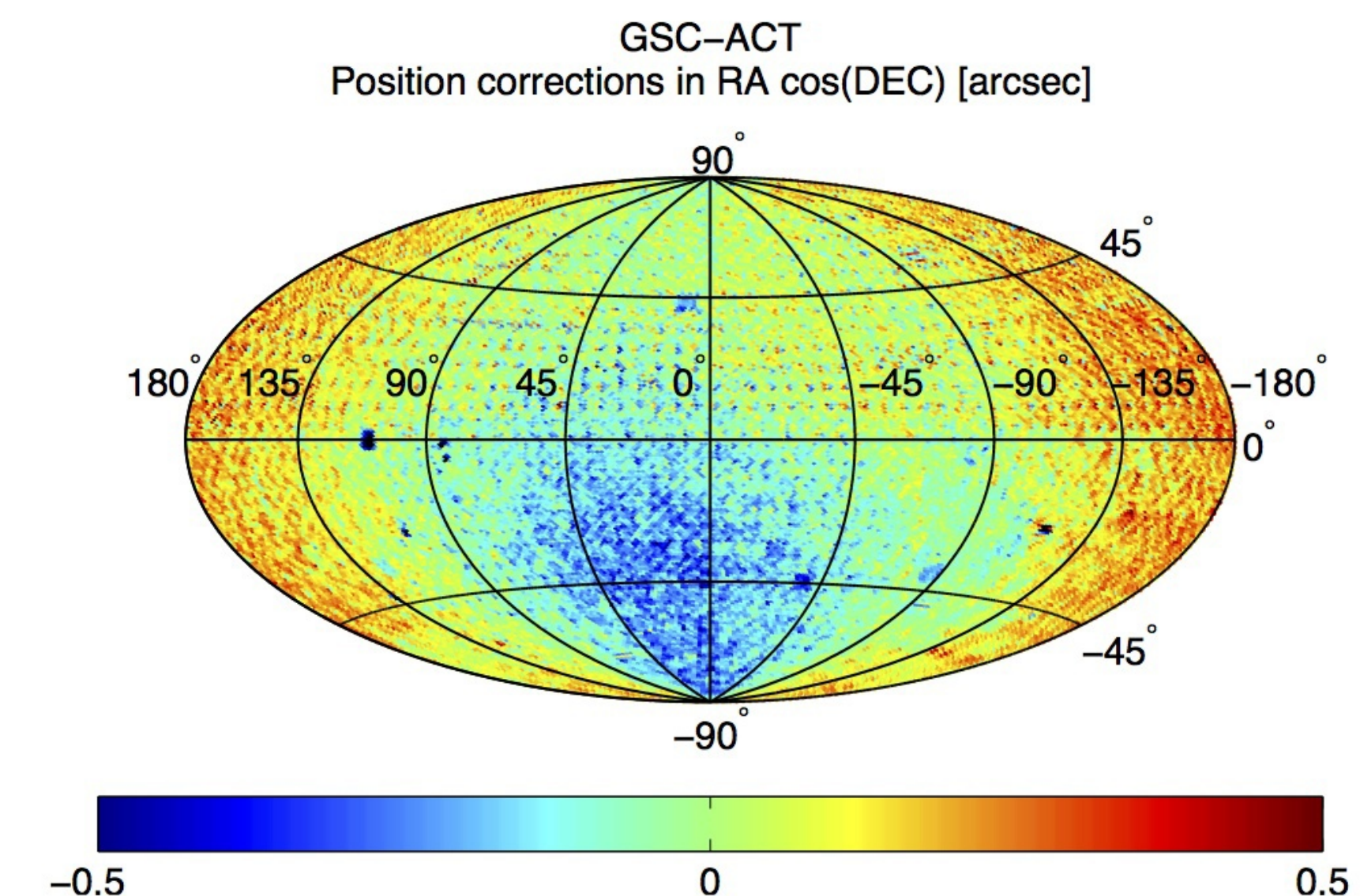}\includegraphics[width=0.6\textwidth]{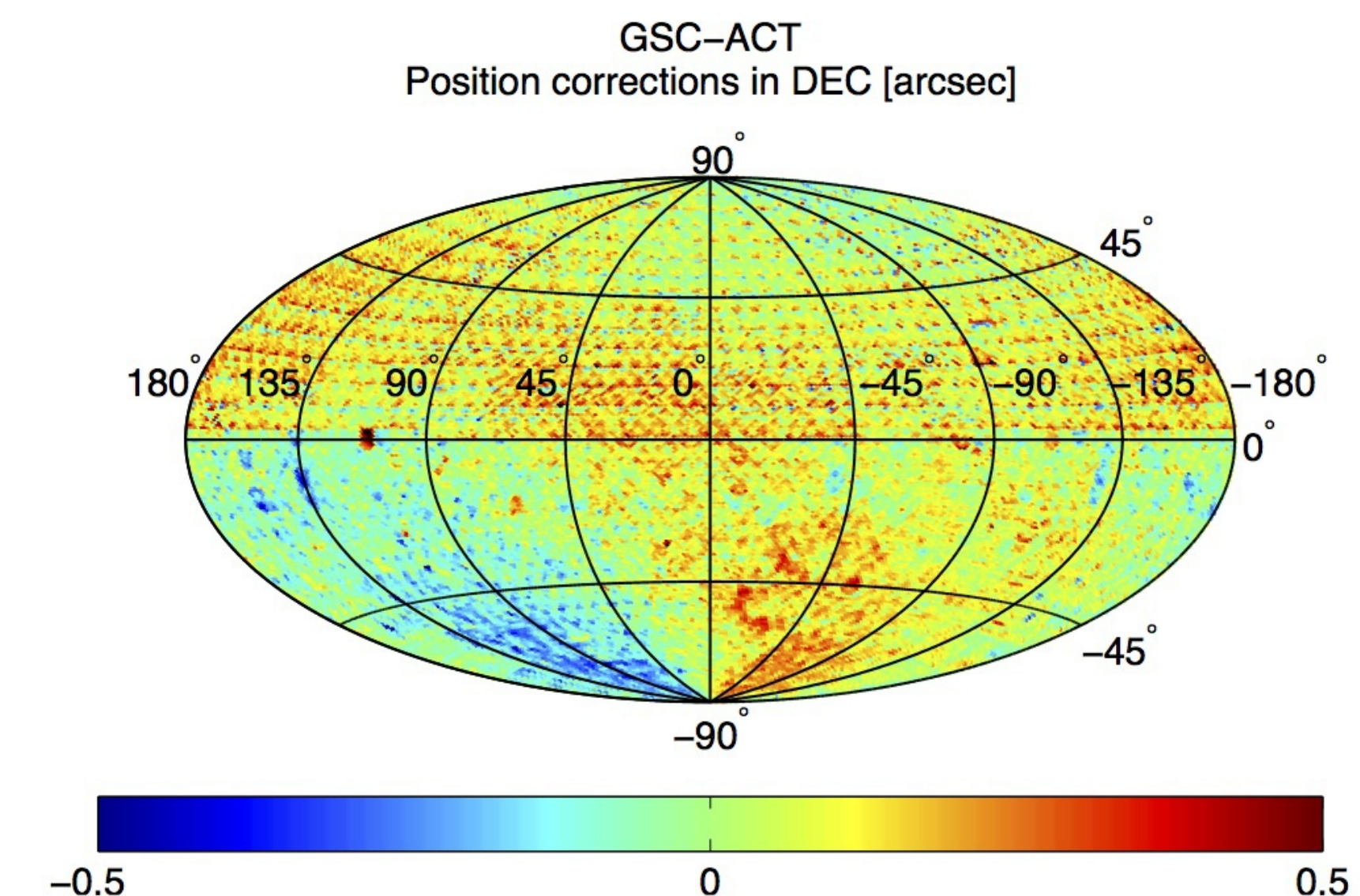}}
\centerline{\includegraphics[width=0.6\textwidth]{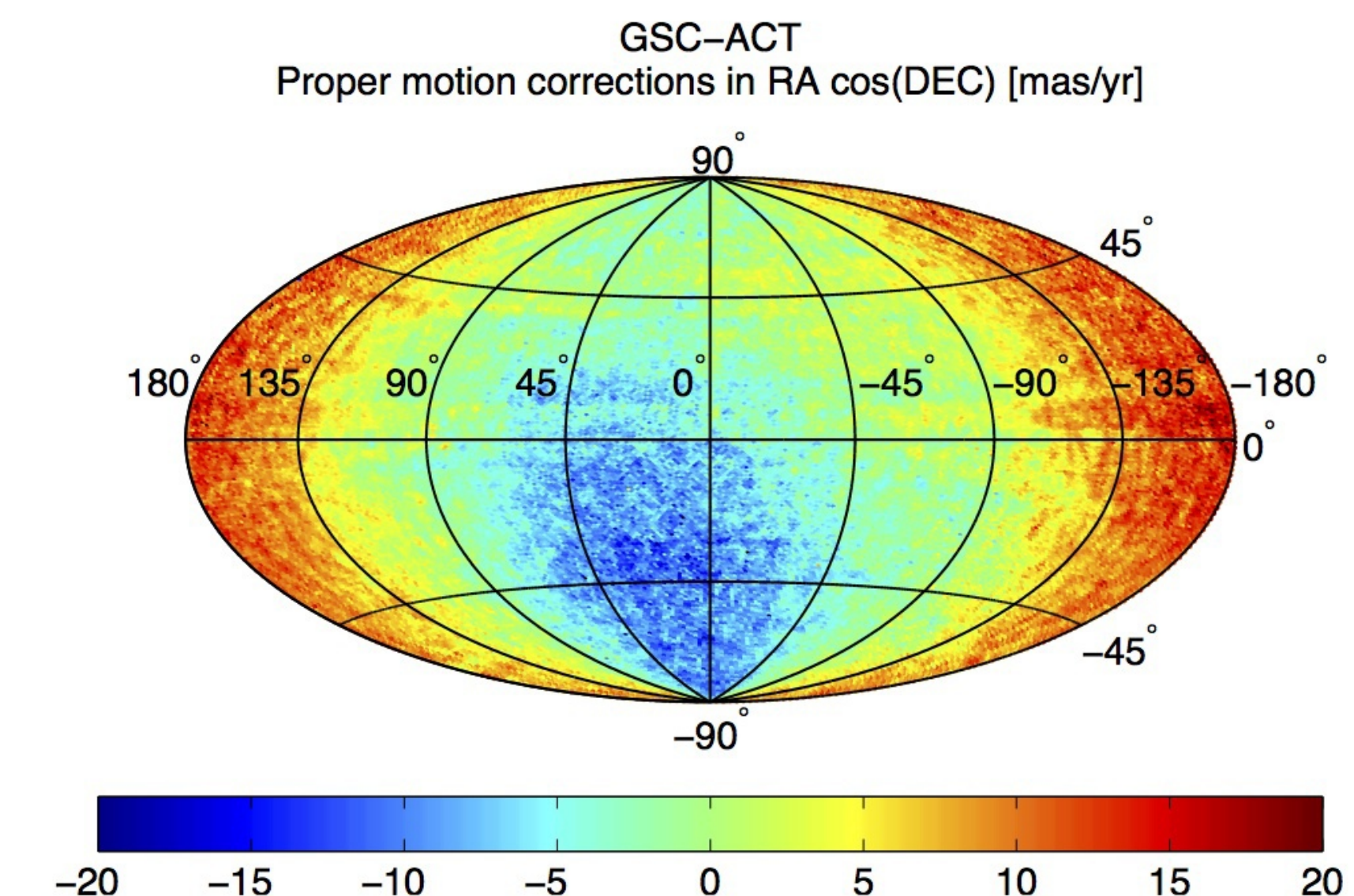}\includegraphics[width=0.6\textwidth]{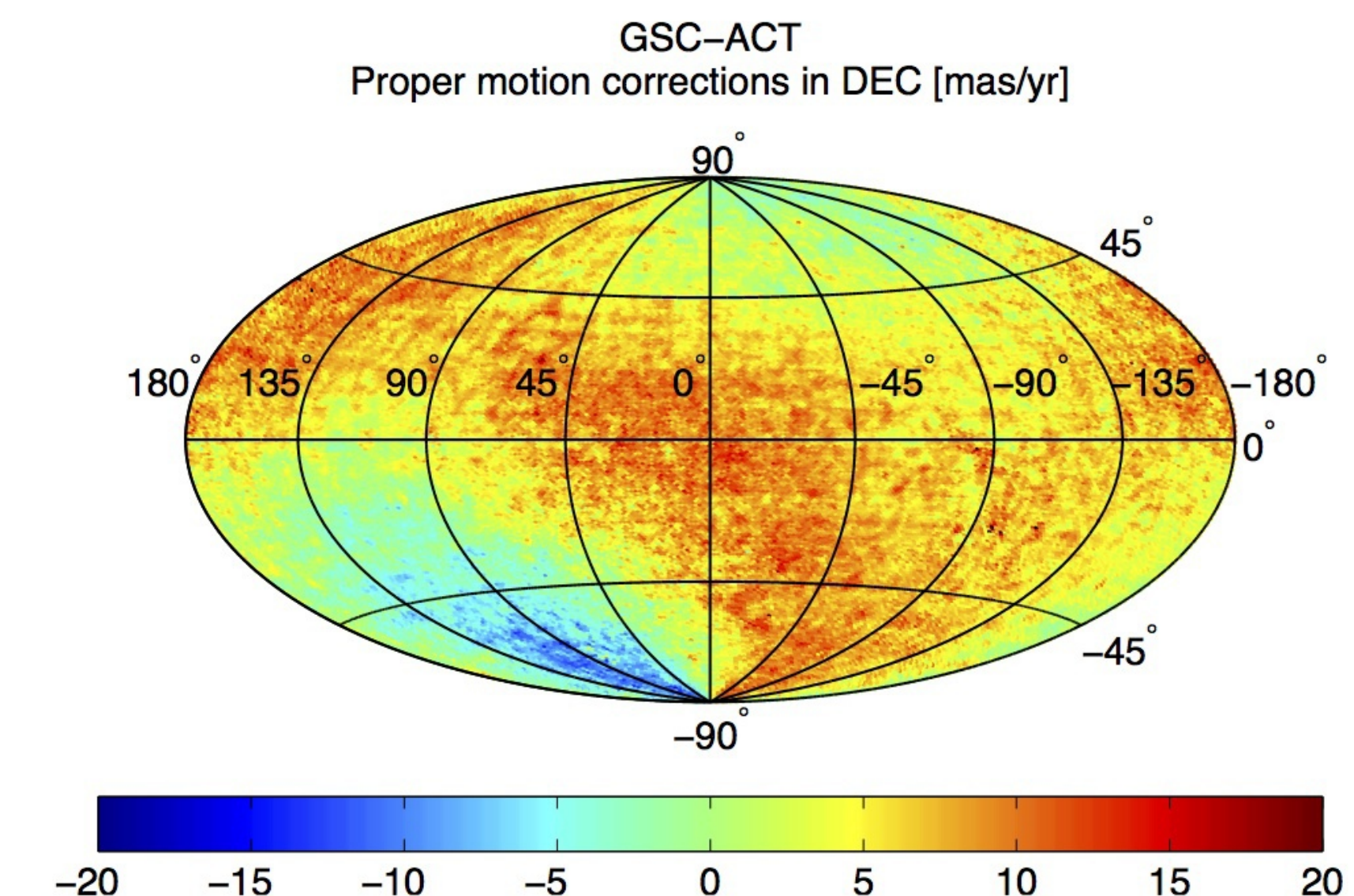}}
\caption{Top: J2000.0 position corrections in right ascension (left)
  and declination (right) for GSC-ACT. Bottom: proper motion corrections
  in right ascension (left) and declination (right) for GSC-ACT.}\label{f:gsc_act}
\end{figure}

GSC-1 (Fig.~\ref{f:gsc_1.1} and Fig.~\ref{f:gsc_1.2}) and GSC-ACT
(Fig.~\ref{f:gsc_act}) catalogs are significantly biased and have no
proper motion information. There is no doubt that the astrometry
reduced with these two catalogs should be corrected.

As shown in Table~\ref{t:catalogs} there are almost 300,000 asteroid
observations submitted with the flag code `z'. These observations were
reduced with one of the GSC catalogs but we do not know which one. It
may be either one of the GSC-1 catalogs or one of the GSC-2 catalogs
\citep{gsc2.2, gsc2.3}. As a consequence, we cannot correct those
observations. It would be very useful if observers could provide the
MPC with the information on which specific GSC catalog they used to
reduce the astrometry.

\subsection{Other catalogs}
\begin{figure}
\centerline{\includegraphics[width=0.6\textwidth]{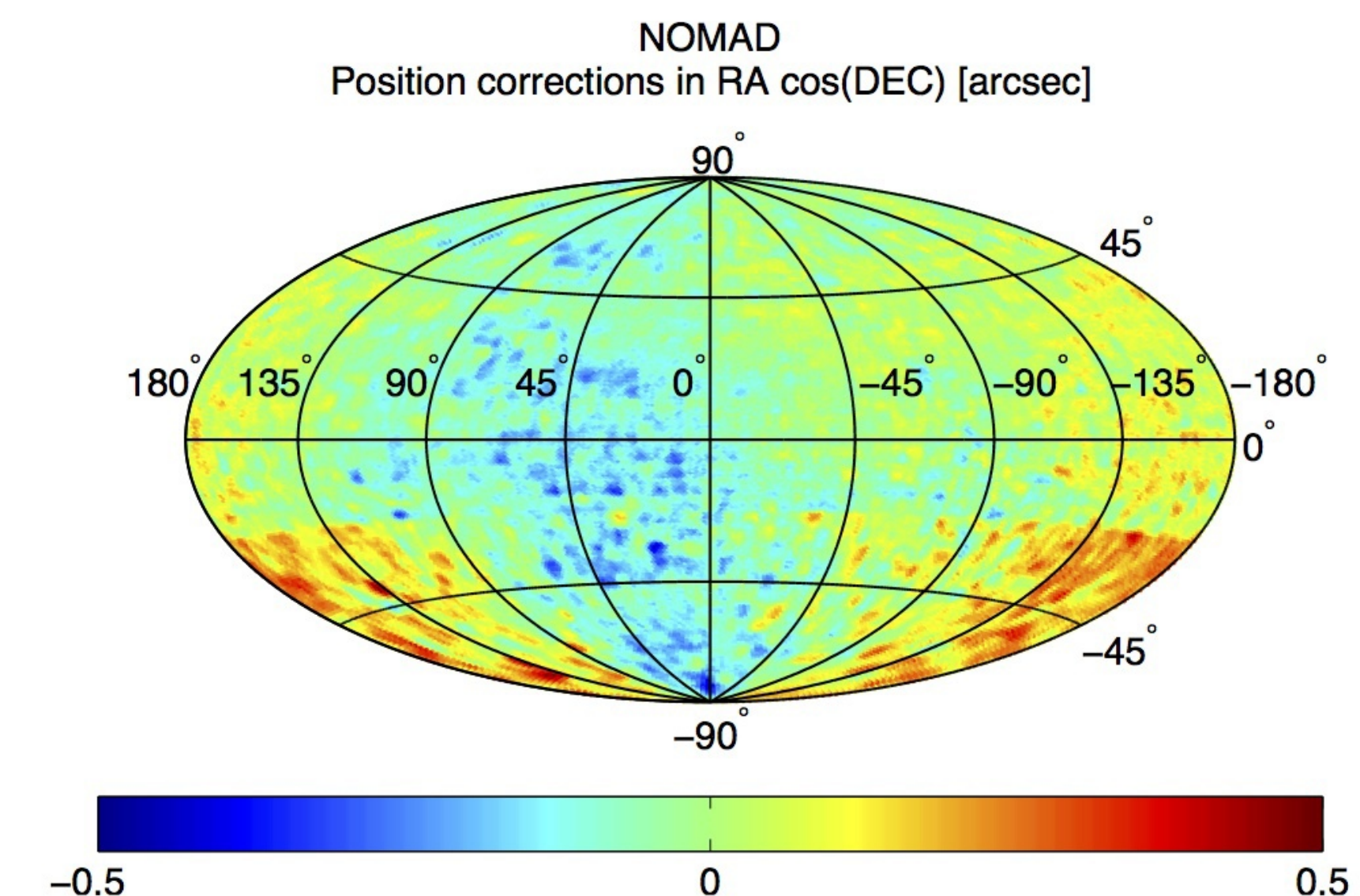}\includegraphics[width=0.6\textwidth]{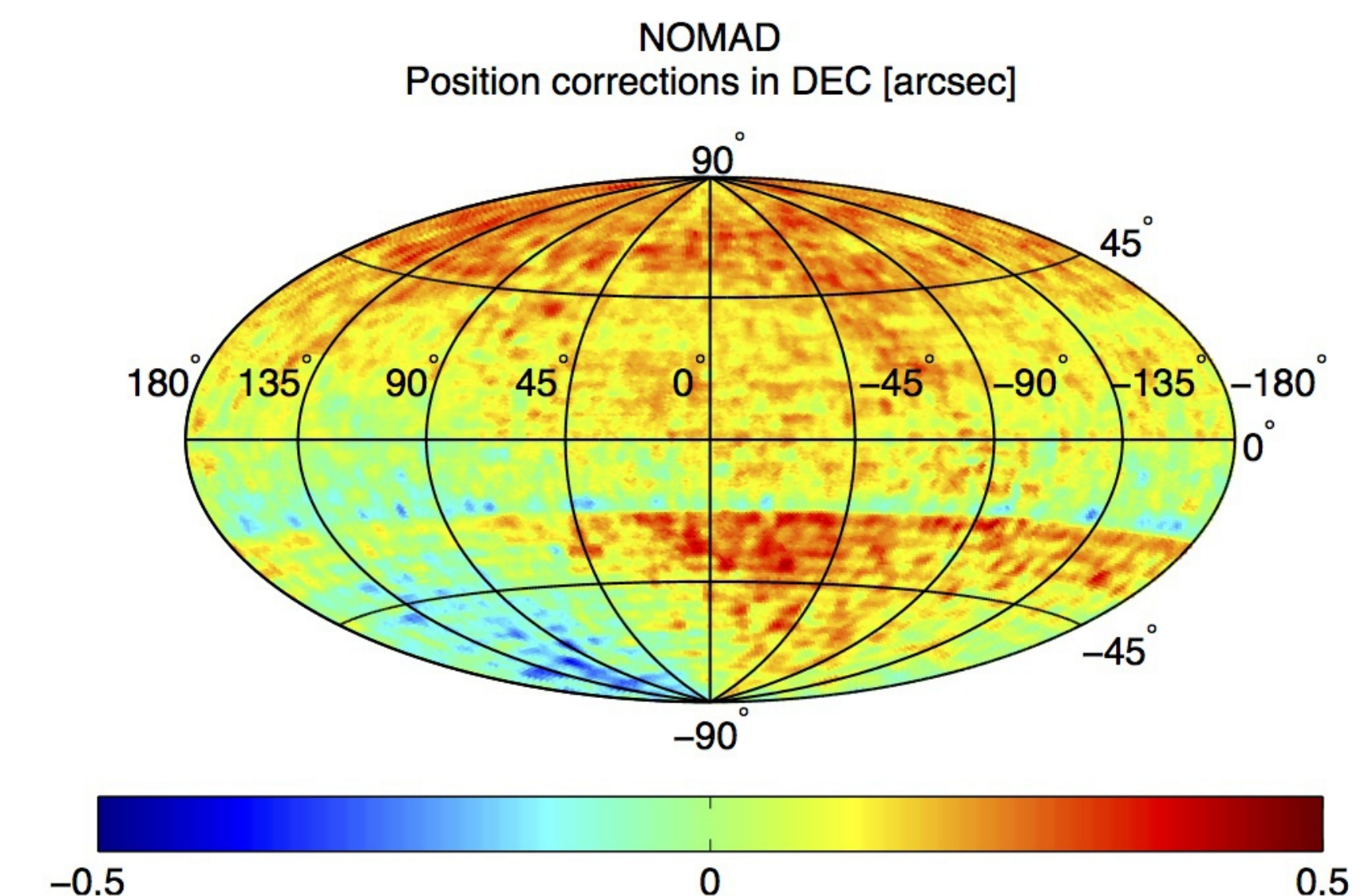}}
\centerline{\includegraphics[width=0.6\textwidth]{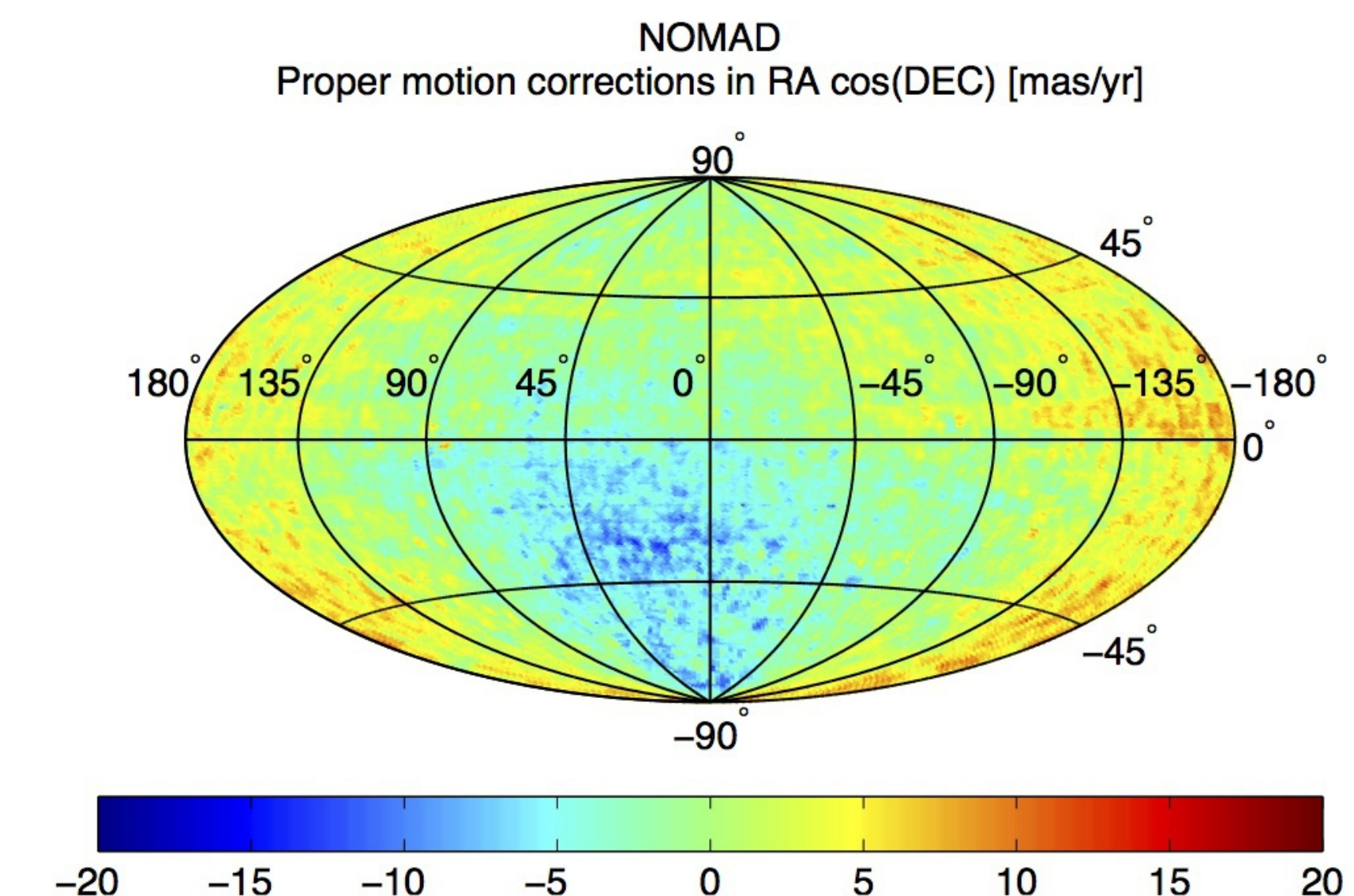}\includegraphics[width=0.6\textwidth]{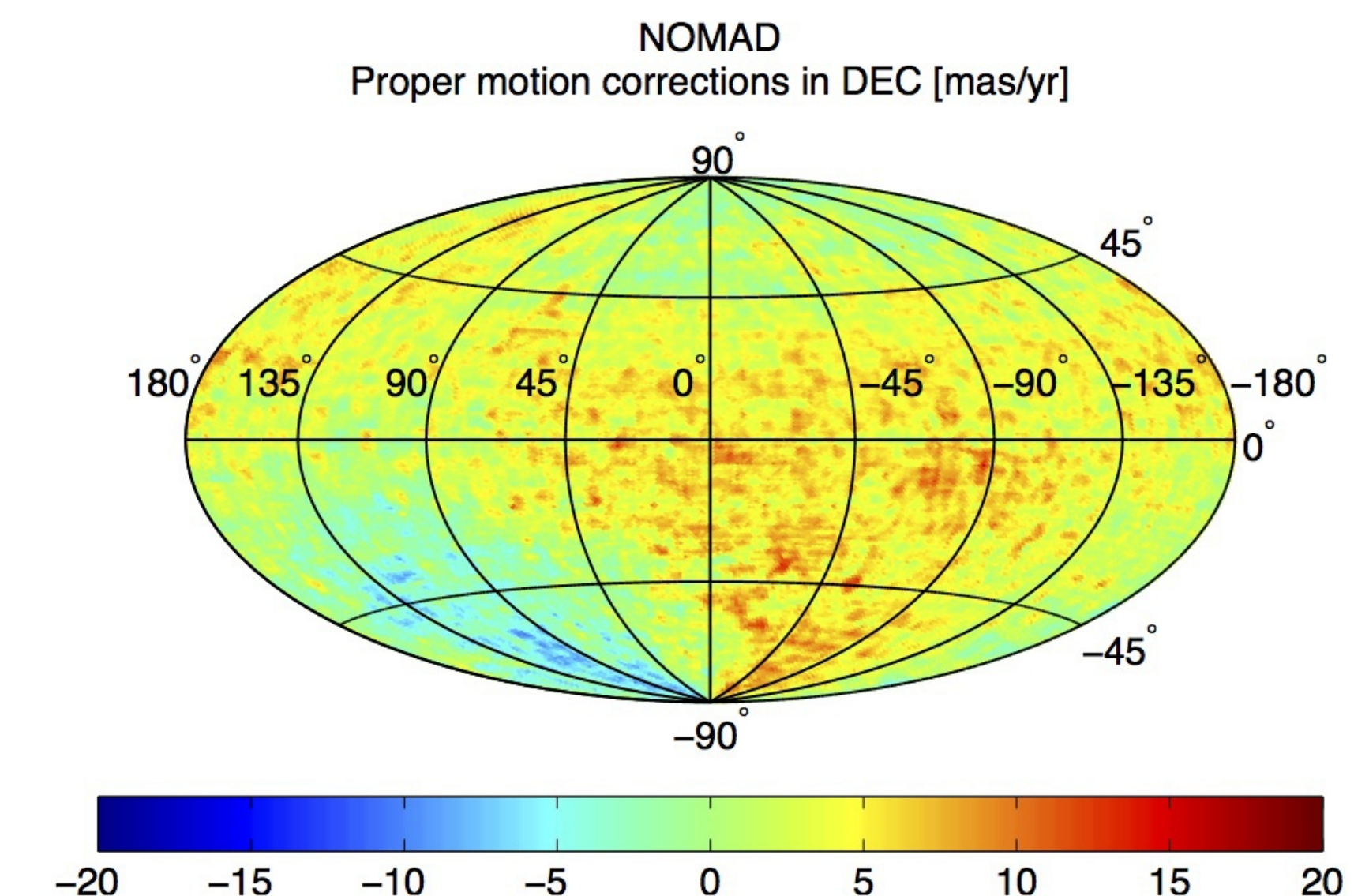}}
\caption{Top: J2000.0 position corrections in right ascension (left)
  and declination (right) for NOMAD. Bottom: proper motion corrections
  in right ascension (left) and declination (right) for NOMAD.}\label{f:nomad}
\end{figure}

For NOMAD (Fig.~\ref{f:nomad}) both position and proper motion
corrections are very similar to those of USNO-B1.0. This is not a
surprise as NOMAD is a merge of a few catalogs, and USNO-B1.0 is the
one with the largest number of stars. Thus, we correct NOMAD for both
positions and proper motions.

\begin{figure}
\centerline{\includegraphics[width=0.6\textwidth]{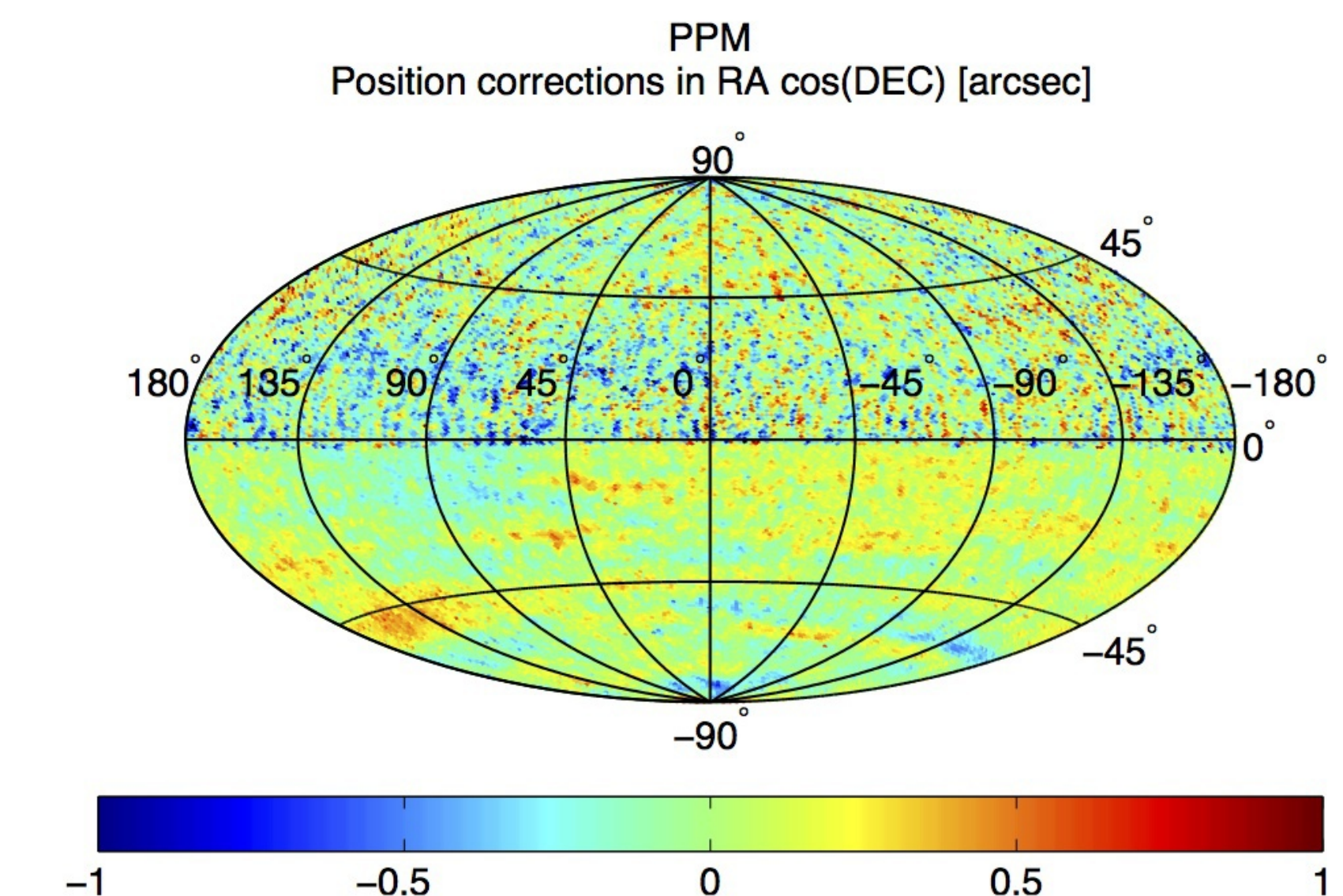}\includegraphics[width=0.6\textwidth]{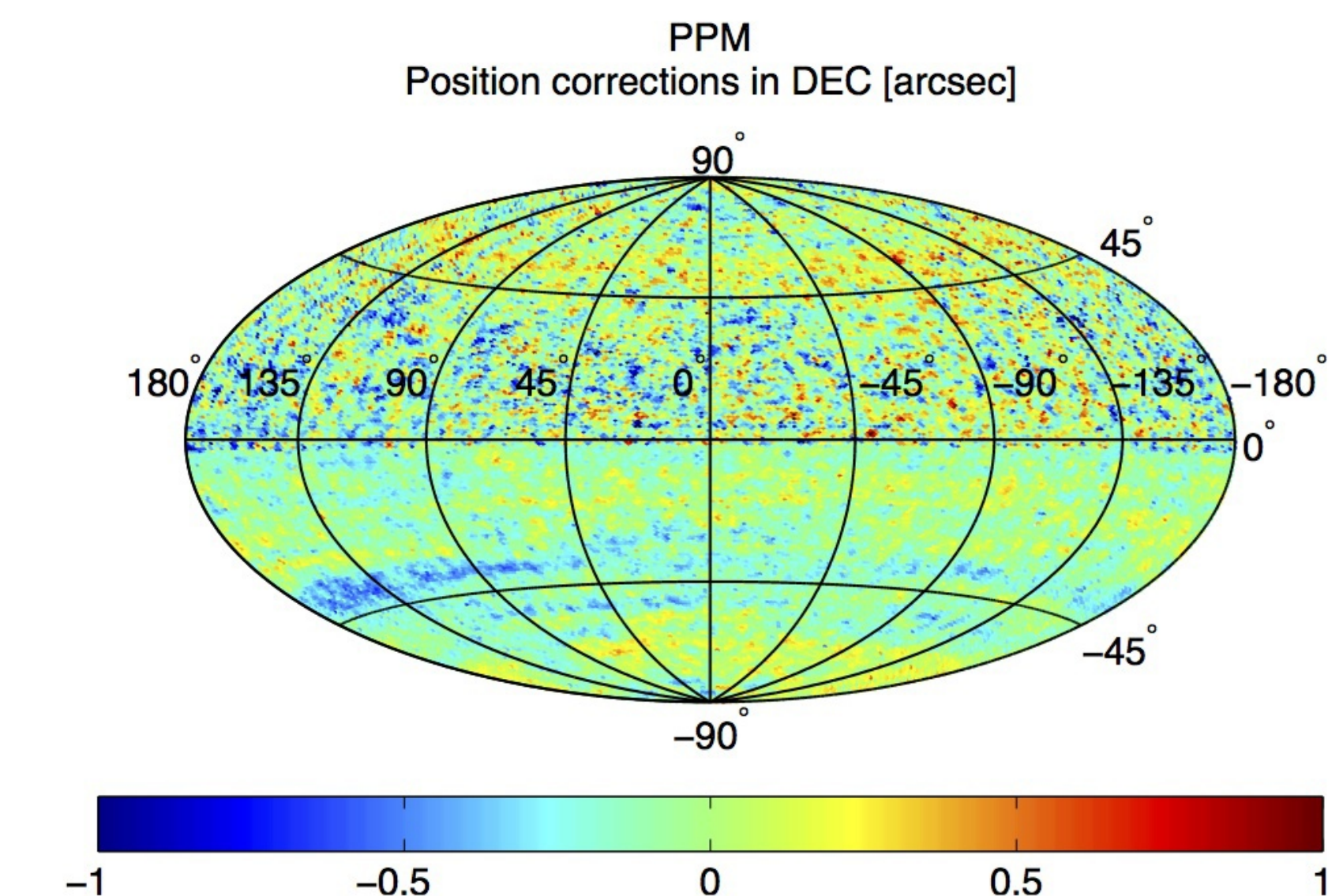}}
\centerline{\includegraphics[width=0.6\textwidth]{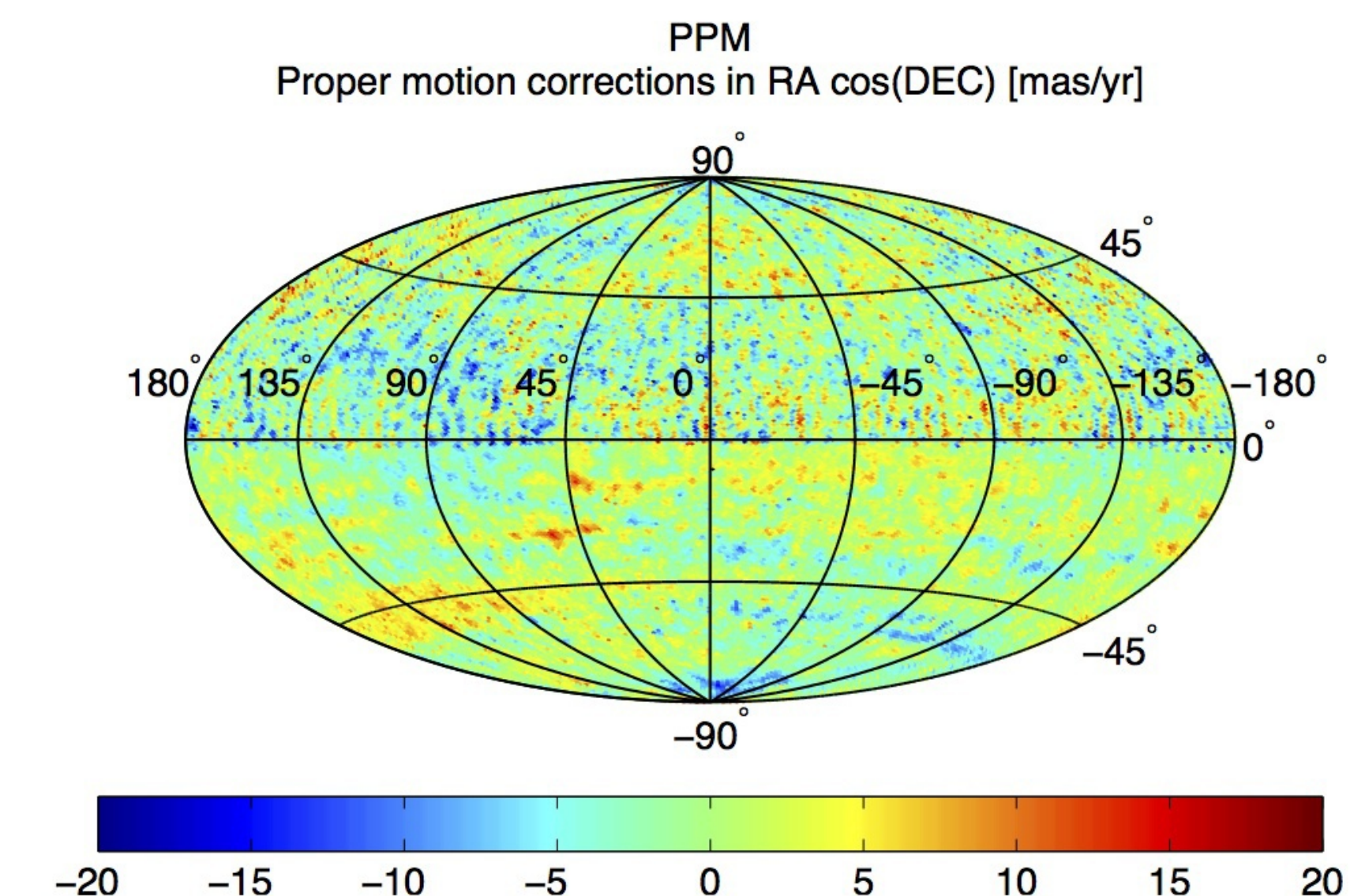}\includegraphics[width=0.6\textwidth]{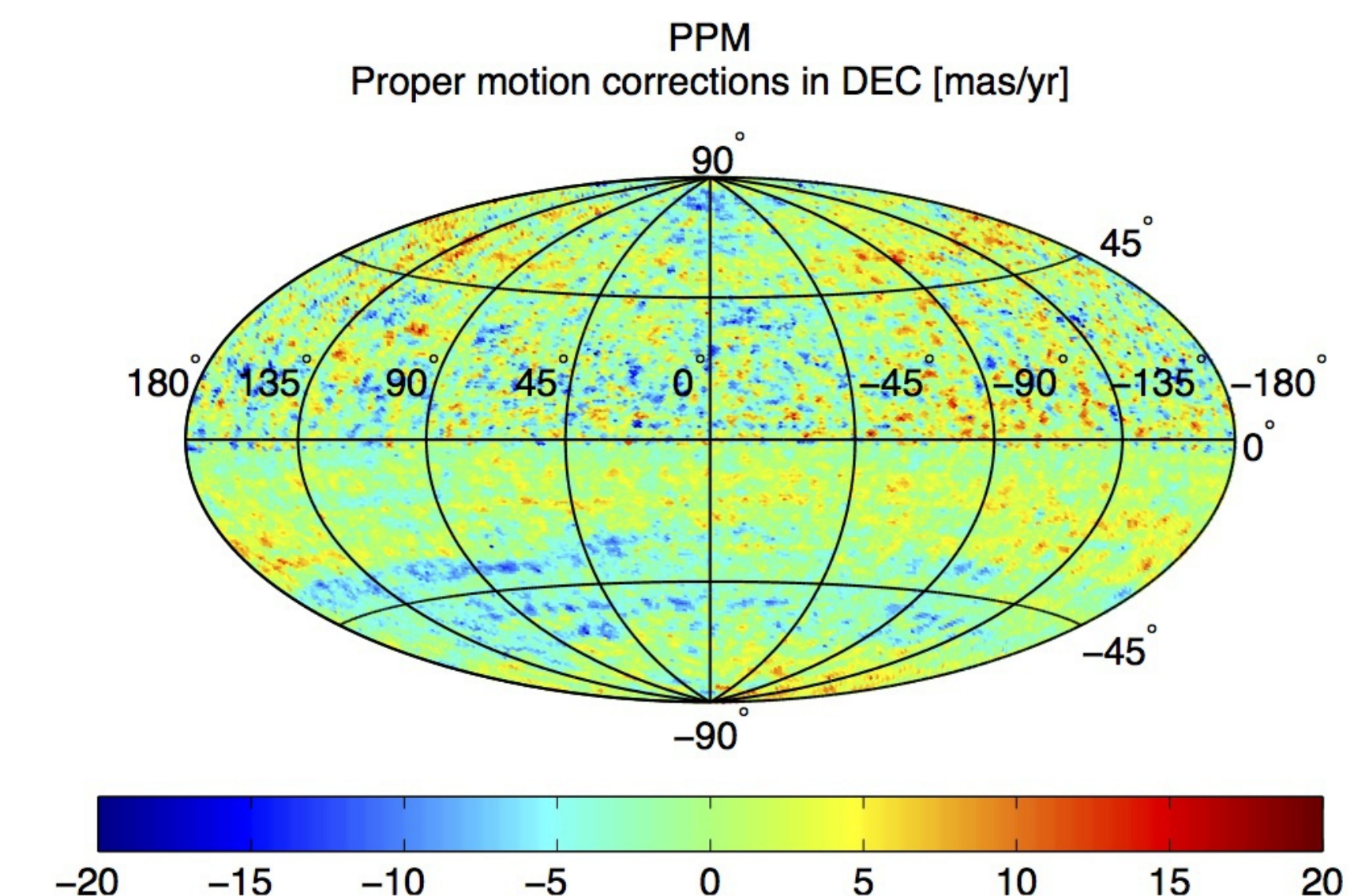}}
\caption{Top: J2000.0 position corrections in right ascension (left)
  and declination (right) for PPM. Bottom: proper motion corrections
  in right ascension (left) and declination (right) for PPM.}\label{f:ppm}
\end{figure}

PPM (Fig.~\ref{f:ppm}) shows significant errors in both positions and
proper motions. There is no doubt that observations reduced with this
catalog should be debiased.

\begin{figure}
\centerline{\includegraphics[width=0.6\textwidth]{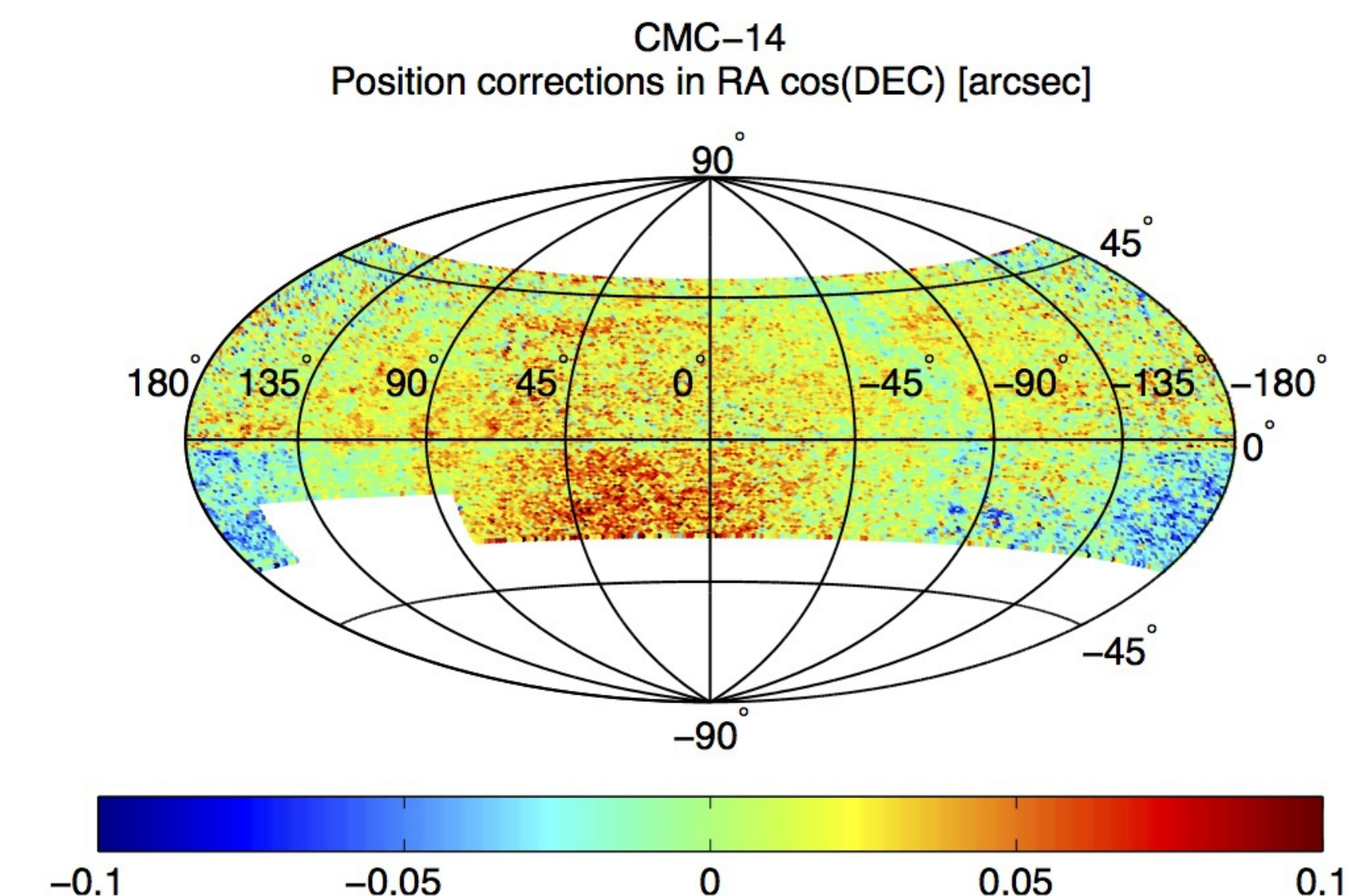}\includegraphics[width=0.6\textwidth]{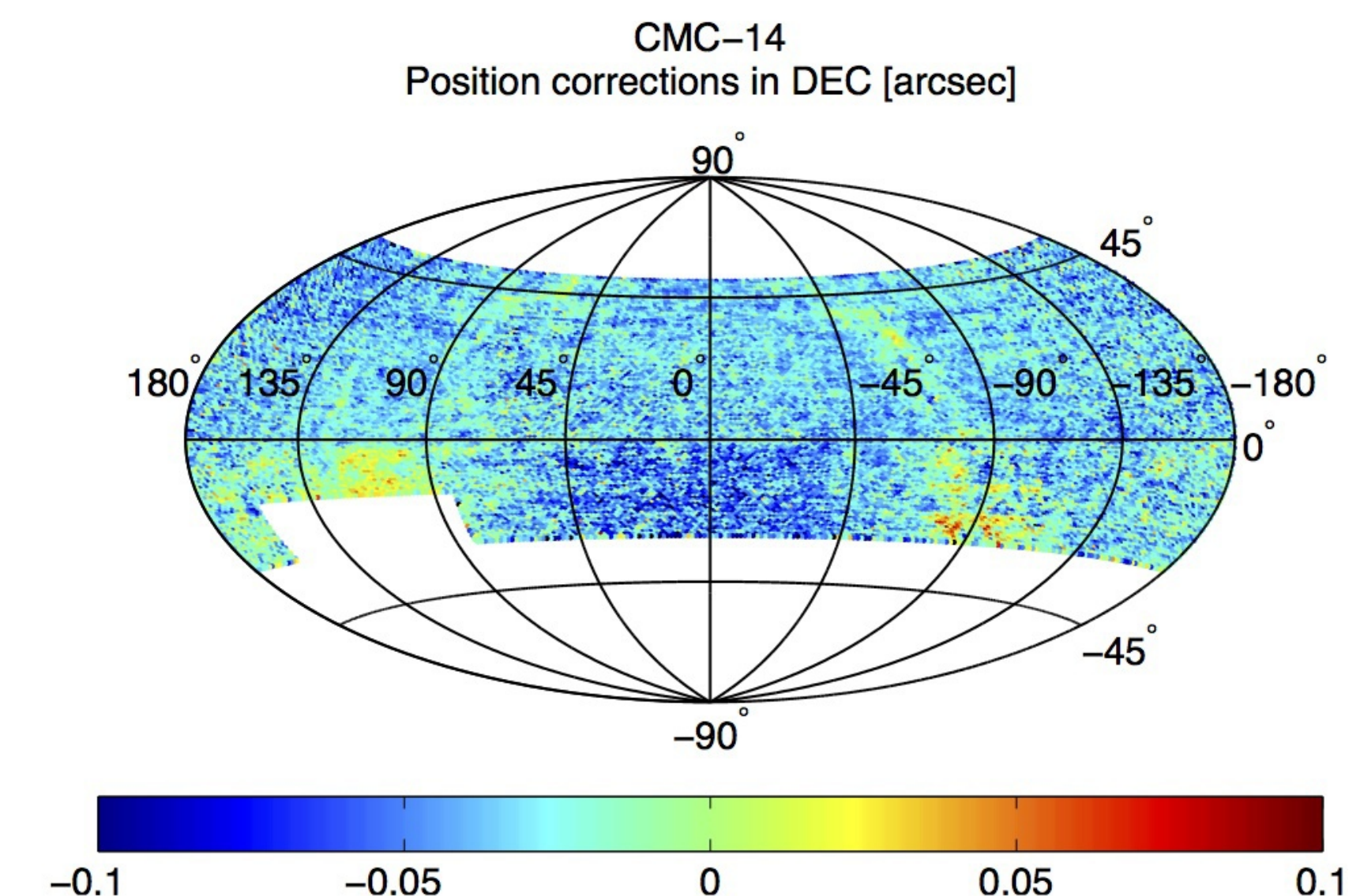}}
\centerline{\includegraphics[width=0.6\textwidth]{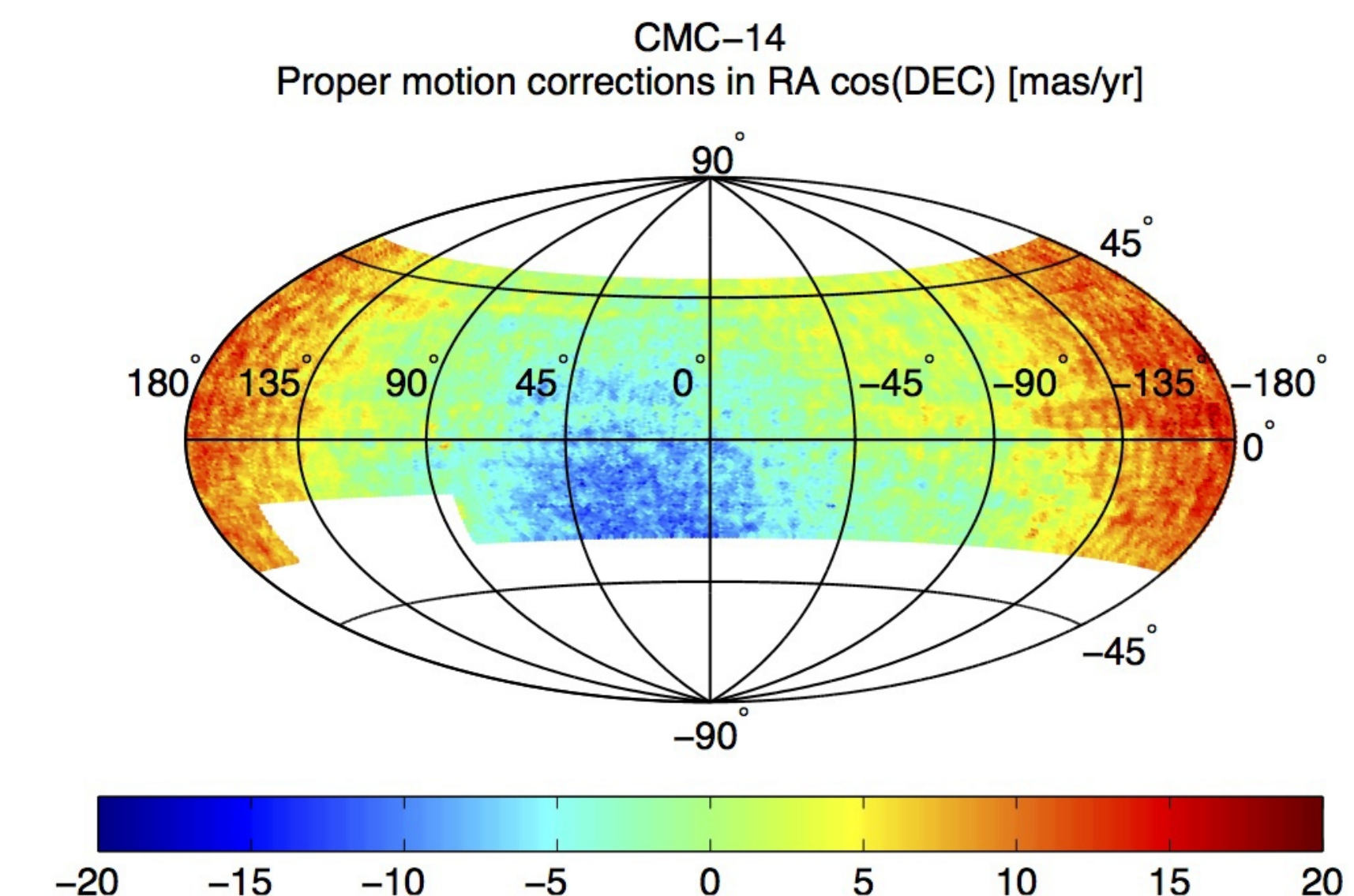}\includegraphics[width=0.6\textwidth]{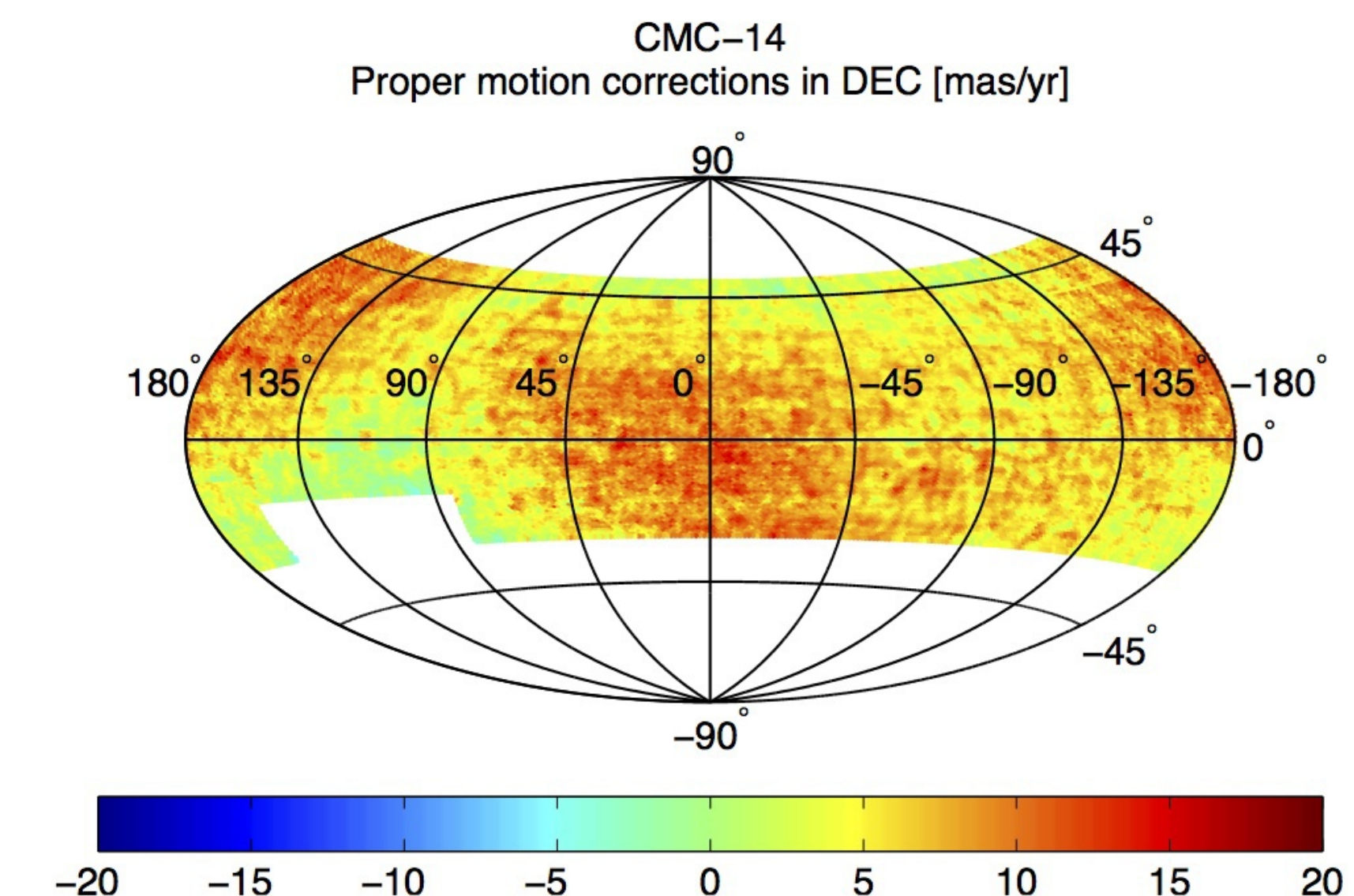}}
\caption{Top: J2000.0 position corrections in right ascension (left)
  and declination (right) for CMC-14. Bottom: proper motion corrections
  in right ascension (left) and declination (right) for CMC-14.}\label{f:cmc14}
\end{figure}

Despite the missing proper motions, CMC-14 (Fig.~\ref{f:cmc14})
provides good star positions. However, position corrections show a
regional dependence correlated to proper motion features. We therefore
corrected all the CMC-14 based astrometry.

\begin{figure}
\centerline{\includegraphics[width=0.6\textwidth]{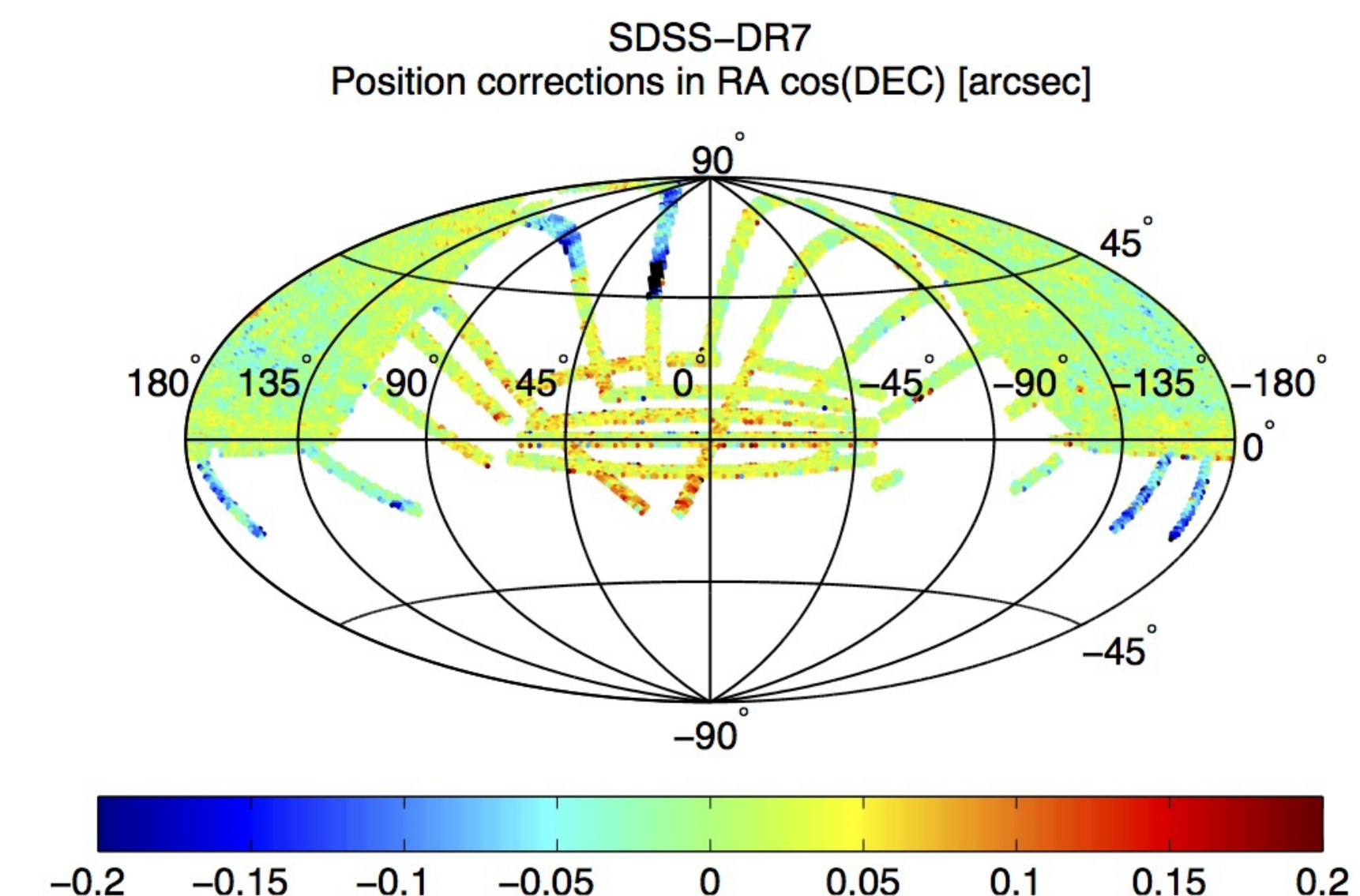}\includegraphics[width=0.6\textwidth]{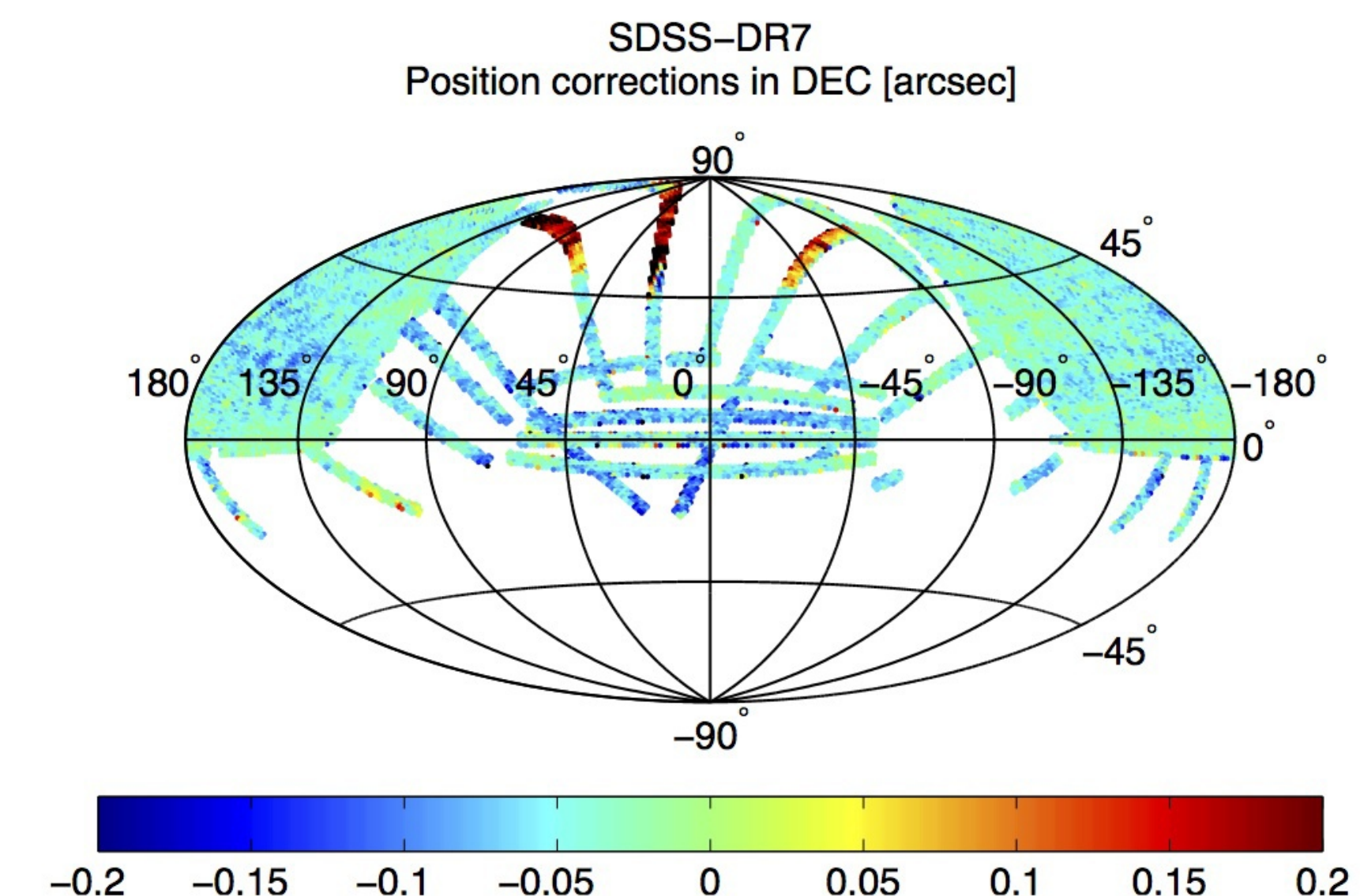}}
\centerline{\includegraphics[width=0.6\textwidth]{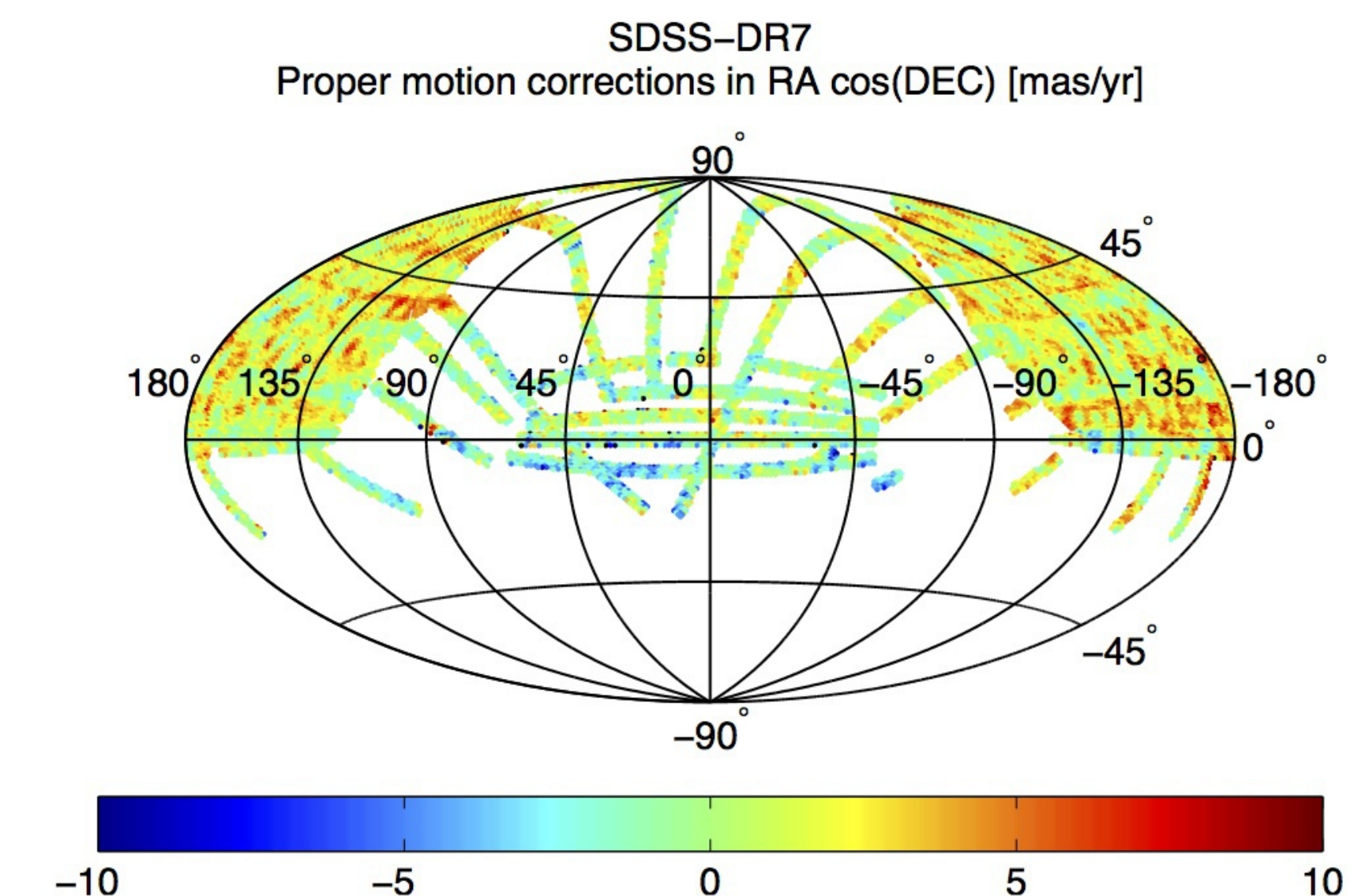}\includegraphics[width=0.6\textwidth]{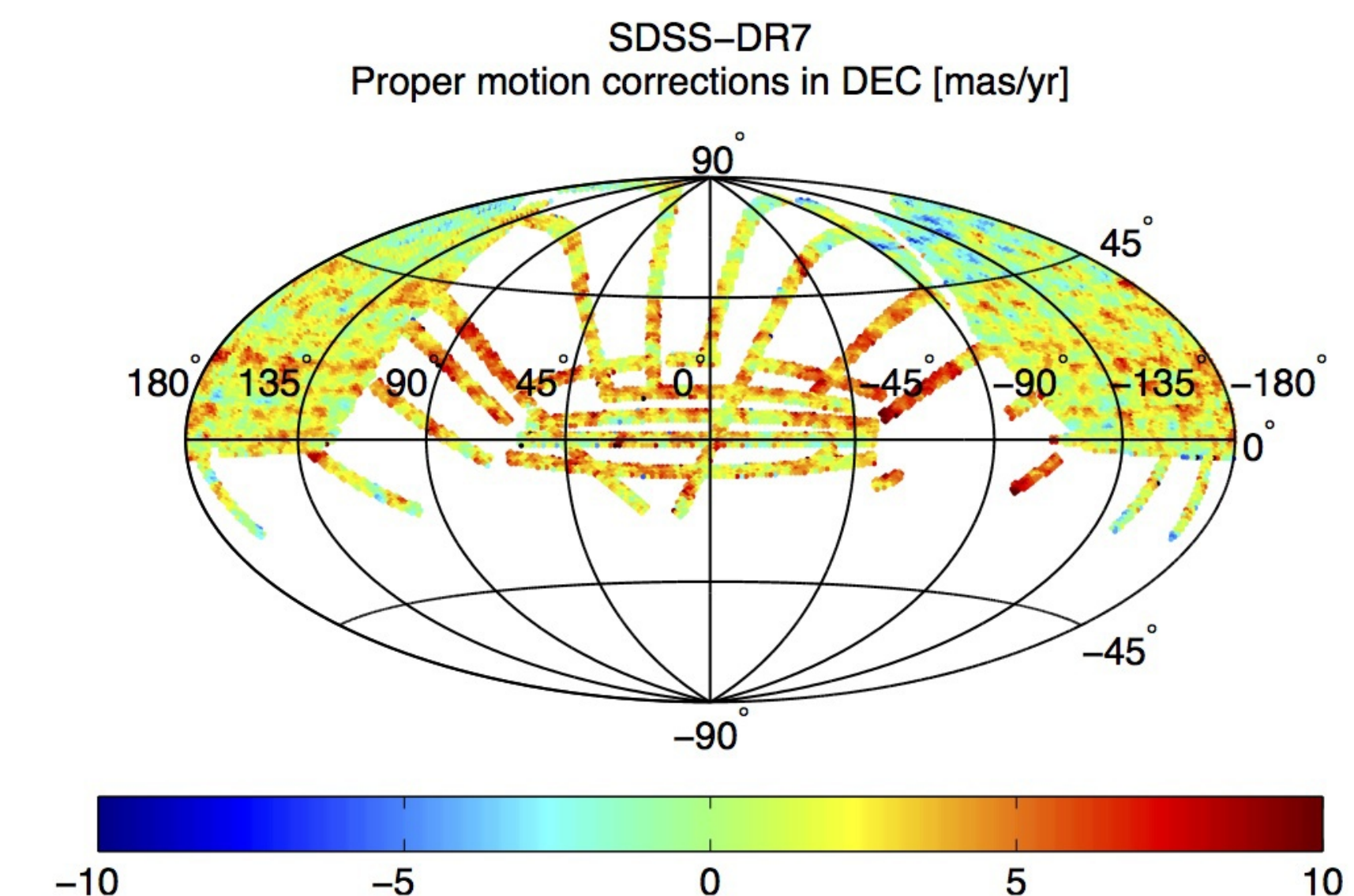}}
\caption{Top: J2000.0 position corrections in right ascension (left)
  and declination (right) for SDSS-DR7. Bottom: proper motion corrections
  in right ascension (left) and declination (right) for SDSS-DR7.}\label{f:sdss7}
\end{figure}

SDSS-DR7 (Fig.~\ref{f:sdss7}) does not seem to be an ideal catalog for
astrometric reduction. As we can see from Fig.~\ref{f:sdss7}, this
catalog does not have uniform coverage of the sky. Moreover, the
position and proper motion errors are significant. We therefore
correct all the SDSS-DR7 based astrometry. It is worth noticing that
all but 10 of the observations reduced using SDSS-DR7 were obtained by
the Palomar Transient Factory survey \citep{ptf}.

\section{Improvement of residual statistcs and ephemeris predictions}

\subsection{Tests with Apophis, Bennu, and Golevka}
We tested the astrometric corrections described in this paper on three
asteroids with the best constrained trajectories: (99942) Apophis,
(101955) Bennu, and (6489) Golevka

\begin{figure}
\centerline{\includegraphics[width=0.8\textwidth]{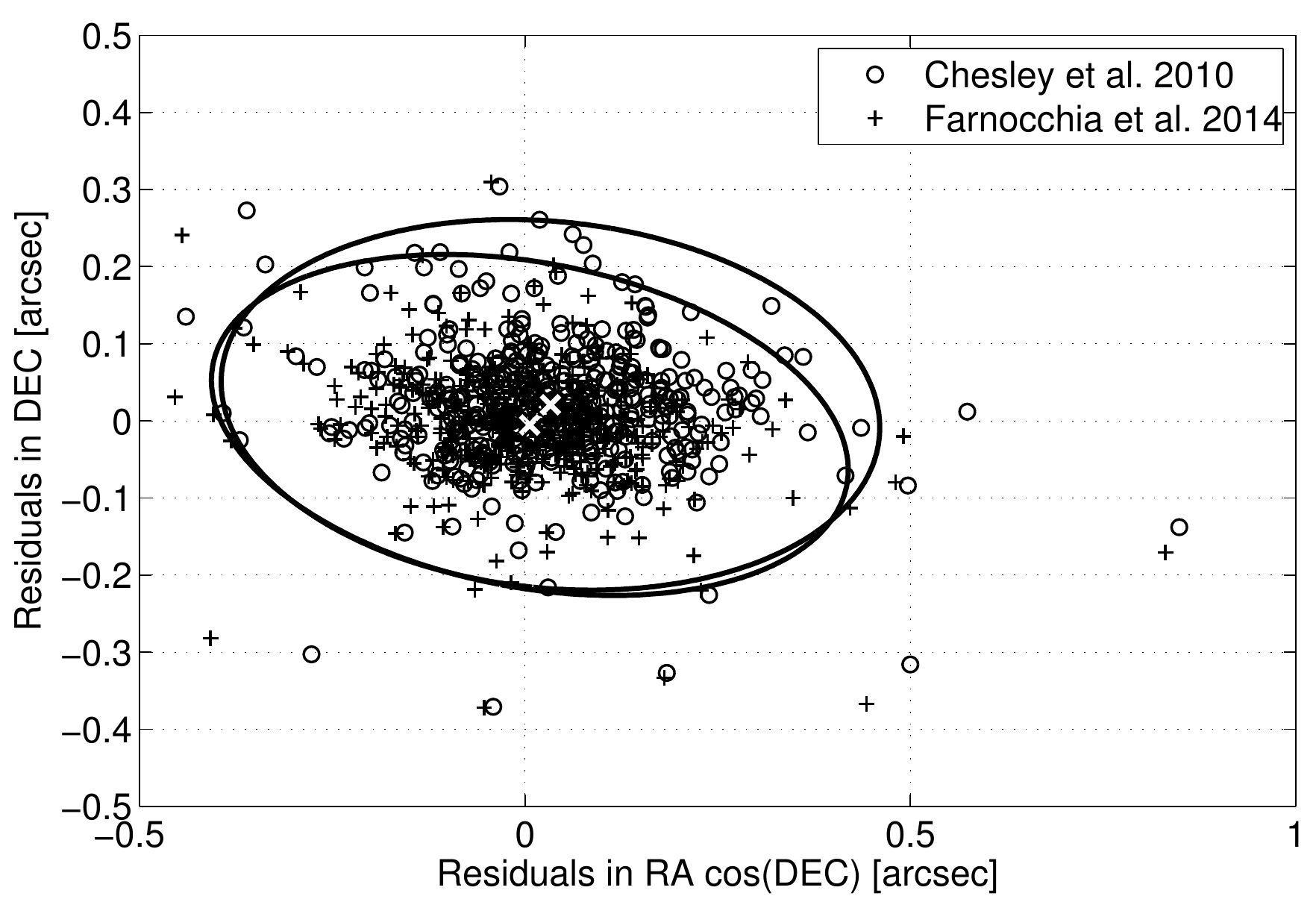}}
\caption{Postfit residuals of the Apophis astrometry from
  \citet{tholen_apo} against the orbital solutions computed by using
  the \citet{cbm10} debiasing scheme (circles) and the one presented
  in this paper (pluses). Mean postfit residuals (white crosses) and
  covariance ellipses at the 3$\sigma$ level are shown for the
  respective datasets.}\label{f:apo_res}
\end{figure}

\citet{tholen_apo} reported over 430 high quality ground-based optical
observations for Apophis. We analyzed the behavior of the postfit
residuals, i.e., against the best fitting orbital solution, for the
\citet{tholen_apo} observations by using the \citet{cbm10} debiasing
scheme and the one presented in this paper. Figure~\ref{f:apo_res}
shows a scatter plot of the postfit residuals in RA and DEC with the two
different schemes. In both cases, the orbital solution is computed by
only using the \citet{tholen_apo} astrometry, Magdalena Ridge and
Pan-STARRS PS1 observations, and radar astrometry \citep[for details
see][]{farnocchia_apophis}. With the \citet{cbm10} scheme, the
\citet{tholen_apo} observations show mean RA/DEC postfit residuals of
$(0.033", 0.021")$. The adoption of the new scheme reduces the RA/DEC
mean postfit residuals to $(0.006", -0.005")$. The clear improvement is mostly
due to the proper motion corrections of 2MASS based astrometry, which
dominates the \citet{tholen_apo} dataset.

Near-Earth asteroids Golevka \citep{golevka}, Bennu \citep{bennu}, and
Apophis \citep{farnocchia_apophis} have exceptionally well constrained
orbits thanks to the availability of three radar
apparitions. Table~\ref{t:chi2} shows the normalized $\chi^2$, i.e.,
the weighted sum of the squared postfit residuals, of the
orbital fit for the \citet{cbm10} debiasing scheme and the one
presented here. For the computation of normalized $\chi^2$ we used
the \citet{cbm10} data weights as well as some manual weights as
described in \citet{bennu} and \citet{farnocchia_apophis}. In both
cases $\chi^2$ improves with the new debiasing scheme, especially for
Golevka. Since the nominal trajectory is already well constrained by
the radar measurements, $\chi^2$ measures how well the optical
observations fit the trajectory. Therefore, the improvement in
$\chi^2$ further suggest that the new debiasing scheme is more
accurate.

\begin{table}
\begin{center}
\begin{tabular}{l|ccc}
  \hline
  Object & $\chi^2$ \citet{cbm10} & $\chi^2$ this paper & $\Delta \chi^2$\\
  \hline
  (99942) Apophis & 231 & 201 & 30\\
  (101955) Bennu & 227 & 224 & 3\\
  (6489) Golevka & 1024 & 1005 & 19\\
  \hline
\end{tabular}
\end{center}
\caption{Normalized $\chi^2$ of the orbital fit for asteroids (99942)
  Apophis, (101955)
  Bennu, and (6489) Golevka. We show the results for
  both the \citet{cbm10}
  debiasing scheme and the one presented in this paper.}
\label{t:chi2}
\end{table}

\subsection{Test with Pan-STARRS PS1 data}
\citet{milani12} found unexpected biases in Pan-STARRS PS1 data and
\citet{tholen_2mass} show clear correlations between the detected
biases and the lack of proper motion in 2MASS, which is the reference
catalog for Pan-STARRS PS1 astrometry. Since the debiasing scheme
presented in this paper corrects for proper motions, the size of
detected biases should decrease significantly.

Table~\ref{t:ps1_res} shows the mean and standard deviation of
Pan-STARRS PS1 residuals in both RA and DEC. There is a modest
improvement in RA and a more significant improvement in DEC.

\begin{table}
\begin{center}
\begin{tabular}{l|cc}
  \hline
  & \multicolumn{2}{c}{Residuals}\\
  & $\text{RA} \cos(\text{DEC})$ & DEC\\
  \hline
  \citet{cbm10}& 0.05'' $\pm $ 0.13''  & 0.06'' $\pm $ 0.12'' \\
  This paper & 0.04'' $\pm$ 0.11'' & -0.01'' $\pm$ 0.11''\\
  \hline
\end{tabular}
\end{center}
\caption{Mean and standard deviation of Pan-STARRS PS1 residuals for
  the \citet{cbm10} debiasing scheme and that of this paper.}
\label{t:ps1_res}
\end{table}

\begin{figure}
\centerline{\includegraphics[width=0.6\textwidth]{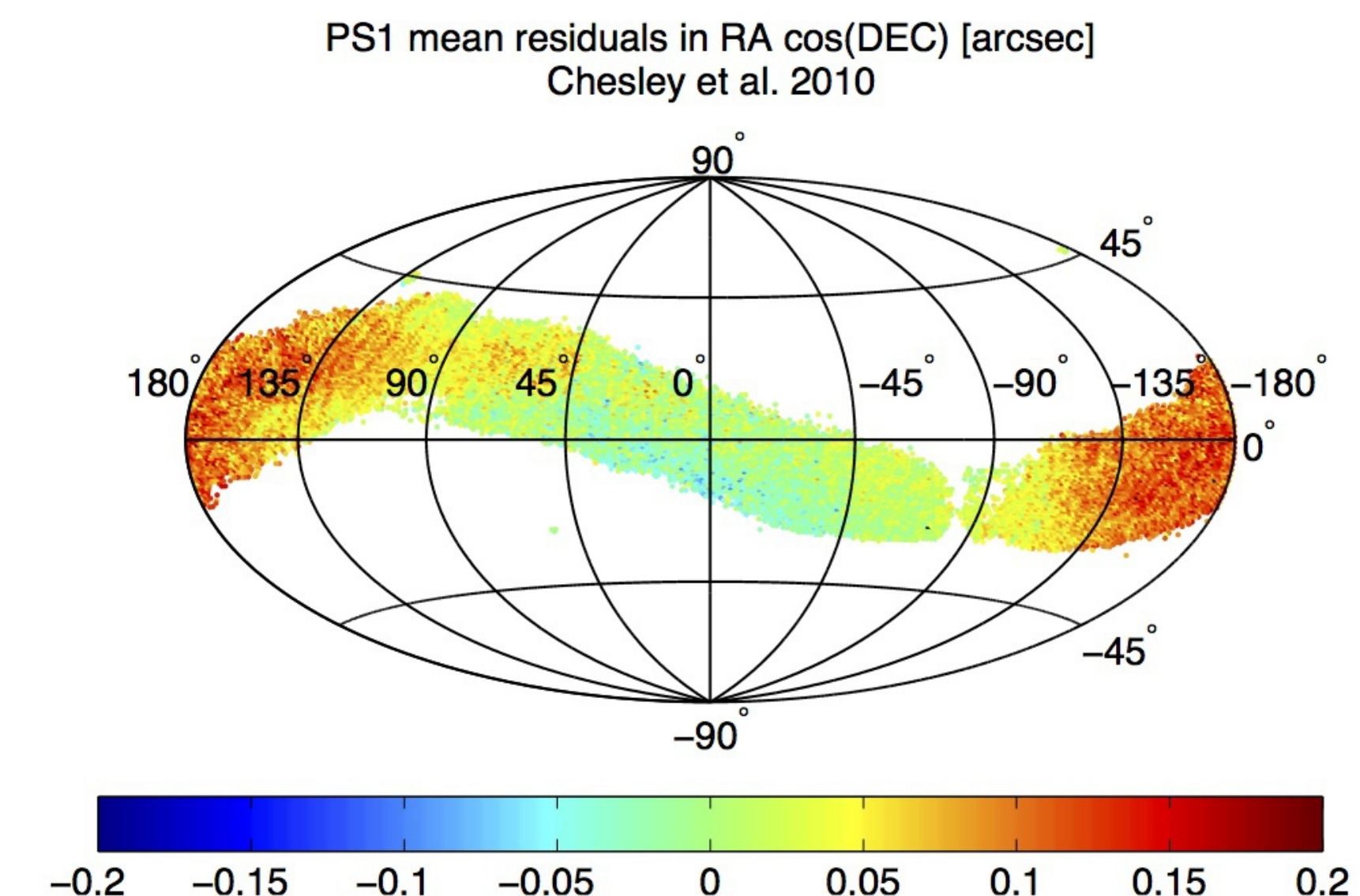}\includegraphics[width=0.6\textwidth]{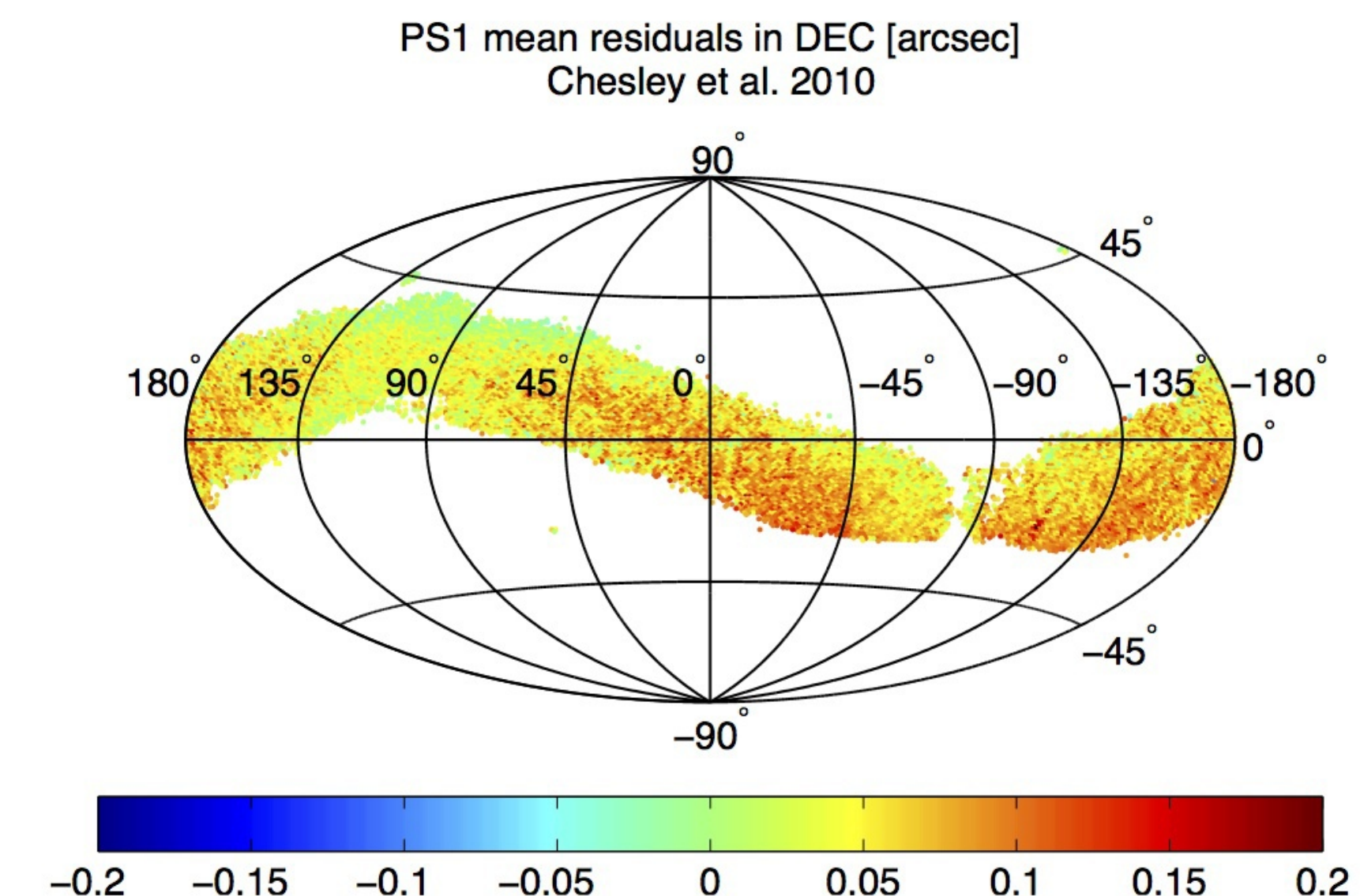}}
\centerline{\includegraphics[width=0.6\textwidth]{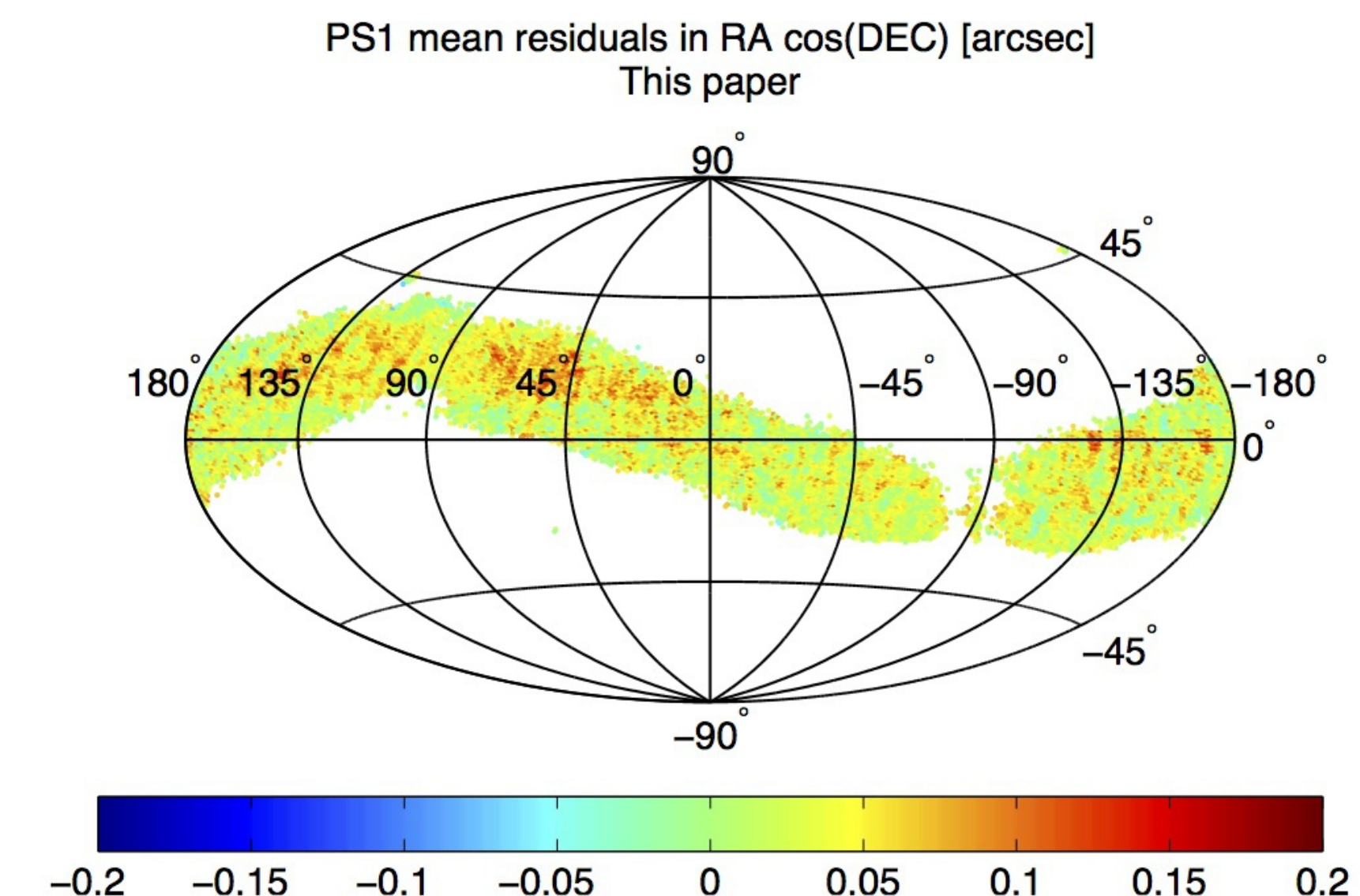}\includegraphics[width=0.6\textwidth]{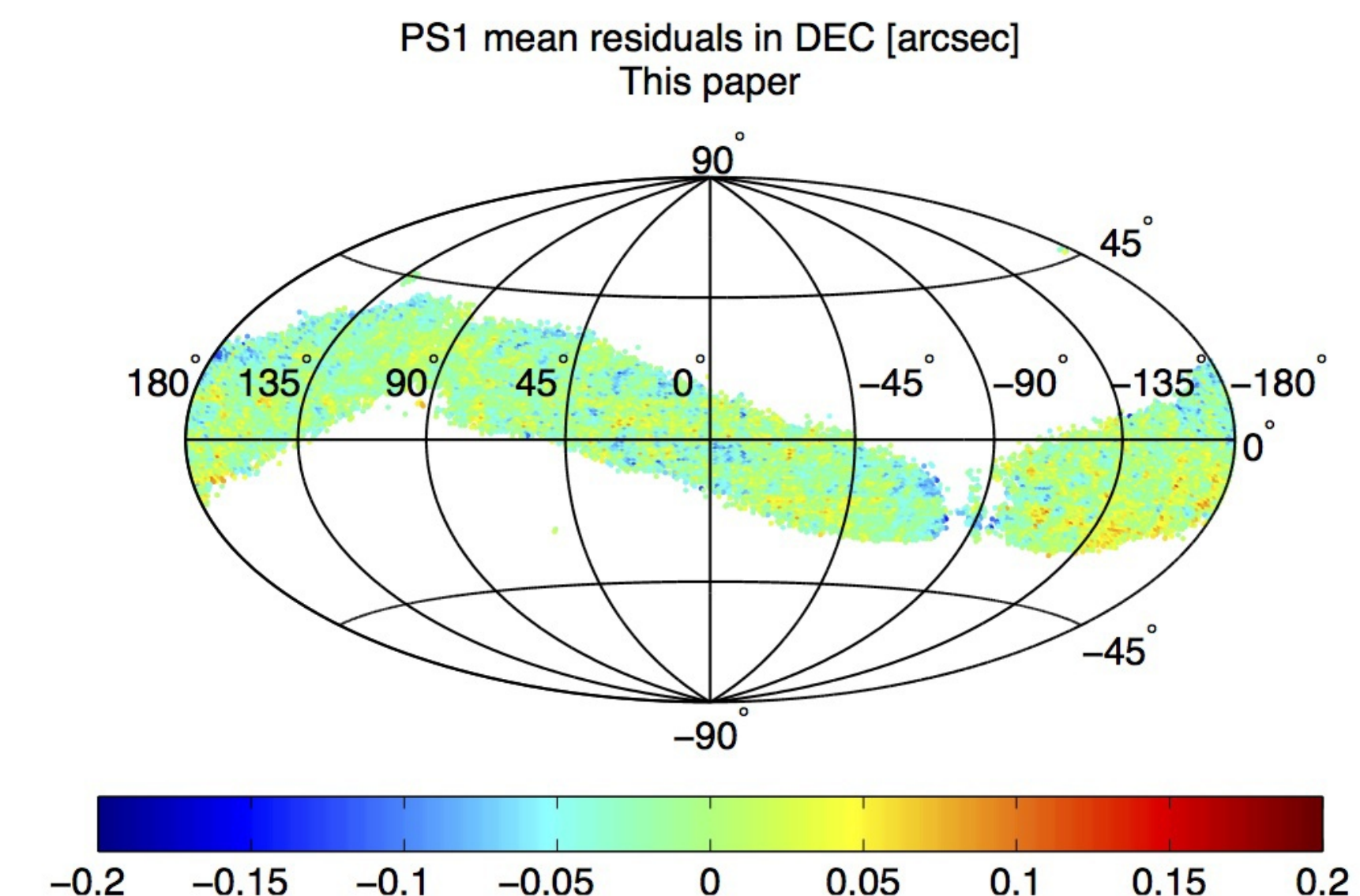}}
\caption{Top: Mean residuals for Pan-STARRS PS1 observations in right
  ascension (right) and declination (left). Top panels are for the
  \citet{cbm10} debiasing scheme, bottom panels are for the debiasing
  scheme described in this paper.}\label{f:ps1}
\end{figure}

Figure~\ref{f:ps1} is a sky map of Pan-STARRS PS1 mean residuals in
the sky and helps to better appreciate the improvement due to the new
debiasing scheme. We only considered those tiles in the sky with at
least 100 observations. Top panels correspond to the \citet{cbm10}
debiasing scheme. We can clearly see the correlation between the found
biases and star proper motions (e.g., see bottom panels of
Fig.~\ref{f:2mass}). The bottom panels show the mean residuals using
the new debiasing scheme. A clear improvement is evident from the
application of the new debiasing scheme. In particular, the clear
regional structure of the systematic error distribution vanished.

\subsection{Prediction test}
\label{s:pred_test}
To validate the new debiasing scheme the most important test is
prediction: the orbits computed with the new scheme have to provide
better predictions. We performed a test similar to that described by
\citet[][Sec. 6]{cbm10}.  We took the same 222 asteroids, but we
considered the last 9 apparitions. For each object we selected
different subsets of the observational arc, propagated to the central
epoch of the 5th apparition, computed the 3-dimensional Cartesian
position, and compared to the solution obtained by using the full
observational dataset, which is considered as the truth.  The
comparison was done consistently, i.e., if the prediction was computed
with the \citet{cbm10} scheme, then the truth was computed with
\citet{cbm10} scheme, and similarly for the scheme presented in this
paper.

\begin{figure}
\centerline{\includegraphics[width=0.9\textwidth]{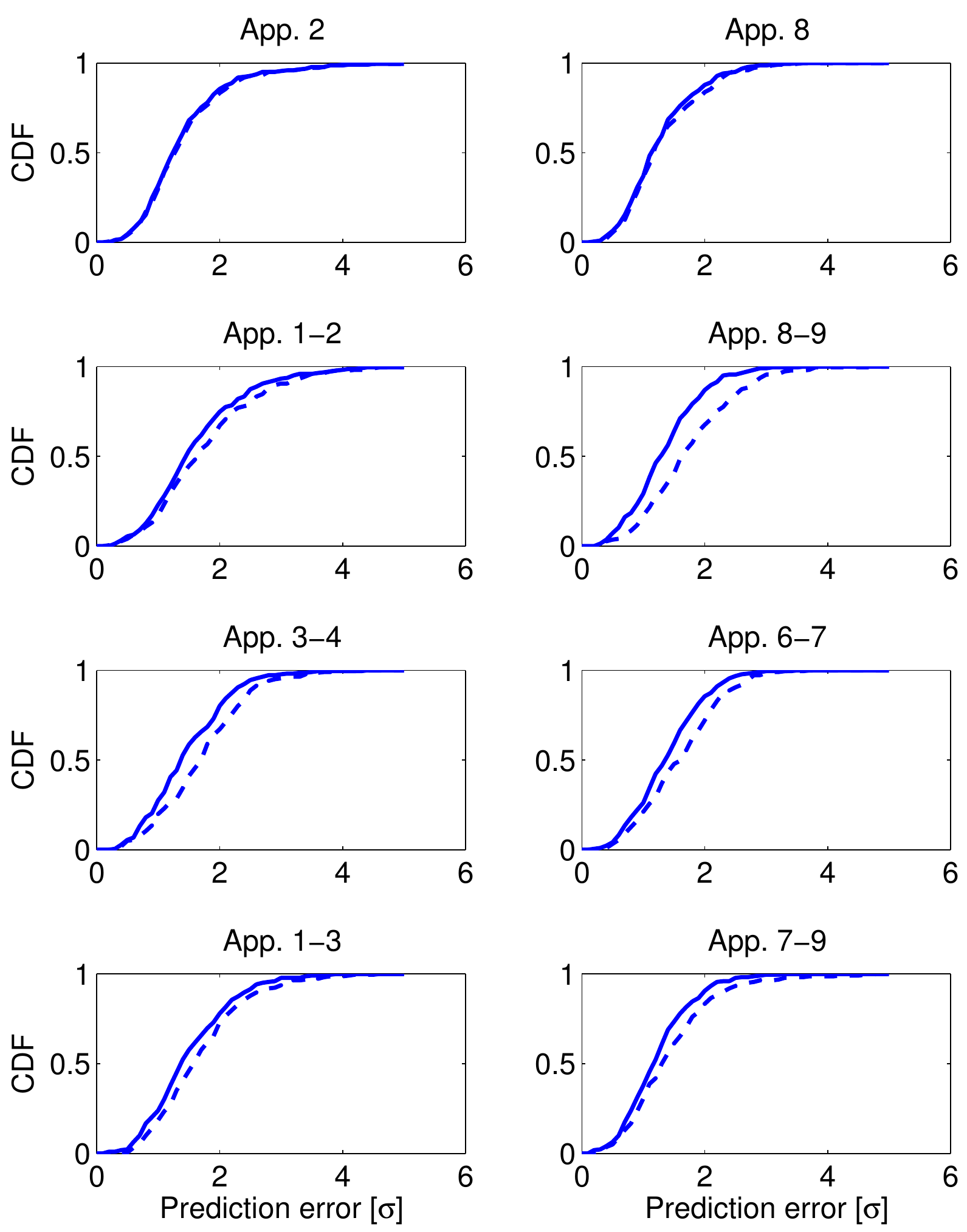}}
\caption{Cumulative distribution of prediction errors as a function of
  the formal prediction $\sigma$ for the debiasing schemes presented
  here (solid line) and the \citet{cbm10} one (dashed line). The
  titles indicate what apparitions were used to compute the
  prediction, e.g., App. 1--3 means that apparition 1, 2, and 3 were
  used.}\label{f:pred_test}
\end{figure}

For each subset of the 9 apparitions, the new debiasing scheme
performed better than that from \citet{cbm10}. As an example,
Fig.~\ref{f:pred_test} shows the cumulative distributions of the
prediction error for predictions made by using different subsets of
the 9 considered apparitions. We can see how the cumulative prediction
error distributions obtained with the new scheme are better than those
obtained with the \citet{cbm10} scheme.

\section{Data weights and correlations}
The computation of an orbit is the result of a least square procedure
\citep[Chap.~5]{orbdet}. It is important that individual observations
are assigned weights that reflect the expected accuracy $\sigma$,
i.e., $w = 1/\sigma^2$. Tables~\ref{t:pho_dt}--\ref{t:spe_cat} list
the weights we have been using in the last few years. The $\sigma$
values for CCD observations of Tables~\ref{t:gen_cat} and
\ref{t:spe_cat} are largely from \citet{cbm10}, with some ad hoc
additions based on our experience. The precedence rule is the
following: Table~\ref{t:spe_cat} has priority over
Tables~\ref{t:pho_dt} and \ref{t:gen_typ}. Table~\ref{t:gen_cat},
which in turn has priority over Note that \citet{cbm10} adjust the CCD
weights by applying their so-called ``safety factor'' of 2 to the
reported $\sigma$ values to provide a more realistic ephemeris
uncertainty, i.e., with a prediction error distribution closer to a
theoretical normal distribution.

\begin{table}
\begin{center}
\begin{tabular}{lcc}
  \hline
Date & $\sigma_{RA}$ & $\sigma_{DEC}$\\
\hline
$< 1890$  & 3.0'' & 3.0''\\
1890--1950 & 2.0'' & 2.0''\\
$> 1950$ & 1.5'' & 1.5''\\
\hline
\end{tabular}
\end{center}
\caption{Weighting rules by date for photographic, A and N-type
  observations (see Table~\ref{t:counts}).}
\label{t:pho_dt}
\end{table}

\begin{table}
\begin{center}
\begin{tabular}{lcc|lcc}
  \hline
Type & $\sigma_{RA}$ & $\sigma_{DEC}$ & Type & $\sigma_{RA}$ & $\sigma_{DEC}$\\
\hline
C, c, n, V, S & 1.0'' & 1.0'' & M & 3'' & 3'' \\ 
H & 0.4'' & 0.4'' & T & 0.5'' & 0.5''\\
E & 0.2'' & 0.2'' & e & 0.75'' & 0.75''\\
\hline
\end{tabular}
\end{center}
\caption{General weighting rules by type (see Table~\ref{t:counts}).}
\label{t:gen_typ}
\end{table}


\begin{table}
\begin{center}
\begin{tabular}{lcc|lcc}
  \hline
Catalog & $\sigma_{RA}$ & $\sigma_{DEC}$ & Catalog & $\sigma_{RA}$ & $\sigma_{DEC}$\\
\hline
c, d & 0.51'' & 0.40'' & m & 0.56'' & 0.57''\\
e, q, r, u & 0.33'' & 0.30'' & w & 0.44'' & 0.36''\\
o, s & 0.50'' & 0.41'' & f, g & 0.73'' & 0.64''\\
a, b & 0.59'' & 0.51'' & L, t & 0.25'' & 0.25''\\
h, i, j, z & 0.45'' & 0.44'' & & &\\
\hline
\end{tabular}
\end{center}
\caption{Specific weighting rules by star catalog for observations
  with MPC type flag 
  C, c, n, or V (see Table~\ref{t:counts}).}
\label{t:gen_cat}
\end{table}

\begin{table}
\begin{center}
\begin{tabular}{lccc|lccc}
\hline
Station & Catalog & $\sigma_{RA}$ & $\sigma_{DEC}$ & Station & Catalog & $\sigma_{RA}$ & $\sigma_{DEC}$\\
\hline 
704 & c, d & 0.62'' & 0.60'' & 608 & c, d & 0.63'' & 0.77''\\
644 & c, d & 0.24'' & 0.28'' & 644 & o, s & 0.18'' & 0.17''\\
703 & c, d & 0.62'' & 0.57'' & 703 & e, r & 0.49'' & 0.46''\\
699 & c, d & 0.47'' & 0.39'' & 699 & o, s & 0.42'' & 0.41''\\
691 & c, d & 0.32'' & 0.34'' & 691 & o, s & 0.25'' & 0.28''\\
G96 & e, r & 0.25'' & 0.21'' & E12 & e, r & 0.41'' & 0.43''\\
F51 & L & 0.15'' & 0.15'' & H01 & t, L & 0.15'' & 0.15''\\ 
568 & t & 0.13'' & 0.13'' & 568 & L & 0.15'' & 0.15''\\
568 & o, s & 0.25'' & 0.25'' & 673 & All & 0.30'' & 0.30''\\                 
683 & e, r & 0.61'' & 0.78'' & 645 & e & 0.15'' & 0.15''\\
689 & g & 0.26'' & 0.32'' & 250 & All & 1.30'' & 1.30''\\                  
C51 & All & 1.00'' & 1.00'' & & & & \\
\hline
\end{tabular}
\end{center}
\caption{Station specific weighting rules by star catalog for
  observations with MPC type flag C,
  c, n, V, or S (see Table~\ref{t:counts}).}
\label{t:spe_cat}
\end{table}

\citet{carpino03}, \citet{cbm10}, and \citet{baer11} show that
asteroid astrometric errors can be correlated, especially for
same-station observations closely spaced in time. The presence of
correlations is not a surprise since still unresolved systematic
errors, such as timing errors, result in correlations. To mitigate the
effect of unresolved systematic errors and correlations we relax the
weights, especially when there are many observations from the same
station on the same night (which we call a batch of observations). Our
strategy is to apply a scale factor $\sqrt{N}$ to each weight, where
$N$ is the number of observations contained in a single-station
batch. We consider as a batch a sequence of observations from the same
station with a time gap smaller than 8 hours between two consecutive
observations. For CCD observations, which have a typical batch of 3--5
observations, this scale factor is close to the safety factor of 2
suggested by \citet{cbm10}. The $\sqrt{N}$ better handles the cases
with a large number of observations in a batch and avoids
down-weighting batches with a lower number of observations. Moreover,
this scale factor is applied to all types of observations thus
mitigating possible correlations for, e.g., old photographic
observations.

\begin{figure}
\centerline{\includegraphics[width=0.9\textwidth]{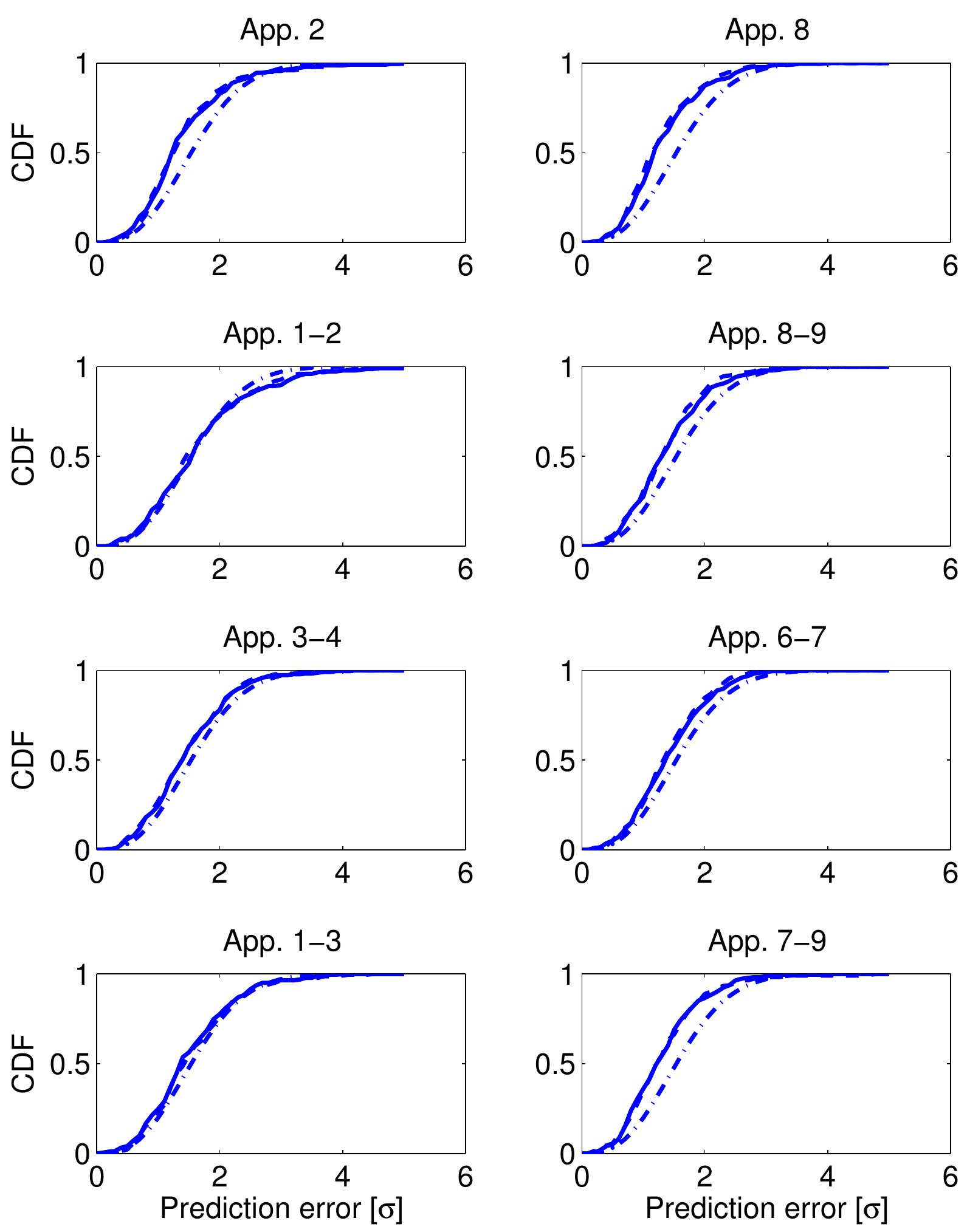}}
\caption{Cumulative distribution of prediction errors as a function of
  the formal prediction $\sigma$ for different weighting
  schemes. Solid line is with the $\sqrt N$ factor, dashed line is
  with the safety factor of 2, and dash-dotted line is a normal cumulative
  distribution.  The titles indicate what apparitions were used to
  compute the prediction, e.g., App. 1--3 means that apparition 1, 2, and 3 were
  used.}\label{f:predw_test}
\end{figure}

To validate the new weighting scheme we performed a test similar to
that of Sec.~\ref{s:pred_test}. In this case we used two different
weighting schemes: the one with a safety factor of 2 and the one that
scales by $\sqrt{N}$. Figure ~\ref{f:predw_test} shows the prediction
error cumulative distributions for both schemes. The two weighting
schemes give very similar results and they both appear to be coarse as
they give uncertainties larger than theoretically expected. A deeper
analysis of the data weights is beyond the scope of this paper, but we
plan to address this issue in the future.

\section{Discussion}
Developing a reliable statistical error model for asteroid astrometric
observations is a complicated task. In this paper we give a
significant contribution by computing star position and proper motion
corrections.

The selection of a reference star catalog was not obvious. We selected
the subset of PPMXL corresponding to 2MASS based stars, therefore
inheriting the good accuracy of 2MASS stars and adding proper motion
information. We decided not to use the whole PPMXL catalog because
more than 50\% of its star positions are derived from USNO-B1.0 and
are not as accurate as desirable. The USNO-B1.0 based stars in PPMXL
can also affect asteroid observations reduced with PPMXL. These
observations cannot be corrected unless we select a reference catalog
independent from PPMXL. A possible solution would be that observers
select 2MASS based stars from PPMXL. However, this approach would
result in an inhomogeneous dataset of PPMXL based astrometry, unless
the corresponding observations are flagged with a separate MPC catalog
code.

\citet{teixeira} question the reliability of proper motions in some of
the main astrometric catalogs, including PPMXL. Although we
acknowledge that this problem has to be fixed, our goal is to improve
the current treatment of asteroid astrometry. The tests discussed in
this paper show that including our position and proper motion
corrections provide better predictions and better orbital fit
statistics. As soon as the GAIA star catalog \citep{GAIA} is
available, we will have a much more reliable reference catalog to
refine our debiasing scheme.

Finally, we presented a data weight scheme to update and somehow
generalize that suggested by \citet{cbm10}. In particular, to properly
mitigate the possible effect of correlated observation errors, we now
account for the number of observations present in a single batch and
scale the data weights accordingly. Since this scheme still is quite
coarse, future work will include a detailed statistical analysis of
the observation errors to produce a more accurate data weighting
scheme.

\section*{Acknowledgments}
We are grateful to D.~G. Monet and M. Micheli for several discussions
that helped in improving the paper. We also thank B.~J. Gray, \v Z.
Ivezi\'c, G. Landais, O. Maliy, B.~J. McLean, P.~A. Ries, S. Roeser,
S. Urban, P.~R. Weissman, and G.~V. Williams for providing us with
some of the catalogs and other useful information.

This research made use of the VizieR catalogue access tool, CDS,
Strasbourg, France \citep{vizier}.

Part of this research was conducted at the Jet Propulsion Laboratory,
California Institute of Technology, under a contract with NASA.

Copyright 2014 California Institute of Technology.

\end{document}